\documentclass[twocolumn,rmp,aps,amsmath,amsfonts,noshowkeys,noshowpacs]{revtex4}
\usepackage[linktocpage=true,colorlinks=true,citecolor=blue]{hyperref}

\newcommand{\cf}{cf.\ }
\newcommand{\vs}{vs.\ } 
\newcommand{\ket}[1]{\ensuremath{|{#1\rangle}}} 
\newcommand{\bra}[1]{\ensuremath{{\langle #1}|}}

\newcommand{\ketbra}[2]{\ensuremath{|{#1 \rangle}{\langle #2}|}}

\begin{document}

\makeatletter
\@addtoreset{equation}{section}
\renewcommand{\theequation}{\@arabic\c@section.\@arabic\c@equation}
\makeatother

\title{Decoherence, the measurement problem, and interpretations of
  quantum mechanics}

\author{Maximilian Schlosshauer} 

\email{MAXL@u.washington.edu}

\affiliation{Department of Physics, University of Washington, Seattle,
  Washington 98195, USA}

\begin{abstract}
  Environment-induced decoherence and superselection have been a
  subject of intensive research over the past two decades, yet their
  implications for the foundational problems of quantum mechanics,
  most notably the quantum measurement problem, have remained a matter
  of great controversy. This paper is intended to clarify key features
  of the decoherence program, including its more recent results, and
  to investigate their application and consequences in the context of
  the main interpretive approaches of quantum mechanics.
\end{abstract}



\maketitle

\tableofcontents

\section{Introduction}

The implications of the decoherence program for the foundations of
quantum mechanics have been the subject of an ongoing debate since the
first precise formulation of the program in the early 1980s. The key
idea promoted by decoherence is the insight that realistic quantum
systems are never isolated, but are immersed in the surrounding
environment and interact continuously with it.  The decoherence
program then studies, entirely within the standard quantum formalism
(i.e., without adding any new elements in the mathematical theory or
its interpretation), the resulting formation of quantum correlations
between the states of the system and its environment and the often
surprising effects of these system-environment interactions. In short,
decoherence brings about a local suppression of interference between
preferred states selected by the interaction with the environment.

\citet{Bub:1997:iq} termed decoherence part of the ``new orthodoxy''
of understanding quantum mechanics---as the working physicist's way of
motivating the postulates of quantum mechanics from physical
principles. Proponents of decoherence called it an ``historical
accident'' \citep[p.~13]{Joos:1999:po} that the implications for
quantum mechanics and for the associated foundational problems were
overlooked for so long.  \citet[][p.~717]{Zurek:2002:ii} suggests
\begin{quote}
  {\small The idea that the ``openness'' of quantum systems might have
  anything to do with the transition from quantum to classical was
  ignored for a very long time, probably because in classical physics
  problems of fundamental importance were always settled in isolated
  systems.}
\end{quote}
When the concept of decoherence was first introduced to the broader
scientific community by Zurek's~\citeyearpar{Zurek:1991:vv} article in
\emph{Physics Today}, it elicited a series of contentious comments
from the readership (see the April 1993 issue of \emph{Physics
  Today}). In response to his critics, \citet[p.~718]{Zurek:2002:ii}
states
\begin{quote} {\small 
    In a field where controversy has reigned for so long this
    resistance to a new paradigm [namely, to decoherence] is no
    surprise.}
\end{quote}
\citet[p.~2]{Omnes:2003:tt} had this assessment:
\begin{quote} {\small
    The discovery of decoherence has already much improved our
    understanding of quantum mechanics. (\dots) [B]ut its foundation,
    the range of its validity and its full meaning are still rather
    obscure. This is due most probably to the fact that it deals with
    deep aspects of physics, not yet fully investigated.}
\end{quote}
In particular, the question whether decoherence provides, or at least
suggests, a solution to the measurement problem of quantum mechanics
has been discussed for several years. For example,
\citet[p.~492]{Anderson:2001:rc} writes in an essay review
\begin{quote} {\small 
    The last chapter (\dots) deals with the quantum measurement
    problem (\dots).  My main test, allowing me to bypass the
    extensive discussion, was a quick, unsuccessful search in the
    index for the word ``decoherence'' which describes the process
    that used to be called ``collapse of the wave function.''}
\end{quote} 
Zurek speaks in various places of the ``apparent'' or ``effective''
collapse of the wave function induced by the interaction with
environment (when embedded into a minimal additional interpretive
framework) and concludes \citep[p.~1793]{Zurek:1998:re}
\begin{quote} {\small 
    A ``collapse'' in the traditional sense is no
    longer necessary.  (\dots) [The] emergence of ``objective
    existence'' [from decoherence] (\dots) significantly reduces and
    perhaps even eliminates the role of the ``collapse'' of the state
    vector.
}\end{quote}
D'Espagnat, who considers the explanation of our experiences (i.e., of
``appearances'') as the only ``sure'' requirement of a physical
theory, states \citep[p.~136]{Espagnat:2000:uy}
\begin{quote} {\small For macroscopic systems, the appearances are
    those of a classical world (no interferences etc.), even in
    circumstances, such as those occurring in quantum measurements,
    where quantum effects take place and quantum probabilities
    intervene (\dots).  Decoherence explains the just mentioned
    appearances and this is a most important result. (\dots) As long
    as we remain within the realm of mere predictions concerning what
    we shall observe (i.e., what will appear to us)---and refrain from
    stating anything concerning ``things as they must be before we
    observe them''---no break in the linearity of quantum dynamics is
    necessary.
}\end{quote}
In his monumental book on the foundations of quantum mechanics (QM),
\citet[p.~791]{Auletta:2000:rv} concludes that
\begin{quote} {\small
the Measurement theory could be part of the interpretation of QM only
to the extent that it would still be an open problem, and we think
that this is largely no longer the case.
}\end{quote}
This is mainly so because, according to
\citet[p.~289]{Auletta:2000:rv},
\begin{quote} {\small
decoherence is able to solve practically all the problems of
Measurement which have been discussed in the previous chapters.
}\end{quote}
On the other hand, even leading adherents of decoherence have
expressed caution or even doubt that decoherence has solved the
measurement problem.  \citet[p.~14]{Joos:1999:po} writes
\begin{quote} {\small Does decoherence solve the measurement problem?
    Clearly not. What decoherence tells us, is that certain objects
    appear classical when they are observed. But what is an
    observation? At some stage, we still have to apply the usual
    probability rules of quantum theory.
}\end{quote}
Along these lines, \citet[p.~5]{Kiefer:1998:rz} warn that
\begin{quote} {\small 
    One often finds explicit or implicit statements
    to the effect that the above processes are equivalent to the
    collapse of the wave function (or even solve the measurement
    problem). Such statements are certainly unfounded.
}\end{quote}
In a response to Anderson's \citeyearpar[p.~492]{Anderson:2001:rc}
comment, \citet[p.~136]{Adler:2001:us} states
\begin{quote} {\small
  I do not believe that either detailed theoretical calculations or
  recent experimental results show that decoherence has resolved the
  difficulties associated with quantum measurement theory.
}\end{quote}
Similarly, \citet[p.~3]{Bacciagaluppi:2003:az} writes
\begin{quote} {\small 
  Claims that \emph{simultaneously} the measurement problem is real
  [and] decoherence solves it are confused at best.
}\end{quote}
Zeh asserts \cite[Ch.~2]{Joos:2003:jh}
\begin{quote} {\small
  Decoherence by itself does not yet solve the measurement
  problem (\dots). This argument is nonetheless found wide-spread in
  the literature. (\dots) It does seem that the measurement problem
  can only be resolved if the Schr\"odinger dynamics (\dots) is
  supplemented by a nonunitary collapse (\dots).
}\end{quote}
The key achievements of the decoherence program, apart from their
implications for conceptual problems, do not seem to be universally
understood either. \citet[p.~1800]{Zurek:1998:re} remarks
\begin{quote} {\small
  [The] eventual diagonality of the density matrix (\dots) is a
  byproduct (\dots) but not the essence of decoherence.  I emphasize
  this because diagonality of [the density matrix] in some basis has
  been occasionally (mis-)interpreted as a key accomplishment of
  decoherence. This is misleading. Any density matrix is diagonal in
  some basis. This has little bearing on the interpretation.
}\end{quote}
These remarks show that a balanced discussion of the key features of
decoherence and their implications for the foundations of quantum
mechanics is overdue. The decoherence program has made great progress
over the past decade, and it would be inappropriate to ignore its
relevance in tackling conceptual problems. However, it is equally
important to realize the limitations of decoherence in providing
consistent and noncircular answers to foundational questions.

An excellent review of the decoherence program has recently been given
by \citet{Zurek:2002:ii}. It deals primarily with the technicalities
of decoherence, although it contains some discussion on how
decoherence can be employed in the context of a relative-state
interpretation to motivate basic postulates of quantum mechanics.  A
helpful first orientation and overview, the entry by
\citet{Bacciagaluppi:2003:yz} in the {\em Stanford Encyclopedia of
  Philosophy} features a relatively short (in comparison to the
present paper) introduction to the role of decoherence in the
foundations of quantum mechanics, including comments on the
relationship between decoherence and several popular interpretations
of quantum theory.  In spite of these valuable recent contributions to
the literature, a detailed and self-contained discussion of the role
of decoherence in the foundations of quantum mechanics seems still to
be lacking. This review article is intended to fill the gap.

To set the stage, we shall first, in Sec.~\ref{sec:mp}, review the
measurement problem, which illustrates the key difficulties that are
associated with describing quantum measurement within the quantum
formalism and that are all in some form addressed by the decoherence
program. In Sec.~\ref{sec:decoherence}, we then introduce and discuss
the main features of the theory of decoherence, with a particular
emphasis on their foundational implications.  Finally, in
Sec.~\ref{sec:interpret}, we investigate the role of decoherence in
various interpretive approaches of quantum mechanics, in particular
with respect to the ability to motivate and support (or disprove)
possible solutions to the measurement problem.

\section{\label{sec:mp}The measurement problem}

One of the most revolutionary elements introduced into physical theory
by quantum mechanics is the superposition principle, mathematically
founded in the linearity of the Hilbert state space. If $\ket{1}$ and
$\ket{2}$ are two states, then quantum mechanics tells us that any
linear combination $\alpha \ket{1} + \beta \ket{2}$ also corresponds
to a possible state. Whereas such superpositions of states have been
experimentally extensively verified for microscopic systems (for
instance, through the observation of interference effects), the
application of the formalism to macroscopic systems appears to lead
immediately to severe clashes with our experience of the everyday
world. A book has never been ever observed to be in a state of being
both ``here'' and ``there'' (i.e., to be in a superposition of
macroscopically distinguishable positions), nor does a Schr\"odinger
cat that is a superposition of being alive and dead bear much
resemblence to reality as we perceive it.  The problem is, then, how
to reconcile the vastness of the Hilbert space of possible states with
the observation of a comparatively few ``classical'' macrosopic
states, defined by having a small number of determinate and robust
properties such as position and momentum. Why does the world appear
classical to us, in spite of its supposed underlying quantum nature,
which would, in principle, allow for arbitrary superpositions?

\subsection{Quantum measurement scheme}

This question is usually illustrated in the context of quantum
measurement where microscopic superpositions are, via quantum
entanglement, amplified into the macroscopic realm and thus lead to
very ``nonclassical'' states that do not seem to correspond to what is
actually perceived at the end of the measurement. In the ideal
measurement scheme devised by \citet{Neumann:1932:gq}, a (typically
microscopic) system $\mathcal{S}$, represented by basis vectors $\{
\ket{s_n} \}$ in a Hilbert space $\mathcal{H}_{\mathcal{S}}$,
interacts with a measurement apparatus $\mathcal{A}$, described by
basis vectors $\{ \ket{a_n} \}$ spanning a Hilbert space
$\mathcal{H}_{\mathcal{A}}$, where the $\ket{a_n}$ are assumed to
correspond to macroscopically distinguishable ``pointer'' positions
that correspond to the outcome of a measurement if $\mathcal{S}$ is in
the state $\ket{s_n}$.\footnote{Note that von Neumann's scheme is in
  sharp contrast to the Copenhagen interpretation, where measurement
  is not treated as a system-apparatus interaction described by the
  usual quantum-mechanical formalism, but instead as an independent
  component of the theory, to be represented entirely in fundamentally
  classical terms.}

Now, if $\mathcal{S}$ is in a (microscopically ``unproblematic'')
superposition $\sum_n c_n \ket{s_n}$, and $\mathcal{A}$ is in the
initial ``ready'' state $\ket{a_r}$, the linearity of the
Schr\"odinger equation entails that the total system $\mathcal{SA}$,
assumed to be represented by the Hilbert product space
$\mathcal{H}_{\mathcal{S}} \otimes \mathcal{H}_{\mathcal{A}}$, evolves
according to
\begin{equation}
\label{eq:measurement1} \bigg( \sum_n c_n \ket{s_n}
\bigg) \ket{a_r} \quad \stackrel{t}{\longrightarrow} \quad
\sum_n c_n \ket{s_n} \ket{a_n}.  
\end{equation}
This dynamical evolution is often referred to as a
\emph{premeasurement} in order to emphasize that the process described
by Eq.~\eqref{eq:measurement1} does not suffice to directly conclude
that a measurement has actually been completed.  This is so for two
reasons.  First, the right-hand side is a \emph{superposition} of
system-apparatus states.  Thus, without supplying an additional
physical process (say, some collapse mechanism) or giving a suitable
interpretation of such a superposition, it is not clear how to
account, given the final composite state, for the definite pointer
positions that are perceived as the result of an actual
measurement---i.e., why do we seem to perceive the pointer to be in
one position $\ket{a_n}$ but not in a superposition of positions? This
is the {\em problem of definite outcomes}.  Second, the expansion of
the final composite state is in general not unique, and therefore the
measured observable is not uniquely defined either. This is the
\emph{problem of the preferred basis}. In the literature, the first
difficulty is typically referred to as the measurement problem, but
the preferred-basis problem is at least equally important, since it
does not make sense even to inquire about specific outcomes if the set
of possible outcomes is not clearly defined. We shall therefore regard
the measurement problem as composed of both the problem of definite
outcomes and the problem of the preferred basis, and discuss these
components in more detail in the following.

\subsection{\label{sec:defoutcomes}The problem of definite outcomes}

\subsubsection{Superpositions and ensembles}

The right-hand side of Eq.~\eqref{eq:measurement1} implies that after
the premeasurement the combined system $\mathcal{SA}$ is left in a
pure state that represents a linear superposition of system-pointer
states. It is a well-known and important property of quantum mechanics
that a superposition of states is fundamentally different from a
classical ensemble of states, where the system actually is in only one
of the states but we simply do not know in which (this is often
referred to as an ``ignorance-interpretable,'' or ``proper''
ensemble).

This can be shown explicitely, especially on microscopic scales, by
performing experiments that lead to the direct observation of
interference patterns instead of the realization of one of the terms
in the superposed pure state, for example, in a setup where electrons
pass individually (one at a time) through a double slit. As is
well known, this experiment clearly shows that, within the standard
quantum-mechanical formalism, the electron must not be described by
either one of the wave functions describing the passage through
a particular slit ($\psi_1$ \emph{or} $\psi_2$), but only by the
superposition of these wave functions ($\psi_1+\psi_2$), since the
correct density distribution $\varrho$ of the pattern on the screen is
not given by the sum of the squared wave functions describing the
addition of individual passages through a single slit ($\varrho =
|\psi_1|^2 + |\psi_2|^2$), but only by the square of the sum of the
individual wave functions ($\varrho = |\psi_1 + \psi_2|^2$).

Put differently, if an ensemble interpretation could be attached to a
superposition, the latter would simply represent an ensemble of more
fundamentally determined states, and based on the additional knowledge
brought about by the results of measurements, we could simply choose a
subensemble consisting of the definite pointer state obtained in the
measurement. But then, since the time evolution has been strictly
deterministic according to the Schr\"odinger equation, we could
backtrack this subensemble in time and thus also specify the
initial state more completely (``postselection''), and
therefore this state necessarily could not be physically identical to
the initially prepared state on the left-hand side of
Eq.~\eqref{eq:measurement1}.

\subsubsection{Superpositions and outcome attribution}

In the standard (``orthodox'') interpretation of quantum mechanics, an
observable corresponding to a physical quantity has a definite value
if and only if the system is in an eigenstate of the observable; if
the system is, however, in a superposition of such eigenstates, as in
Eq.~\eqref{eq:measurement1}, it is, according to the orthodox
interpretation, meaningless to speak of the state of the system
as having any definite value of the observable at all.  (This is
frequently referred to as the so-called eigenvalue-eigenstate
link, or ``{e-e} link'' for short.) The {e-e} link, however, is by
no means forced upon us by the structure of quantum mechanics or by
empirical constraints \citep{Bub:1997:iq}.  The concept of (classical)
``values'' that can be ascribed through the e-e link based on
observables and the existence of exact eigenstates of these
observables has therefore frequently been either weakened or
altogether abandonded.  For instance, outcomes of measurements are
typically registered in position space (pointer positions, etc.), but
there exist no exact eigenstates of the position operator, and the
pointer states are never exactly mutually orthogonal. One might then
(explicitely or implicitly) promote a ``fuzzy'' e-e link, or give up
the concept of observables and values entirely and directly interpret
the time-evolved wave functions (working in the Schr\"odinger picture)
and the corresponding density matrices.  Also, if it is regarded as
sufficient to explain our perceptions rather than describe the
``absolute'' state of the entire universe (see the argument below),
one might only require that the (exact or fuzzy) e-e link hold in a
``relative'' sense, i.e., for the state of the rest of the universe
relative to the state of the observer.

Then, to solve the problem of definite outcomes, some interpretations
(for example, modal interpretations and relative-state
interpretations) interpret the final-state superposition in such a way
as to explain the existence, or at least the subjective perception, of
``outcomes'' even if the final composite state has the form of a
superposition. Other interpretations attempt to solve the measurement
problem by modifying the strictly unitary Schr\"odinger dynamics. Most
prominently, the orthodox interpretation postulates a collapse
mechanism that transforms a pure-state density matrix into an
ignorance-interpretable ensemble of individual states (a ``proper
mixture'').  Wave-function collapse theories add stochastic terms to
the Schr\"odinger equation that induce an effective (albeit only
approximate) collapse for states of macroscopic systems
\citep{Pearle:1979:rq,Gisin:1984:qs,Ghirardi:1986:ud,Pearle:1999:cr},
while other authors suggested that collapse occurs at the level of the
mind of a conscious observer \citep{Wigner:1963:yt,Stapp:1993:mm}.
Bohmian mechanics, on the other hand, upholds a unitary time evolution
of the wavefunction, but introduces an additional dynamical law that
explicitely governs the always-determinate positions of all particles
in the system.

\subsubsection{\label{sec:objsubj}Objective \vs subjective definiteness}

In general, (macroscopic) definiteness---and thus a solution to the
problem of outcomes in the theory of quantum measurement---can be
achieved either on an \emph{ontological} (objective) or an
\emph{observational} (subjective) level.  Objective definiteness aims
at ensuring ``actual'' definiteness in the macroscopic realm, whereas
subjective definiteness only attempts to explain why the macroscopic
world appears to be definite---and thus does not make any
claims about definiteness of the underlying physical reality (whatever
this reality might be). This raises the question of the significance
of this distinction with respect to the formation of a satisfactory
theory of the physical world. It might appear that a solution to the
measurement problem based on ensuring subjective, but not objective,
definiteness is merely good ``for all practical
purposes''---abbreviated, rather disparagingly, as ``FAPP'' by
\citet{Bell:1990:po}---and thus not capable of solving the
``fundamental'' problem that would seem relevant to the construction
of the ``precise theory'' that Bell demanded so vehemently.

It seems to the author, however, that this critism is not justified,
and that subjective definiteness should be viewed on a par with
objective definitess with respect to a satisfactory solution to the
measurement problem. We demand objective definiteness because we
experience definiteness on the subjective level of observation, and it
should not be viewed as an \emph{a priori} requirement for a physical
theory.  If we knew independently of our experience that definiteness
existed in nature, subjective definiteness would presumably follow as
soon as we had employed a simple model that connected the ``external''
physical phenomena with our ``internal'' perceptual and cognitive
apparatus, where the expected simplicity of such a model can be
justified by referring to the presumed identity of the physical laws
governing external and internal processes.  But since knowledge is
based on experience, that is, on observation, the existence of
objective definiteness could only be derived from the observation of
definiteness. And, moreover, observation tells us that definiteness is
in fact not a universal property of nature, but rather a property of
macroscopic objects, where the borderline to the macroscopic realm is
difficult to draw precisely; mesoscopic interference experiments have
demonstrated clearly the blurriness of the boundary. Given the lack of
a precise definition of the boundary, any demand for fundamental
definiteness on the objective level should be based on a much deeper
and more general commitment to a definiteness that applies to
every physical entity (or system) across the board, regardless
of spatial size, physical property, and the like.

Therefore, if we realize that the often deeply felt commitment to a
general objective definiteness is only based on our experience of
macroscopic systems, and that this definiteness in fact fails in an
observable manner for microscopic and even certain mesoscopic systems,
the author sees no compelling grounds on which objective definiteness
must be demanded as part of a satisfactory physical theory, provided
that the theory can account for subjective, observational definiteness
in agreement with our experience. Thus the author suggests that the
same legitimacy be attributed to proposals for a solution of the
measurement problem that achieve ``only'' subjective but not objective
definiteness---after all, the measurement problem arises solely from a
clash of our experience with certain implications of the quantum
formalism.  D'Espagnat \citeyearpar[pp.~134--135]{Espagnat:2000:uy}
has advocated a similar viewpoint:
\begin{quote} {\small 
    The fact that we perceive such ``things'' as macroscopic objects
    lying at distinct places is due, partly at least, to the structure
    of our sensory and intellectual equipment.  We should not,
    therefore, take it as being part of the body of sure knowledge
    that we have to take into account for defining a quantum state.
    (\dots) In fact, scientists most righly claim that the purpose of
    science is to describe human experience, not to describe ``what
    really is''; and as long as we only want to describe human
    experience, that is, as long as we are content with being able to
    predict what will be observed in all possible circumstances
    (\dots) we need not postulate the existence---in some absolute
    sense---of unobserved (i.e., not yet observed) objects lying at
    definite places in ordinary 3-dimensional space.}
\end{quote}

\subsection{The preferred-basis problem \label{sec:pbprob}}

The second difficulty associated with quantum measurement is known as
the preferred-basis problem, which demonstrates that the measured
observable is in general not uniquely defined by
Eq.~\eqref{eq:measurement1}. For any choice of system states $\{
\ket{s_n} \}$, we can find corresponding apparatus states $\{
\ket{a_n} \}$, and vice versa, to equivalently rewrite the final state
emerging from the premeasurement interaction, i.e., the right-hand
side of Eq.~\eqref{eq:measurement1}.  In general, however, for some
choice of apparatus states the corresponding new system states will
not be mutually orthogonal, so that the observable associated with
these states will not be Hermitian, which is usually not desired
\citep[however, not forbidden---see the discussion
by][]{Zurek:2002:ii}.  Conversely, to ensure distinguishable outcomes,
we must, in general, require the (at least approximate) orthogonality
of the apparatus (pointer) states, and it then follows from the
biorthogonal decomposition theorem that the expansion of the final
premeasurement system-apparatus state of Eq.~\eqref{eq:measurement1},
\begin{equation}
\ket{\psi} = \sum_n c_n \ket{s_n}
\ket{a_n}, 
\end{equation}
is unique, but only if all coefficients $c_n$ are distinct. Otherwise, we
can in general rewrite the state in terms of different state vectors,
\begin{equation} 
\ket{\psi} = \sum_n c_n^\prime \ket{s^\prime_n}
\ket{a^\prime_n}, 
\end{equation}
such that the same postmeasurement state seems to correspond to two
different measurements, that is, of the observables $\widehat{A} =
\sum_n \lambda_n \ket{s_n} \bra{s_n}$ and
$\widehat{B} = \sum_n \lambda_n^\prime \ket{s^\prime_n}
\bra{s^\prime_n}$ of the system, respectively,
although in general $\widehat{A}$ and $\widehat{B}$ do not
commute.

As an example, consider a Hilbert space
$\mathcal{H}=\mathcal{H}_1\otimes \mathcal{H}_2$ where $\mathcal{H}_1$
and $\mathcal{H}_2$ are two-dimensional spin spaces with states
corresponding to spin up or spin down along a given axis. Suppose we
are given an entangled spin state of the Einstein-Podolsky-Rosen form
\cite{Einstein:1935:dr}
\begin{equation} \label{eq:spin1} 
\ket{\psi} = \frac{1}{\sqrt{2}} (\ket{z+}_1
\ket{z-}_2 - \ket{z-}_1 \ket{z+}_2), 
\end{equation}
where $\ket{z\pm}_{1,2}$ represents the eigenstates of the observable
$\sigma_z$ corresponding to spin up or spin down along the $z$ axis of
the two systems 1 and 2. The state $\ket{\psi}$ can however
equivalently be expressed in the spin basis corresponding to any other
orientation in space. For example, when using the eigenstates
$\ket{x\pm}_{1,2}$ of the observable $\sigma_x$ (which represents a
measurement of the spin orientation along the $x$ axis) as basis
vectors, we get
\begin{equation} \label{eq:spin2} 
\ket{\psi} = \frac{1}{\sqrt{2}} (\ket{x+}_1
\ket{x-}_2 - \ket{x-}_1 \ket{x+}_2). 
\end{equation}
Now suppose that system~2 acts as a measuring device for the spin of
system~1.  Then Eqs.~\eqref{eq:spin1} and \eqref{eq:spin2} imply that
the measuring device has established a correlation with both the $z$
and the $x$ spin of system~1. This means that, if we interpret the
formation of such a correlation as a measurement in the spirit of the
von Neumann scheme (without assuming a collapse), our apparatus
(system~2) could be considered as having measured also the $x$ spin
once it has measured the $z$ spin, and vice versa---in spite of the
noncommutativity of the corresponding spin observables $\sigma_z$ and
$\sigma_x$. Moreover, since we can rewrite Eq.~\eqref{eq:spin1} in
infinitely many ways, it appears that once the apparatus has measured
the spin of system~1 along one direction, it can also be regarded as
having measured the spin along any other direction, again in apparent
contradiction with quantum mechanics due to the noncommutativity of
the spin observables corresponding to different spatial orientations.

It thus seems that quantum mechanics has nothing to say about which
observable(s) of the system is (are) being recorded, via the formation
of quantum correlations, by the apparatus. This can be stated in a
general theorem \citep{Zurek:1981:dd,Auletta:2000:rv}: When quantum
mechanics is applied to an isolated composite object consisting
of a system $\mathcal{S}$ and an apparatus $\mathcal{A}$, it cannot
determine which observable of the system has been measured---in
obvious contrast to our experience of the workings of measuring
devices that seem to be ``designed'' to measure certain quantities.

\subsection{The quantum-to-classical transition and decoherence}

In essence, as we have seen above, the measurement problem deals with
the transition from a quantum world, described by essentially
arbitrary linear superpositions of state vectors, to our perception of
``classical'' states in the macroscopic world, that is, a
comparatively small subset of the states allowed by the quantum-mechanical
superposition principle, having only a few, but determinate and robust,
properties, such as position, momentum, etc. The question of why and
how our experience of a ``classical'' world emerges from quantum
mechanics thus lies at the heart of the foundational problems of
quantum theory.

Decoherence has been claimed to provide an explanation for this
\emph{quantum-to-classical transition} by appealing to the ubiquitous
immersion of virtually all physical systems in their environment
(``environmental monitoring'').  This trend can also be read off
nicely from the titles of some papers and books on decoherence, for
example, ``The emergence of classical properties through interaction
with the environment'' \citep{Joos:1985:iu}, ``Decoherence and the
transition from quantum to classical'' \citep{Zurek:1991:vv}, and
``Decoherence and the appearance of a classical world in quantum
theory'' \citep{Joos:2003:jh}. We shall critically investigate in this
paper to what extent the appeal to decoherence for an explanation of
the quantum-to-classical transition is justified.

\section{\label{sec:decoherence}The decoherence program}

As remarked earlier, the theory of decoherence is based on a study of
the effects brought about by the interaction of physical systems with
their environment. In classical physics, the environment is usually
viewed as a kind of disturbance, or noise, that perturbs the system
under consideration in such a way as to negatively influence the study
of its ``objective'' properties. Therefore science has established the
idealization of isolated systems, with experimental physics aiming at
eliminating any outer sources of disturbance as much as possible in
order to discover the ``true'' underlying nature of the system under
study.

The distinctly nonclassical phenomenon of quantum entanglement,
however, has demonstrated that the correlations between two systems
can be of fundamental importance and can lead to properties that are
not present in the individual systems.\footnote{Broadly speaking,
  this means that the (quantum-mechanical) whole is different
  from the sum of its parts.} The earlier view of phenomena arising
from quantum entanglement as ``paradoxa'' has generally been replaced
by the recognition of entanglement as a fundamental property of
nature.

The decoherence program\footnote{For key ideas and concepts, see
  \citet{Zeh:1970:yt,Kubler:1973:ux,Zeh:1973:wq,Zeh:1995:jg,%
Zeh:1996:gy,Zeh:1999:qr,Joos:1985:iu,Zurek:1981:dd,Zurek:1982:tv,%
Zurek:1991:vv,Zurek:1993:pu,Zurek:2002:ii,Joos:2003:jh}.}  is based on
the idea that such quantum correlations are ubiquitous; that nearly
every physical system must interact in some way with its environment
(for example, with the surrounding photons that then create the visual
experience within the observer), which typically consists of a large
number of degrees of freedom that are hardly ever fully controlled.
Only in very special cases of typically microscopic (atomic)
phenomena, so goes the claim of the decoherence program, is the
idealization of isolated systems applicable so that the predictions of
linear quantum mechanics (i.e., a large class of superpositions of
states) can actually be observationally confirmed. In the majority of
the cases accessible to our experience, however, interaction with the
environment is so dominant as to preclude the observation of the
``pure'' quantum world, imposing effective superselection rules
\citep{Wick:1952:pp,Wick:1970:iz,Galindo:1962:tl,Wightman:1995:ug,Cisnerosy:1998:kz,Giulini:2000:ry}
onto the space of observable states that lead to states corresponding
to the ``classical'' properties of our experience. Interference
between such states gets locally suppressed and is thus claimed to
become inaccessible to the observer.

Probably the most surprising aspect of decoherence is the
effectiveness of the system-environment interactions. Decoherence
typically takes place on extremely short time scales and requires the
presence of only a minimal environment \citep{Joos:1985:iu}. Due to
the large number of degrees of freedom of the environment, it is
usually very difficult to undo system-environment entanglement, which
has been claimed as a source of our impression of irreversibility in
nature \citep[see, for
example,][]{Zurek:1982:tv,Zurek:2002:ii,Zurek:1994:om,Kiefer:1998:rz,Zeh:2001:tt}.
In general, the effect of decoherence increases with the size of the
system (from microscopic to macroscopic scales), but it is important
to note that there exist, admittedly somewhat exotic, examples for
which the decohering influence of the environment can be sufficiently
shielded to lead to mesoscopic and even macroscopic superpositions.
One such example would be the case of superconducting quantum
interference devices (SQUIDs), in which superpositions of macroscopic
currents become observable \cite{Friedman:2000:rr,Wal:2000:om}.
Conversely, some microscopic systems (for instance, certain chiral
molecules that exist in different distinct spatial configurations) can
be subject to remarkably strong decoherence.

The decoherence program has dealt with the following two main
consequences of environmental interaction:
\begin{enumerate}
  
\item[(1)] {\em Environment-induced decoherence.} The fast local
  suppression of interference between different states of the system.
  However, since only unitary time evolution is employed, global phase
  coherence is not actually destroyed---it becomes absent from the
  local density matrix that describes the system alone, but remains
  fully present in the total system-environment
  composition.\footnote{Note that the persistence of coherence in the
    total state is important to ensure the possibility of describing
    special cases in which mesoscopic or macrosopic superpositions have
    been experimentally realized.} We shall discuss
  environment-induced local decoherence in more detail in
  Sec.~\ref{sec:interf}.
  
\item[(2)] {\em Environment-induced superselection.} The selection of
  preferred sets of states, often referred to as ``pointer states,''
  that are robust (in the sense of retaining correlations over time)
  in spite of their immersion in the environment. These states are
  determined by the form of the interaction between the system and its
  environment and are suggested to correspond to the ``classical''
  states of our experience. We shall consider this mechanism in
  Sec.~\ref{sec:einsel}.

\end{enumerate}
Another, more recent aspect of the decoherence program, termed
\emph{enviroment-assisted invariance} or ``envariance,'' was
introduced by \citet{Zurek:2002:ii,Zurek:2003:rv,Zurek:2003:pl} and
further developed in \citet{Zurek:2004:yb}.  In particular, Zurek used
envariance to explain the emergence of probabilities in quantum
mechanics and to derive Born's rule based on certain assumptions. We
shall review envariance and Zurek's derivation of the Born rule in
Sec.~\ref{sec:envar}.

Finally, let us emphasize that decoherence arises from a direct
application of the quantum mechanical formalism to a description of
the interaction of a physical system with its environment. By itself,
decoherence is therefore neither an interpretation nor a modification
of quantum mechanics.  Yet the implications of decoherence need to be
interpreted in the context of the different interpretations of quantum
mechanics.  Also, since decoherence effects have been studied
extensively in both theoretical models and experiments \citep[for a
survey, see, for example,][]{Joos:2003:jh,Zurek:2002:ii}, their
existence can be taken as a well-confirmed fact.

\subsection{Resolution into subsystems\label{sec:division}}

Note that decoherence derives from the presupposition of the existence
and the possibility of a division of the world into ``system(s)'' and
``environment.''  In the decoherence program, the term ``environment''
is usually understood as the ``remainder'' of the system, in the sense
that its degrees of freedom are typically not (cannot be, do not need
to be) controlled and are not directly relevant to the observation
under consideration (for example, the many microsopic degrees of
freedom of the system), but that nonetheless the environment includes
``all those degrees of freedom which contribute significantly to the
evolution of the state of the apparatus''
\citep[][p.~1520]{Zurek:1981:dd}.

This system--environment dualism is generally associated with quantum
entanglement, which always describes a correlation between parts of
the universe. As long as the universe is not resolved into individual
subsystems, there is no measurement problem: the state vector
$\ket{\Psi}$ of the entire universe\footnote{If we dare to postulate
  this total state---see counterarguments by \citet{Auletta:2000:rv}.}
evolves deterministically according to the Schr\"odinger equation
$i\hbar \frac{\partial}{\partial t} \ket{\Psi} = \widehat{H}
\ket{\Psi}$, which poses no interpretive difficulty. Only when we
decompose the total Hilbert-state space $\mathcal{H}$ of the universe
into a product of two spaces $\mathcal{H}_1 \otimes \mathcal{H}_2$,
and accordingly form the joint-state vector
$\ket{\Psi}=\ket{\Psi_1}\ket{\Psi_2}$, and want to ascribe an
individual state (besides the joint state that describes a
correlation) to one of the two systems (say, the apparatus), does the
measurement problem arise.  \citet[p.~718]{Zurek:2002:ii} puts it
like this:
\begin{quote} {\small
    In the absence of systems, the problem of interpretation seems to
    disappear.  There is simply no need for ``collapse'' in a universe
    with no systems. Our experience of the classical reality does not
    apply to the universe as a whole, seen from the outside, but to
    the systems within it.}
\end{quote}
Moreover, terms like ``observation,'' ``correlation,'' and
``interaction'' will naturally make little sense without a division
into systems. Zeh has suggested that the locality of the observer
defines an observation in the sense that any observation arises from
the ignorance of a part of the universe; and that this also defines
the ``facts'' that can occur in a quantum system.
\citet[pp.~45--46]{Landsman:1995:oi} argues similarly:
\begin{quote} {\small 
    The essence of a ``measurement,'' ``fact'' or ``event'' in quantum
    mechanics lies in the non-observation, or irrelevance, of a
    certain part of the system in question. (\dots) A world without
    parts declared or forced to be irrelevant is a world without
    facts.}
\end{quote}
However, the assumption of a decomposition of the universe into
subsystems---as necessary as it appears to be for the emergence of the
measurement problem and for the definition of the decoherence
program---is definitely nontrivial. By definition, the universe as a
whole is a closed system, and therefore there are no ``unobserved
degrees of freedom'' of an external environment which would allow for
the application of the theory of decoherence to determine the space of
quasiclassical observables of the universe in its entirety.  Also,
there exists no general criterion for how the total Hilbert space is
to be divided into subsystems, while at the same time much of what is
called a property of the system will depend on its correlation with
other systems. This problem becomes particularly acute if one would
like decoherence not only to motivate explanations for the subjective
perception of classicality \citep[as in Zurek's ``existential
interpretation,'' see][and Sec.~\ref{sec:everett}
below]{Zurek:1993:pu,Zurek:1998:re,Zurek:2002:ii}, but moreover to
allow for the definition of quasiclassical ``macrofacts.''
\citet[p.~1820]{Zurek:1998:re} admits this severe conceptual
difficulty:
\begin{quote} {\small
    In particular, one issue which has been often taken for granted is
    looming big, as a foundation of the whole decoherence program. It
    is the question of what are the ``systems'' which play such a
    crucial role in all the discussions of the emergent classicality.
    (\dots) [A] compelling explanation of what are the
    systems---how to define them given, say, the overall Hamiltonian
    in some suitably large Hilbert space---would be undoubtedly most
    useful.}
\end{quote}
A frequently proposed idea is to abandon the notion of an ``absolute''
resolution and instead postulate the intrinsic relativity of the
distinct state spaces and properties that emerge through the
correlation between these relatively defined spaces \citetext{see, for
  example, the proposals, unrelated to decoherence, of
  \citealp{Everett:1957:rw,Mermin:1998:ii,Mermin:1998:wi}; and
  \citealp{Rovelli:1996:rq}}.  This relative view of systems and
correlations has counterintuitive, in the sense of nonclassical,
implications. However, as in the case of quantum entanglement, these
implications need not be taken as paradoxa that demand further
resolution.  Accepting some properties of nature as counterintuitive
is indeed a satisfactory path to take in order to arrive at a
description of nature that is as complete and objective as is allowed
by the range of our experience (which is based on inherently local
observations).

\subsection{The concept of reduced density matrices \label{sec:redmat}}

Since reduced density matrices are a key tool of decoherence, it will
be worthwile to briefly review their basic properties and
interpretation in the following.  The concept of reduced density
matrices emerged in the earliest days of quantum mechanics
\citetext{\citealp{Landau:1927:uy,Neumann:1932:gq,Furry:1936:pp}; for
  some historical remarks, see \citealp{Pessoa:1998:yl}}. In the
context of a system of two entangled systems in a pure state of the
Einstein-Podolsky-Rosen-type,
\begin{equation} \label{eq:epr}
\ket{\psi} = \frac{1}{\sqrt{2}}(\ket{+}_1\ket{-}_2 -
\ket{-}_1\ket{+}_2),
\end{equation}
it had been realized early that for an observable $\widehat{O}$ that
pertains only to system 1, $\widehat{O}=\widehat{O}_1 \otimes
\widehat{I}_2$, the pure-state density matrix $\rho =
\ket{\psi}\bra{\psi}$ yields, according to the trace rule $\langle
\widehat{O} \rangle = \text{Tr} (\rho \widehat{O})$ and given the
usual Born rule for calculating probabilities, exactly the same
statistics as the reduced density matrix $\rho_1$ obtained by
tracing over the degrees of freedom of system 2 (i.e., the states
$\ket{+}_2$ and $\ket{-}_2$),
\begin{equation} \label{eq:epr-rho}
\rho_1 = \text{Tr}_2 \ket{\psi}\bra{\psi} = {_2\langle +} | \psi 
\rangle \langle \psi | + \rangle_2 + {_2\langle -} | \psi 
\rangle \langle \psi | - \rangle_2,
\end{equation}
since it is easy to show that, for this observable $\widehat{O}$,
\begin{equation}
\langle \widehat{O} \rangle_{\psi} = \text{Tr} (\rho
\widehat{O}) = \text{Tr}_1 (\rho_1 \widehat{O}_1). 
\end{equation}
This result holds in general for any pure state $\ket{\psi} = \sum_i
\alpha_i \ket{\phi_i}_1 \ket{\phi_i}_2 \cdots \ket{\phi_i}_N$ of a
resolution of a system into $N$ subsystems, where the $\{
\ket{\phi_i}_j \}$ are assumed to form orthonormal basis sets in their
respective Hilbert spaces $\mathcal{H}_j$, $j=1 \cdots N$.  For any
observable $\widehat{O}$ that pertains only to system $j$,
$\widehat{O}=\widehat{I}_1 \otimes \widehat{I}_2 \otimes \cdots
\otimes \widehat{I}_{j-1} \otimes \widehat{O}_j \otimes
\widehat{I}_{j+1} \otimes \cdots \otimes \widehat{I}_N$, the
statistics of $\widehat{O}$ generated by applying the trace rule will
be identical regardless of whether we use the pure-state density matrix
$\rho = \ket{\psi} \bra{\psi}$ or the reduced density matrix $\rho_j =
\text{Tr}_{1, \hdots, j-1, j+1, \hdots, N} \ket{\psi} \bra{\psi}$,
since again $\langle \widehat{O} \rangle = \text{Tr} (\rho
\widehat{O}) = \text{Tr}_j (\rho_j \widehat{O}_j)$.

The typical situation in which the reduced density matrix arises is
this: Before a premeasurement-type interaction, the observer knows
that each individual system is in some (unknown) pure state.  After
the interaction, i.e., after the correlation between the systems is
established, the observer has access to only one of the systems, say,
system 1; everything that can be known about the state of the
composite system must therefore be derived from measurements on system
1, which will yield the possible outcomes of system 1 and their
probability distribution. All information that can be extracted by the
observer is then, exhaustively and correctly, contained in the reduced
density matrix of system 1, assuming that the Born rule for quantum
probabilities holds.

Let us return to the Einstein-Podolsky-Rosen-type example,
Eqs.~\eqref{eq:epr} and \eqref{eq:epr-rho}. If we assume that the
states of system 2 are orthogonal, ${_2\langle +} | - \rangle_2 = 0$,
$\rho_1$ becomes diagonal,
\begin{equation}
\rho_1 = \text{Tr}_2 \ket{\psi}\bra{\psi} = \frac{1}{2} (\ket{+}
\bra{+})_1 +  \frac{1}{2}(\ket{-} \bra{-})_1.
\end{equation}
But this density matrix is formally identical to the density matrix
that would be obtained if system 1 were in a mixed state, i.e., in
either one of the two states $\ket{+}_1$ and $\ket{-}_1$ with equal
probabilties---as opposed to the superposition $\ket{\psi}$, in which
both terms are considered present, which could in principle be
confirmed by suitable interference experiments.  This implies that a
measurement of an observable that only pertains to system 1 cannot
discriminate between the two cases, pure vs mixed state.\footnote{As
  discussed by \citet[pp.~208--210]{Bub:1997:iq}, this result also
  holds for any observable of the composite system that factorizes
  into the form $\widehat{O}=\widehat{O}_1 \otimes \widehat{O}_2$,
  where $\widehat{O}_1$ and $\widehat{O}_2$ do not commute with the
  projection operators $(\ket{\pm}\bra{\pm})_1$ and
  $(\ket{\pm}\bra{\pm})_2$, respectively.}

However, note that the formal identification of the reduced density
matrix with a mixed-state density matrix is easily misinterpreted as
implying that the state of the system can be viewed as mixed too
\citep[see also the discussion by][]{Espagnat:1988:cf}.  Density
matrices are only a calculational tool for computing the probability
distribution of a set of possible outcomes of measurements; they do
not specify the state of the system.\footnote{In this context we note
  that any nonpure density matrix can be written in many different
  ways, demonstrating that any partition in a particular ensemble of
  quantum states is arbitrary.}  Since the two systems are entangled
and the total composite system is still described by a superposition,
it follows from the standard rules of quantum mechanics that no
individual definite state can be attributed to one of the systems. The
reduced density matrix looks like a mixed-state density matrix
because, if one actually measured an observable of the system, one
would expect to get a definite outcome with a certain probability; in
terms of measurement statistics, this is equivalent to the situation
in which the system is in one of the states from the set of possible
outcomes from the beginning, that is, before the measurement.  As
\citet[p.~432]{Pessoa:1998:yl} puts it, ``taking a partial trace
amounts to the statistical version of the projection postulate.''

\subsection{A modified von Neumann measurement scheme}

Let us now reconsider the von Neumann model for ideal
quantum-mechanical measurement, Eq.~\eqref{eq:measurement2}, but now
with the environment included. We shall denote the environment by
$\mathcal{E}$ and represent its state before the measurement
interaction by the initial state vector $\ket{e_0}$ in a Hilbert space
$\mathcal{H}_{\mathcal{E}}$. As usual, let us assume that the state
space of the composite object system-apparatus-environment is given
by the tensor product of the individual Hilbert spaces,
$\mathcal{H}_{\mathcal{S}} \otimes \mathcal{H}_{\mathcal{A}} \otimes
\mathcal{H}_{\mathcal{E}}$. The linearity of the Schr\"odinger
equation then yields the following time evolution of the entire system
$\mathcal{SAE}$,
\begin{eqnarray} \label{eq:measurement2} 
\bigg( \sum_n c_n \ket{s_n}
\bigg) \ket{a_r}\ket{e_0} 
\, &\stackrel{(1)}{\longrightarrow}& \, \bigg( \sum_n c_n
\ket{s_n} \ket{a_n} \bigg) \ket{e_0} \nonumber \\
\, & \stackrel{(2)}{\longrightarrow} & \, \sum_n c_n \ket{s_n}
\ket{a_n} \ket{e_n}, 
\end{eqnarray}
where the $\ket{e_n}$ are the states of the environment associated
with the different pointer states $\ket{a_n}$ of the measuring
apparatus. Note that while for two subsystems, say, $\mathcal{S}$ and
$\mathcal{A}$, there always exists a diagonal (``Schmidt'')
decomposition of the final state of the form $\sum_n c_n \ket{s_n}
\ket{a_n}$, for three subsystems (for example, $\mathcal{S}$,
$\mathcal{A}$, and $\mathcal{E}$), a decomposition of the form $\sum_n
c_n \ket{s_n} \ket{a_n} \ket{e_n}$ is not always possible. This
implies that the total Hamiltonian that induces a time evolution of
the above kind, Eq.~\eqref{eq:measurement2}, must be of a special
form.\footnote{For an example of such a Hamiltonian, see the model of
  \citet{Zurek:1981:dd,Zurek:1982:tv} and its outline in
  Sec.~\ref{sec:zurekmodel} below. For a critical comment regarding
  limitations on the form of the evolution operator and the
  possibility of a resulting disagreement with experimental evidence,
  see \citet{Pessoa:1998:yl}.}

Typically, the $\ket{e_n}$ will be product states of many microsopic
subsystem states $\ket{\varepsilon_n}_i$ corresponding to the
individual parts that form the environment, i.e.,
$\ket{e_n}=\ket{\varepsilon_n}_1 \ket{\varepsilon_n}_2
\ket{\varepsilon_n}_3 \cdots$. We see that a nonseparable and in most
cases, for all practical purposes, irreversible (due to the enormous
number of degrees of freedom of the environment) correlation has been
established between the states of the system--apparatus combination
$\mathcal{SA}$ and the different states of the environment
$\mathcal{E}$. Note that Eq.~\eqref{eq:measurement2} also implies that
the environment has recorded the state of the system---and,
equivalently, the state of the system-apparatus composition. The
environment, composed of many subsystems, thus acts as an amplifying,
higher-order measuring device.

\subsection{Decoherence and local suppression of interference \label{sec:interf}}

Interaction with the environment typically leads to a rapid vanishing
of the diagonal terms in the local density matrix describing the
probability distribution for the outcomes of measurements on the
system. This effect has become known as environment-induced
decoherence, and it has also frequently been claimed to imply at
least a partial solution to the measurement problem.

\subsubsection{General formalism}

In Sec.~\ref{sec:redmat}, we have already introduced the concept of
local (or reduced) density matrices and pointed out some caveats on
their interpretation.  In the context of the decoherence program,
reduced density matrices arise as follows. Any observation will
typically be restricted to the system-apparatus component,
$\mathcal{SA}$, while the many degrees of freedom of the environment
$\mathcal{E}$ remain unobserved. Of course, typically some degrees of
freedom of the environment will always be included in our observation
(e.g., some of the photons scattered off the apparatus) and we shall
accordingly include them in the ``observed part $\mathcal{SA}$ of the
universe.''  The crucial point is that there still remains a
comparatively large number of environmental degrees of freedom that will
not be observed directly.

Suppose then that the operator $\widehat{O}_\mathcal{SA}$ represents
an observable of $\mathcal{SA}$ only.  Its expectation value $\langle
\widehat{O}_\mathcal{SA} \rangle $ is given by
\begin{equation} \label{eq:global-to-local} 
\langle
\widehat{O}_{\mathcal{SA}} \rangle = \text{Tr}
(\widehat{\rho}_\mathcal{SAE}
[\widehat{O}_\mathcal{SA} \otimes
\widehat{I}_\mathcal{E}]) = \text{Tr}_\mathcal{SA}
(\widehat{\rho}_\mathcal{SA}
\widehat{O}_\mathcal{SA}), 
\end{equation}
where the density matrix $\widehat{\rho}_{\mathcal{SAE}}$ of the total
$\mathcal{SAE}$ combination,
\begin{equation} \widehat{\rho}_{\mathcal{SAE}} = \sum_{mn} c_m c_n^*
\ket{s_m} \ket{a_m} \ket{e_m}
\bra{s_n} \bra{a_n} \bra{e_n}, \end{equation}
has, for all practical purposes of statistical prediction, been
replaced by the local (or reduced) density matrix
$\widehat{\rho}_{\mathcal{SA}}$, obtained by ``tracing out
the unobserved degrees of the environment,'' that is,
\begin{equation} \label{eq:rho-S+A} \widehat{\rho}_\mathcal{SA} =
\mathrm{Tr}_\mathcal{E} (\widehat{\rho}_\mathcal{SAE})
= \sum_{mn} c_m c_n^* \ket{s_m} \ket{a_m} \bra{s_n}\bra{a_n}
\langle e_n | e_m \rangle.  
\end{equation}
So far, $\widehat{\rho}_{\mathcal{SA}}$ contains characteristic
interference terms $\ket{s_m} \ket{a_m} \bra{s_n}\bra{a_n}$, $m \not=
n$, since we cannot assume from the outset that the basis vectors
$\ket{e_m}$ of the environment are necessarily mutually orthogonal,
i.e., that $\langle e_n | e_m \rangle = 0$ if $m \not= n$. Many
explicit physical models for the interaction of a system with the
environment (see Sec.~\ref{sec:zurekmodel} below for a simple
example), however, have shown that due to the large number of
subsystems that compose the environment, the pointer states
$\ket{e_n}$ of the environment rapidly approach orthogonality,
$\langle e_n | e_m \rangle (t) \rightarrow \delta_{n,m}$, such that
the reduced density matrix $\widehat{\rho}_{\mathcal{SA}}$ becomes
approximately orthogonal in the ``pointer basis'' $\{ \ket{a_n} \}$;
that is,
\begin{eqnarray} \label{eq:rho-S+A-diagonal} 
\widehat{\rho}_{\mathcal{SA}} \,
\stackrel{t}{\longrightarrow} \, \widehat{\rho}^{\,\,d}_{\mathcal{SA}}
& \approx &
\sum_n |c_n|^2 \ket{s_n} \ket{a_n}
\bra{s_n} \bra{a_n} \nonumber \\ &=& \sum_n |c_n|^2
\widehat{P}^{(\mathcal{S})}_n \otimes \widehat{P}^{(\mathcal{A})}_n. 
\end{eqnarray}
Here, $\widehat{P}^{(\mathcal{S})}_n$ and
$\widehat{P}^{(\mathcal{A})}_n$ are the projection operators onto the
eigenstates of $\mathcal{S}$ and $\mathcal{A}$, respectively.
Therefore the interference terms have vanished in this local
representation, i.e., phase coherence has been locally lost. This is
precisely the effect referred to as environment-induced decoherence.
The decohered local density matrices describing the probability
distribution of the outcomes of a measurement on the system-apparatus
combination is formally (approximately) identical to the corresponding
mixed-state density matrix. But as we pointed out in
Sec.~\ref{sec:redmat}, we must be careful in interpreting this state
of affairs, since full coherence is retained in the total density
matrix $\rho_{\mathcal{SAE}}$.

\subsubsection{\label{sec:zurekmodel}An exactly solvable two-state model for decoherence}

To see how the approximate mutual orthogonality of the environmental
state vectors arises, let us discuss a simple model first introduced
by \citet{Zurek:1982:tv}.  Consider a system $\mathcal{S}$ with two
spin states $\{ \ket{\Uparrow}, \ket{\Downarrow} \}$ that interacts
with an environment $\mathcal{E}$ described by a collection of $N$
other two-state spins represented by $\{ \ket{\uparrow_k},
\ket{\downarrow_k} \}$, $k=1\cdots N$. The self-Hamiltonians
$\widehat{H}_{\mathcal{S}}$ and $\widehat{H}_{\mathcal{E}}$ and the
self-interaction Hamiltonian $\widehat{H}_\mathcal{EE}$ of the
environment are taken to be equal to zero.  Only the interaction
Hamiltonian $\widehat{H}_\mathcal{SE}$ that describes the coupling of
the spin of the system to the spins of the environment is assumed to
be nonzero and of the form
\begin{equation} 
\widehat{H}_\mathcal{SE} = ( \ketbra{\Uparrow}{\Uparrow} -
\ketbra{\Downarrow}{\Downarrow} ) \otimes \sum_k g_k (
\ketbra{\uparrow_k}{\uparrow_k} - \ketbra{\downarrow_k}{\downarrow_k} )
\bigotimes_{k' \not= k}  \widehat{I}_{k'}, 
\end{equation}
where the $g_k$ are coupling constants and $\widehat{I}_{k} =
(\ketbra{\uparrow_k}{\uparrow_k} +
\ketbra{\downarrow_k}{\downarrow_k})$ is the identity operator for the
$k$th environmental spin. Applied to the initial state before the
interaction is turned on,
\begin{equation}
\ket{\psi(0)} = ( a\ket{\Uparrow} + b\ket{\Downarrow} ) \bigotimes_{k=1}^N
\, ( \alpha_k \ket{\uparrow_k} + \beta_k \ket{\downarrow_k} ),
\end{equation}
this Hamiltonian yields a time evolution of the state given by
\begin{equation} 
\ket{\psi(t)} = a\ket{\Uparrow}\ket{E_{\Uparrow}(t)} +
b\ket{\Downarrow}\ket{E_{\Downarrow}(t)},
\end{equation}
where the two environmental states $\ket{E_{\Uparrow}(t)}$ and
$\ket{E_{\Downarrow}(t)}$ are
\begin{equation} 
\ket{E_{\Uparrow}(t)} = \ket{E_{\Downarrow}(-t)} 
= \bigotimes_{k=1}^N\,
 ( \alpha_k e^{ig_kt}\ket{\uparrow_k} + \beta_k
e^{-ig_kt}\ket{\downarrow_k} ). \phantom{xxxx} 
\end{equation}
The reduced density matrix $\rho_{\mathcal{S}}(t)=\text{Tr}_{\mathcal{E}} (\ket{\psi(t)}
\bra{\psi(t)})$ is then
\begin{eqnarray}
\rho_{\mathcal{S}}(t) &=& |a|^2
\ketbra{\Uparrow}{\Uparrow} + |b|^2 \ketbra{\Downarrow}{\Downarrow}
\nonumber \\ && \, + z(t)ab^*\ketbra{\Uparrow}{\Downarrow} +
z^*(t)a^*b\ketbra{\Downarrow}{\Uparrow},
\end{eqnarray}
where the interference coefficient $z(t)$ which determines the weight
of the off-diagonal elements in the reduced density matrix is given by
\begin{equation} 
z(t) = \langle E_{\Uparrow}(t) | E_{\Downarrow}(t) \rangle 
=\prod_{k=1}^N (|\alpha_k|^2 e^{ig_kt} + |\beta_k|^2 e^{-ig_kt} ),
\end{equation} 
and thus
\begin{equation} \
|z(t)|^2 = \prod_{k=1}^N \{ 1 + [(|\alpha_k|^2 - |\beta_k|^2)^2
-1] \sin^2 2g_kt \}. 
\end{equation} 
At $t=0$, $z(t)=1$, i.e., the interference terms are fully present, as
expected. If $|\alpha_k|^2=0$ or 1 for each $k$, i.e., if the
environment is in an eigenstate of the interaction Hamiltonian
$\widehat{H}_\mathcal{SE}$ of the type $\ket{\uparrow_1} \ket{\uparrow_2}
\ket{\downarrow_3} \cdots \ket{\uparrow_N}$, and/or if $2g_kt = m\pi$
($m=0,1,\hdots$), then $z(t)^2 \equiv 1$ so coherence is retained over
time.  However, under realistic circumstances, we can typically assume
a random distribution of the initial states of the environment (i.e., of
coefficients $\alpha_k$, $\beta_k$) and of the coupling coefficients
$g_k$. Then, in the long-time average, 
\begin{equation}
\langle |z(t)|^2 \rangle_{t \rightarrow \infty} 
\simeq 2^{-N} \prod_{k=1}^N [1 + (|\alpha_k|^2 -
|\beta_k|^2)^2] \, \stackrel{N\rightarrow \infty}{\longrightarrow} \, 0,
\end{equation} 
so the off-diagonal terms in the reduced density matrix become
strongly damped for large $N$. 

It can also be shown directly that, given very general assumptions
about the distribution of the couplings $g_k$ (namely, requiring their
initial distribution to have finite variance), $z(t)$ exhibits a
Gaussian time dependence of the form $z(t) \propto e^{iAt}
e^{-B^2t^2/2}$, where $A$ and $B$ are real constants
\citep{Zurek:2003:om}. For the special case in which $\alpha_k=\alpha$
and $g_k=g$ for all $k$, this behavior of $z(t)$ can be immediately
seen by first rewriting $z(t)$ as the binomial expansion
\begin{eqnarray} 
z(t) &=& (|\alpha|^2 e^{igt} + |\beta|^2 e^{-igt} )^N \nonumber \\ &=& \sum_{l=0}^N
\binom{N}{l} |\alpha|^{2l} |\beta|^{2(N-l)} e^{ig(2l-N)t}.
\end{eqnarray} 
For large $N$, the binomial distribution can then be approximated by a
Gaussian,
\begin{equation}
\binom{N}{l} |\alpha|^{2l} |\beta|^{2(N-l)} \approx
\frac{e^{-(l-N|\alpha|^2)^2/(2N|\alpha|^2 |\beta|^2)}}{\sqrt{2\pi N |\alpha|^2 |\beta|^2}},
\end{equation} 
in which case $z(t)$ becomes
\begin{equation}
z(t) = \sum_{l=0}^N \frac{e^{-(l-N|\alpha|^2)^2/(2N|\alpha|^2
 |\beta|^2)}}{\sqrt{2\pi N |\alpha|^2 |\beta|^2}} 
 e^{ig(2l-N)t}, 
\end{equation} 
that is, $z(t)$ is the Fourier transform of an (approximately) Gaussian
distribution and is therefore itself (approximately) Gaussian.

Detailed model calculations, in which the environment is typically
represented by a more sophisticated model consisting of a collection
of harmonic oscillators
\citep{Zurek:1993:qq,Zurek:2002:ii,Joos:2003:jh,Caldeira:1983:on,Unruh:1989:rc,Hu:1992:om},
have shown that the damping occurs on extremely short decoherence time
scales $\tau_D$, which are typically many orders of magnitude shorter
than the thermal relaxation. Even microscopic systems such as large
molecules are rapidly decohered by the interaction with thermal
radiation on a time scale that is much shorter than any practical
observation could resolve; for mesoscopic systems such as dust
particles, the 3K cosmic microwave background radiation is sufficient
to yield strong and immediate decoherence
\citep{Joos:1985:iu,Zurek:1991:vv}.

Within $\tau_D$, $|z(t)|$ approaches zero and remains close to zero,
fluctuating with an average standard deviation of the random-walk type
$\sigma \sim \sqrt{N}$ \citep{Zurek:1982:tv}. However, the multiple
periodicity of $z(t)$ implies that coherence, and thus the purity of
the reduced density matrix, will reappear after a certain time
$\tau_r$, which can be shown to be very long and of the Poincar\'e
type with $\tau_r \sim N!$. For macroscopic environments of realistic
but finite sizes, $\tau_r$ can exceed the lifetime of the universe
\citep{Zurek:1982:tv}, but nevertheless always remains finite.

From a conceptual point of view, recurrence of coherence is of little
relevance. The recurrence time could only be infinitely long in the
hypothetical case of an infinitely large environment. In this
situation, off-diagonal terms in the reduced density matrix would be
irreversibly damped and lost in the limit $t \rightarrow \infty$,
which has sometimes been regarded as describing a physical collapse of
the state vector \citep{Hepp:1972:pa}. But the assumption of infinite
sizes and times is never realized in nature \citep{Bell:1975:oi}, nor
can information ever be truly lost (as achieved by a ``true'' state
vector collapse) through unitary time evolution---full coherence is
always retained at all times in the total density matrix
$\rho_\mathcal{SAE}(t)=\ket{\psi(t)} \bra{\psi(t)}$.

We can therefore state the general conclusion that, except for
exceptionally well-isolated and carefully prepared microsopic and
mesoscopic systems, the interaction of the system with the environment
causes the off-diagonal terms of the local density matrix, expressed
in the pointer basis and describing the probability distribution of
the possible outcomes of a measurement on the system, to become
extremely small in a very short period of time, and that this process
is irreversible for all practical purposes.

\subsection{Environment-induced superselection \label{sec:einsel}}

Let us now turn to the second main consequence of the interaction with
the environment, namely, the environment-induced selection of stable
preferred-basis states.  We discussed in Sec.~\ref{sec:pbprob} the
fact that the quantum-mechanical measurement scheme as represented by
Eq.~\eqref{eq:measurement1} does not uniquely define the expansion of
the postmeasurement state and thereby leaves open the question of
which observable can be considered as having been measured by the
apparatus.  This situation is changed by the inclusion of the
environment states in Eq.~\eqref{eq:measurement2}, for the following
two reasons:

\begin{enumerate} 

\item[(1)] {\em Environment-induced superselection of a preferred basis.}
  The interaction between the apparatus and the environment singles
  out a set of mutually commuting observables.
  
\item[(2)] {\em The existence of a tridecompositional uniqueness
    theorem} \citep{Elby:1994:tz,Clifton:1995:po,Bub:1997:iq}. If a
  state $\ket{\psi}$ in a Hilbert space $\mathcal{H}_1 \otimes
  \mathcal{H}_2 \otimes \mathcal{H}_3$ can be decomposed into the
  diagonal (``Schmidt'') form $\ket{\psi} = \sum_i \alpha_i
  \ket{\phi_i}_1 \ket{\phi_i}_2 \ket{\phi_i}_3$, the expansion is
  unique provided that the $\{ \ket{\phi_i}_1 \}$ and $\{
  \ket{\phi_i}_2 \}$ are sets of linearly independent, normalized
  vectors in $\mathcal{H}_1$ and $\mathcal{H}_2$, respectively, and
  that $\{ \ket{\phi_i}_3 \}$ is a set of mutually noncollinear
  normalized vectors in $\mathcal{H}_3$.  This can be generalized to
  an $N$-decompositional uniqueness theorem, in which $N \ge 3$.  Note
  that it is not always possible to decompose an arbitrary pure state
  of more than two systems ($N \ge 3$) into the Schmidt form
  $\ket{\psi} = \sum_i \alpha_i \ket{\phi_i}_1 \ket{\phi_i}_2 \cdots
  \ket{\phi_i}_N$, but if the decomposition exists, its uniqueness is
  guaranteed.

\end{enumerate}

The tridecompositional uniqueness theorem ensures that the expansion
of the final state in Eq.~\eqref{eq:measurement2} is unique, which
fixes the ambiguity in the choice of the set of possible outcomes. It
demonstrates that the inclusion of (at least) a third ``system'' (here
referred to as the environment) is necessary to remove the basis
ambiguity.

Of course, given any pure state in the composite Hilbert space
$\mathcal{H}_1 \otimes \mathcal{H}_2 \otimes \mathcal{H}_3$, the
tridecompositional uniqueness theorem neither tells us whether a
Schmidt decomposition exists nor specifies the unique expansion itself
(provided the decomposition is possible), and since the precise states
of the environment are generally not known, an additional criterion is
needed that determines what the preferred states will be.

\subsubsection{Stability criterion and pointer basis}

The decoherence program has attempted to define such a criterion based
on the interaction with the environment and the idea of robustness and
preservation of correlations. The environment thus plays a double role
in suggesting a solution to the preferred-basis problem: it selects a
preferred pointer basis, and it guarantees its uniqueness via the
tridecompositional uniqueness theorem.

In order to motivate the basis superselection approach proposed by the
decoherence program, we note that in step (2) of
Eq.~\eqref{eq:measurement2} we tacitly assumed that interaction with
the environment does not disturb the established correlation between
the state of the system, $\ket{s_n}$, and the corresponding pointer
state $\ket{a_n}$. This assumption can be viewed as a generalization
of the concept of ``faithful measurement'' to the realistic case in which
the environment is included.  Faithful measurement in the usual sense
concerns step (1), namely, the requirement that the measuring
apparatus $\mathcal{A}$ act as a reliable ``mirror'' of the states of
the system $\mathcal{S}$ by forming only correlations of the form
$\ket{s_n}\ket{a_n}$ but not $\ket{s_m} \ket{a_n}$ with $m \not= n$.
But since any realistic measurement process must include the
inevitable coupling of the apparatus to its environment, the
measurement could hardly be considered faithful as a whole if the
interaction with the environment disturbed the correlations between
the system and the apparatus.\footnote{For fundamental limitations on
  the precision of von Neumann measurements of operators that do not
  commute with a globally conserved quantity, see the
  Wigner-Araki-Yanase theorem \citep{Wigner:1952:gb,Araki:1960:ff}.}

It was therefore first suggested by \citet{Zurek:1981:dd} that the
preferred pointer basis be taken as the basis which ``contains a
reliable record of the state of the system $\mathcal{S}$''
\citep[p.~1519]{Zurek:1981:dd}, i.e., the basis in which the
system-apparatus correlations $\ket{s_n}\ket{a_n}$ are left
undisturbed by the subsequent formation of correlations with the
environment (the {\em stability criterion}). One can then find a
sufficient criterion for dynamically stable pointer states that
preserve the system--apparatus correlations in spite of the
interaction of the apparatus with the environment by requiring all
pointer state projection operators $\widehat{P}^{(\mathcal{A})}_n =
\ket{a_n} \bra{a_n}$ to commute with the apparatus-environment
interaction Hamiltonian $\widehat{H}_\mathcal{AE}$,\footnote{For
  simplicity, we assume here that the environment $\mathcal{E}$
  interacts directly only with the apparatus $\mathcal{A}$, but not
  with the system $\mathcal{S}$.}
\begin{equation} \label{eq:commut}
[\widehat{P}^{(\mathcal{A})}_n, \,
\widehat{H}_\mathcal{AE}] = 0 \qquad \text{for all $n$.}
\end{equation}
This implies that any correlation of the measured system (or any
other system, for instance an observer) with the eigenstates of a
\emph{preferred apparatus observable},
\begin{equation} \label{eq:prefobs} \widehat{O}_\mathcal{A} = \sum_n \lambda_n
\widehat{P}^{(\mathcal{A})}_n, \end{equation}
is preserved, and that the states of the environment reliably mirror
the pointer states $\widehat{P}^{(\mathcal{A})}_n$. In this case, the
environment can be regarded as carrying out a nondemolition
measurement on the apparatus. The commutativity requirement,
Eq.~\eqref{eq:commut}, is obviously fulfilled if
$\widehat{H}_\mathcal{AE}$ is a function of $\widehat{O}_\mathcal{A}$,
$\widehat{H}_\mathcal{AE}=\widehat{H}_\mathcal{AE}(\widehat{O}_\mathcal{A})$.
Conversely, system-apparatus correlations in which the states of the
apparatus are not eigenstates of an observable that commutes with
$\widehat{H}_\mathcal{AE}$ will in general be rapidly destroyed by the
interaction.

Put the other way around, this implies that the environment determines
through the form of the interaction Hamiltonian
$\widehat{H}_\mathcal{AE}$, a preferred apparatus observable
$\widehat{O}_\mathcal{A}$, Eq.~\eqref{eq:prefobs}, and thereby also
the states of the system that are measured by the apparatus, that is,
reliably recorded through the formation of dynamically stable quantum
correlations. The tridecompositional uniqueness theorem then
guarantees the uniqueness of the expansion of the final state
$\ket{\psi} = \sum_n c_n \ket{s_n} \ket{a_n} \ket{e_n}$ (where no
constraints on the $c_n$ have to be imposed) and thereby the
uniqueness of the preferred pointer basis.

Other criteria similar to the commutativity requirement,
Eq.~\eqref{eq:commut}, have been suggested for the selection of the
preferred pointer basis because it turns out that in realistic cases
the simple relation of Eq.~\eqref{eq:commut} can usually only be
fulfilled approximately \citep{Zurek:1993:qq,Zurek:1993:pu}. More
general criteria, for example, have been based on the von Neumann
entropy $-\text{Tr} \rho^2_\Psi(t) \ln \rho^2_\Psi(t)$, or the purity
$\text{Tr}\rho^2_\Psi(t)$, with the goal of finding the most robust
states or the states which become least entangled with the environment
in the course of the evolution
\citep{Zurek:1993:qq,Zurek:1993:pu,Zurek:1998:re,Zurek:2002:ii}.
Pointer states are obtained by extremizing the measure (i.e.,
minimizing entropy, or maximizing purity, etc.) over the initial state
$\ket{\Psi}$ and requiring the resulting states to be robust when
varying the time $t$.  Application of this method leads to a ranking
of the possible pointer states with respect to their ``classicality,''
i.e., their robustness with respect to interaction with the
environment, and thus allows for the selection of preferred pointer
basis in terms of the ``most classical'' pointer states \citep[the
``predictability sieve''; see][]{Zurek:1993:pu,Zurek:1993:qq}.
Although the proposed criteria differ somewhat and other meaningful
criteria are likely to be suggested in the future, it is hoped that in
the macrosopic limit the resulting stable pointer states obtained from
different criteria will turn out to be very similar
\citep{Zurek:2002:ii}. For some toy models (in particular, for
harmonic-oscillator models that lead to coherent states as pointer
states), this has already been verified explicitly \citep[see, for
example,][]{Kubler:1973:ux,Zurek:1993:pu,Diosi:2000:yn,Joos:2003:jh,Eisert:2003:ib}.

\subsubsection{Selection of quasiclassical properties}

System-environment interaction Hamiltonians frequently describe a
scattering process of surrounding particles (photons, air molecules,
etc.) interacting with the system under study. Since the force laws
describing such processes typically depend on some power of distance
(such as $\propto r^{-2}$ in Newton's or Coulomb's force law), the
interaction Hamiltonian will usually commute with the position basis,
such that, according the commutativity requirement of
Eq.~\eqref{eq:commut}, the preferred basis will be in position space.
The fact that position is frequently the determinate property of our
experience can then be explained by referring to the dependence of
most interactions on distance
\citep{Zurek:1981:dd,Zurek:1982:tv,Zurek:1991:vv}.

This holds, in particular, for mesoscopic and macroscopic systems, as
demonstrated, for instance, by the pioneering study of
\citet{Joos:1985:iu}, in which surrounding photons and air molecules
are shown to continuously ``measure'' the spatial structure of dust
particles, leading to rapid decoherence into an apparent (improper)
mixture of wave packets that are sharply peaked in position space.
Similar results sometimes even hold for microscopic systems (usually
found in energy eigenstates; see below) when they occur in distinct
spatial structures that couple strongly to the surrounding medium. For
instance, chiral molecules such as sugar are always observed to be in
chirality eigenstates (left-handed and right-handed) which are
superpositions of different energy eigenstates
\citep{Harris:1981:rc,Zeh:1999:qr}.  This is explained by the fact
that the spatial structure of these molecules is continuously
``monitored'' by the environment, for example, through the scattering
of air molecules, which gives rise to a much stronger coupling than
could typically be achieved by a measuring device that was intended to
measure, say, parity or energy; furthermore, any attempt to prepare
such molecules in energy eigenstates would lead to immediate
decoherence into environmentally stable (``dynamically robust'')
chirality eigenstates, thus selecting position as the preferred basis.

On the other hand, it is well known that many systems, especially in the
microsopic domain, are typically found in energy eigenstates, even if
the interaction Hamiltonian depends on a different observable than
energy, e.g., position. \citet{Paz:1999:vv} have shown that this
situation arises when the predominant frequencies  present in the
environment are significantly lower than the intrinsic frequencies of
the system, that is, when the separation between the energy states of
the system is greater than the largest energies available in the
environment. Then, the environment will only be able to monitor
quantities that are constants of motion. In the case of
nondegeneracy, this will be energy, thus leading to the
environment-induced superselection of energy eigenstates for the
system.

Another example of environment-induced superselection that has been
studied is related to the fact that only eigenstates of the charge
operator are observed, but never superpositions of different charges.
The existence of the corresponding superselection rules was first only
postulated \citep{Wick:1952:pp,Wick:1970:iz}, but could subsequently be
explained in the framework of decoherence by referring to the
interaction of the charge with its own Coulomb (far) field, which takes
the role of an ``environment,'' leading to immediate decoherence of
charge superpositions into an apparent mixture of charge eigenstates
\citep{Giulini:1995:zh,Giulini:2000:ry}.

In general, three different cases have typically been distinguished
\citep[for example, in][]{Paz:1999:vv} for the kind of pointer
observable emerging from an interaction with the environment,
depending on the relative strengths of the system's self-Hamiltonian
$\widehat{H}_{\mathcal{S}}$ and of the system-environment interaction
Hamiltonian $\widehat{H}_\mathcal{SE}$:
\begin{enumerate} 
  
\item[(1)] When the dynamics of the system are dominated by
  $\widehat{H}_\mathcal{SE}$, i.e., the interaction with the
  environment, the pointer states will be eigenstates of
  $\widehat{H}_\mathcal{SE}$ (and thus typically eigenstates of
  position). This case corresponds to the typical quantum measurement
  setting; see, for example, the model of
  \citet{Zurek:1981:dd,Zurek:1982:tv}, which is outlined in
  Sec.~\ref{sec:zurekmodel} above.
  
\item[(2)] When the interaction with the environment is weak and
  $\widehat{H}_\mathcal{S}$ dominates the evolution of the system
  (that is, when the environment is ``slow'' in the above sense), a
  case that frequently occurs in the microscopic domain, pointer
  states will arise that are energy eigenstates of
  $\widehat{H}_\mathcal{S}$ \citep{Paz:1999:vv}.
  
\item[(3)] In the intermediate case, when the evolution of the system is
  governed by $\widehat{H}_\mathcal{SE}$ and $\widehat{H}_\mathcal{S}$
  in roughly equal strength, the resulting preferred states will
  represent a ``compromise'' between the first two cases; for
  instance, the frequently studied model of quantum Brownian motion
  has shown the emergence of pointer states localized in phase space,
  i.e., in both position and momentum
  \citep{Zurek:1993:qq,Zurek:2002:ii,Joos:2003:jh,Unruh:1989:rc,Eisert:2003:ib}.
  
\end{enumerate}

\subsubsection{\label{sec:einsel-mp}Implications for the
  preferred-basis problem}

The decoherence program proposes that the preferred basis be selected
by the requirement that correlations be preserved in spite of the
interaction with the environment, and thus be chosen through the form
of the system-environment interaction Hamiltonian. This seems
certainly reasonable, since only such ``robust'' states will in
general be observable---and, after all, we solely seek an explanation
for our experience (see the discussion in Sec.~\ref{sec:objsubj}).
Although only particular examples have been studied \citep[for a
survey and references, see, for
example,][]{Joos:2003:jh,Blanchard:2000:fq,Zurek:2002:ii}, the results
thus far suggest that the selected properties are in agreement with
our observation: for mesoscopic and macroscopic objects the
distance-dependent scattering interaction with surrounding air
molecules, photons, etc., will in general give rise to immediate
decoherence into spatially localized wave packets and thus select
position as the preferred basis. On the other hand, when the
environment is comparably ``slow,'' as is frequently the case for
microsopic systems, environment-induced superselection will typically
yield energy eigenstates as the preferred states.

The clear merit of the approach of environment-induced superselection
lies in the fact that the preferred basis is not chosen in an \emph{ad
  hoc} manner simply to make our measurement records determinate or to
match our experience of which physical quantities are usually
perceived as determinate (for example, position). Instead the
selection is motivated on physical, observer-free grounds, that is,
through the system-environment interaction Hamiltonian. The vast
space of possible quantum-mechanical superpositions is reduced so much
because the laws governing physical interactions depend only on a few
physical quantities (position, momentum, charge, and the like), and
the fact that precisely these are the properties that appear
determinate to us is explained by the dependence of the preferred
basis on the form of the interaction. The appearance of
``classicality'' is therefore grounded in the structure of the
physical laws---certainly a highly satisfying and reasonable approach.

The above argument in favor of the approach of environment-induced
superselection could, of course, be considered as inadequate on a
fundamental level: All physical laws are discovered and formulated by
us, so they can contain only the determinate quantities of our
experience. These are the only quantities we can perceive and thus
include in a physical law. Thus the derivation of determinacy from the
structure of our physical laws might seem circular.  However, we argue
again that it suffices to demand a subjective solution to the
preferred-basis problem---that is, to provide an answer to the
question of why we perceive only such a small subset of properties
as determinate, not whether there really are determinate
properties (on an ontological level) and what they are (\cf the
remarks in Sec.~\ref{sec:objsubj}).

We might also worry about the generality of this approach. One would
need to show that any such environment-induced superselection leads,in
fact, to precisely those properties that appear determinate to us. But
this would require precise knowledge of the system and the interaction
Hamiltonian.  For simple toy models, the relevant Hamiltonians can be
written down explicitly. In more complicated and realistic cases, this
will in general be very difficult, if not impossible, since the form
of the Hamiltonian will depend on the particular system or apparatus
and the monitoring environment under consideration, where, in
addition, the environment is not only difficult to define precisely,
but also constantly changing, uncontrollable, and, in essence,
infinitely large.

But the situation is not as hopeless as it might sound, since we know
that the interaction Hamiltonian will, in general, be based on the set
of known physical laws which, in turn, employ only a relatively small
number of physical quantities. So as long as we assume the stability
criterion and consider the set of known physical quantities as
complete, we can automatically anticipate that the preferred basis
will be a member of this set. The remaining, yet very relevant,
question is then which subset of these properties will be chosen in a
specific physical situation (for example, will the system preferably
be found in an eigenstate of energy or position?), and to what extent
this will match the experimental evidence. To give an answer, one will
usually need a more detailed knowledge of the interaction Hamiltonian
and of its relative strength with respect to the self-Hamiltonian of
the system in order to verify the approach.  Besides, as mentioned in
Sec.~\ref{sec:einsel}, there exist other criteria than the
commutativity requirement, and whether they all lead to the same
determinate properties is a question that has not yet been fully
explored.

Finally, a fundamental conceptual difficulty of the decoherence-based
approach to the preferred-basis problem is the lack of a general
criterion for what defines the systems and the ``unobserved'' degrees
of freedom of the environment (see the discussion in
Sec.~\ref{sec:division}). While in many laboratory-type situations,
the division into system and environment might seem straightforward,
it is not clear {\em a priori} how quasiclassical observables can be
defined through environment-induced superselection on a larger and
more general scale, when larger parts of the universe are considered
where the split into subsystems is not suggested by some specific
system-apparatus-surroundings setup.

To summarize, environment-induced superselection of a preferred basis
(i) proposes an explanation for why a particular pointer basis gets
chosen at all---by arguing that it is only the pointer basis that
leads to stable, and thus perceivable, records when the interaction of
the apparatus with the environment is taken into account; and (ii)
argues that the preferred basis will correspond to a subset of the set
of the determinate properties of our experience, since the governing
interaction Hamiltonian will depend solely on these quantities. But it
does not tell us precisely what the pointer basis will be in any given
physical situation, since it will usually be hardly possible to write
down explicitly the relevant interaction Hamiltonian in realistic
cases.  This also means that it will be difficult to argue that any
proposed criterion based on the interaction with the environment will
always and in all generality lead to exactly those properties that we
perceive as determinate.

More work remains to be done, therefore, to fully explore the general
validity and applicability of the approach of environment-induced
superselection. But since the results obtained thus far from toy
models have been in promising agreement with empirical data, there is
little reason to doubt that the decoherence program has proposed a
very valuable criterion for explaining the emergence of preferred states
and their robustness. The fact that the approach is derived from
physical principles should be counted additionally in its favor.

\subsubsection{Pointer basis vs instantaneous Schmidt states \label{sec:schmidt}}

The so-called Schmidt basis, obtained by diagonalizing the
(reduced) density matrix of the system at each instant of time, has been
frequently studied with respect to its ability to yield a preferred
basis \citep[see, for
example,][]{Zeh:1973:wq,Albrecht:1992:rz,Albrecht:1993:pq}, having led
some to consider the Schmidt basis states as describing
``instantaneous pointer states'' \citep{Albrecht:1992:rz}.  However,
as it has been emphasized \citep[for example, by][]{Zurek:1993:pu},
any density matrix is diagonal in some basis, and this basis will in
general not play any special interpretive role. Pointer states that
are supposed to correspond to quasiclassical stable observables must
be derived from an explicit criterion for classicality (typically, the
stability criterion); the simple mathematical diagonalization
procedure of the instantaneous density matrix will generally not
suffice to determine a quasiclassical pointer basis \citep[see the
studies by][]{Barvinsky:1995:pa,Kent:1997:oz}.

In a more refined method, one refrains from computing instantaneous
Schmidt states and instead allows for a characteristic decoherence
time $\tau_D$ to pass during which the reduced density matrix
decoheres (a process that can be described by an appropriate master
equation) and becomes approximately diagonal in the stable pointer
basis, the basis that is selected by the stability criterion.  Schmidt
states are then calculated by diagonalizing the decohered density
matrix.  Since decoherence usually leads to rapid diagonality of the
reduced density matrix in the stability-selected pointer basis to a
very good approximation, the resulting Schmidt states are typically
very similar to the pointer basis except when the pointer states are
very nearly degenerate. The latter situation is readily illustrated by
considering the approximately diagonalized decohered density matrix
\begin{equation} 
\rho = \left( \begin{array}{cc}
    1/2 + \delta & \omega^* \\
    \omega & 1/2 - \delta
\end{array} \right),
\end{equation}
where $|\omega| \ll 1$ (strong decoherence) and $\delta \ll 1$
\citep[near-degeneracy;][]{Albrecht:1993:pq}. If decoherence led to
exact diagonality, $\omega = 0$, the eigenstates would be, for any
fixed value of $\delta$, proportional to $(0,1)$ and $(1,0)$
(corresponding to the ``ideal'' pointer states).  However, for fixed
$\omega > 0$ (approximate diagonality) and $\delta \rightarrow 0$
(degeneracy), the eigenstates become proportional to $(\pm
|\omega| / \omega, 1)$, which implies that, in the case of
degeneracy, the Schmidt decomposition of the reduced density matrix can
yield preferred states that are very different from the stable pointer
states, even if the decohered, rather than the instantaneous, reduced
density matrix is diagonalized.

In summary, it is important to emphasize that stability (or a similar
criterion) is the relevant requirement for the emergence of a
preferred quasiclassical basis, which cannot, in general, be achieved
by simply diagonalizing the instantaneous reduced density matrix.
However, the eigenstates of the decohered reduced density matrix will,
in many situations, approximate the quasiclassical stable pointer
states well, especially when these pointer states are sufficiently
nondegenerate.

\subsection{Envariance, quantum probabilities, and the Born rule \label{sec:envar}}

In the following, we shall review an interesting and promising
approach introduced recently by
\citet{Zurek:2002:ii,Zurek:2003:rv,Zurek:2003:pl,Zurek:2004:yb} that
aims to explain the emergence of quantum probabilities and to deduce
the Born rule based on a mechanism termed ``environment-assisted
invariance,'' or ``envariance'' for short, a particular symmetry
property of entangled quantum states. The original exposition of
\citet{Zurek:2002:ii} was followed up by several articles by other
authors, who assessed the approach, pointed out more clearly the
assumptions entering into the derivation, and presented variants of
the proof \cite{Schlosshauer:2003:ms,Barnum:2003:yb,Mohrhoff:2004:tv}.
An expanded treatment of envariance and quantum probabilities that
addresses some of the issues discussed in these papers and that offers
an interesting outlook on further implications of envariance can be
found in \citet{Zurek:2004:yb}. In our outline of the theory of
envariance, we shall follow this most recent treatment, as it spells
out the derivation and the required assumptions more explicitly and
in greater detail and clarity than in Zurek's earlier
\citeyearpar{Zurek:2002:ii,Zurek:2003:rv,Zurek:2003:pl} papers
\citep[\cf also the remarks of][]{Schlosshauer:2003:ms}.

We include a discussion of Zurek's proposal here for two reasons.
First, the derivation is based on the inclusion of an environment
$\mathcal{E}$, entangled with the system $\mathcal{S}$ of interest to
which probabilities of measurement outcomes are to be assigned, and
thus it matches well the spirit of the decoherence program. Second,
and more importantly, despite the contributions of decoherence to
explaining the emergence of subjective classicality from quantum
mechanics, a consistent derivation of classicality \citep[including a
motivation for some of the axioms of quantum mechanics, as suggested
by][]{Zurek:2002:ii} requires the separate derivation of the Born
rule.  The decoherence program relies heavily on the concept of
reduced density matrices and the related formalism and interpretation
of the trace operation, see Eq.~\eqref{eq:global-to-local}, which
\emph{presuppose} Born's rule.  Therefore decoherence itself cannot be
used to derive the Born rule \citetext{as was tried, for example, by
  \citealp{Deutsch:1999:tz} and \citealp{Zurek:1998:re}} since otherwise
the argument would be rendered circular
\citep{Zeh:1996:gy,Zurek:2002:ii}.
 
There have been various attempts in the past to replace the postulate
of the Born rule by a derivation. Gleason's
\citeyearpar{Gleason:1957:zp} theorem has shown that if one imposes
the condition that for any orthonormal basis of a given Hilbert space
the sum of the probabilities associated with each basis vector must
add up to one, the Born rule is the only possibility for the
calculation of probabilities. However, Gleason's proof provides little
insight into the physical meaning of the Born probabilities,
and therefore various other attempts, typically based on a relative
frequencies approach (i.e., on a counting argument), have been made
towards a derivation of the Born rule in a no-collapse (and usually
relative-state) setting \citep[see, for
example,][]{Everett:1957:rw,DeWitt:1973:pz,Hartle:1968:gg,DeWitt:1971:pz,%
  Graham:1973:ww,Geroch:1984:yt,Farhi:1989:uh,Deutsch:1999:tz}.
However, it was pointed out that these approaches fail due to the use
of circular arguments
\citep{Stein:1984:uu,Kent:1990:nm,Squires:1990:lz,Barnum:2000:oz}; \cf
also \citet{Wallace:2003:zr} and \citet{Saunders:2002:tz}.

Zurek's recently developed theory of envariance provides a promising
new strategy for deriving, given certain assumptions, the Born rule in
a manner that avoids the circularities of the earlier approaches.  We
shall outline the concept of envariance in the following and show how
it can lead to Born's rule.

\subsubsection{Environment-assisted invariance}

Zurek introduces his definition of envariance as follows. Consider a
composite state $\ket{\psi_\mathcal{SE}}$ (where, as usual,
$\mathcal{S}$ refers to the ``system'' and $\mathcal{E}$ to some
``environment'') in a Hilbert space given by the tensor product
$\mathcal{H}_\mathcal{S} \otimes \mathcal{H}_\mathcal{E}$, and a pair
of unitary transformations $\widehat{U}_\mathcal{S} =
\widehat{u}_\mathcal{S} \otimes \widehat{I}_\mathcal{E}$ and
$\widehat{U}_\mathcal{E} = \widehat{I}_\mathcal{S} \otimes
\widehat{u}_\mathcal{E}$ acting on $\mathcal{S}$ and $\mathcal{E}$,
respectively.  If $\ket{\psi_\mathcal{SE}}$ is invariant under the
combined application of $\widehat{U}_\mathcal{S}$ and
$\widehat{U}_\mathcal{E}$,
\begin{equation} \label{eq:envar}
\widehat{U}_\mathcal{E} (\widehat{U}_\mathcal{S} 
\ket{\psi_\mathcal{SE}}) = \ket{\psi_\mathcal{SE}},
\end{equation}
$\ket{\psi_\mathcal{SE}}$ is called \emph{envariant under
  $\widehat{u}_\mathcal{S}$}. In other words, the change in
$\ket{\psi_\mathcal{SE}}$ induced by acting on $\mathcal{S}$ via
$\widehat{U}_\mathcal{S}$ can be undone by acting on $\mathcal{E}$ via
$\widehat{U}_\mathcal{E}$.  Note that envariance is a distinctly
quantum feature, absent from pure classical states, and a consequence
of quantum entanglement.

The main argument of Zurek's derivation is based on a study of a
composite pure state in the diagonal Schmidt decomposition
\begin{equation} \label{eq:schmidt}
\ket{\psi_\mathcal{SE}} = \frac{1}{\sqrt{2}} \big( e^{i\varphi_1} \ket{s_1} \ket{e_1}
+ e^{i\varphi_2} \ket{s_1} \ket{e_1} \big),
\end{equation}
where the $\{\ket{s_k}\}$ and $\{\ket{e_k}\}$ are sets of orthonormal
basis vectors that span the Hilbert spaces $\mathcal{H}_\mathcal{S}$
and $\mathcal{H}_\mathcal{E}$, respectively.  The case of
higher-dimensional state spaces can be treated similarly, and a
generalization to expansion coefficients of different magnitudes can be
made by application of a standard counting argument
\citep{Zurek:2003:rv,Zurek:2004:yb}.  The Schmidt states $\ket{s_k}$
are identified with the outcomes, or ``events''
\citep[][p.~12]{Zurek:2003:pl}, to which probabilities are to be
assigned.

Zurek now states three simple assumptions, called ``facts''
\citetext{\citealp{Zurek:2004:yb}, p.~4; see also the discussion in
  \citealp{Schlosshauer:2003:ms}}:

\begin{enumerate}

\item[(A1)] A unitary transformation of the form $\cdots \otimes
  \widehat{I}_\mathcal{S} \otimes \cdots$ does not alter the state of
  $\mathcal{S}$.
  
\item[(A2)] All measurable properties of $\mathcal{S}$, including
  probabilities of outcomes of measurements on $\mathcal{S}$, are
  fully determined by the state of $\mathcal{S}$.
  
\item[(A3)] The state of $\mathcal{S}$ is completely specified by the global
  composite state vector $\ket{\psi_\mathcal{SE}}$.

\end{enumerate}

Given these assumptions, one can show that the state of $\mathcal{S}$
and any measurable properties of $\mathcal{S}$ cannot be affected by
envariant transformations. The proof goes as follows.  The effect of
an envariant transformation $\widehat{u}_\mathcal{S} \otimes
\widehat{I}_\mathcal{E}$ acting on $\ket{\psi_\mathcal{SE}}$ can be
undone by a corresponding ``countertransformation''
$\widehat{I}_\mathcal{S} \otimes \widehat{u}_\mathcal{E}$ that
restores the original state vector $\ket{\psi_\mathcal{SE}}$. Since it
follows from (A1) that the latter transformation has left the state of
$\mathcal{S}$ unchanged, but (A3) implies that the final state of
$\mathcal{S}$ (after the transformation and countertransformation) is
identical to the initial state of $\mathcal{S}$, the first
transformation $\widehat{u}_\mathcal{S} \otimes
\widehat{I}_\mathcal{E}$ cannot have altered the state of
$\mathcal{S}$ either. Thus, using assumption (A2), it follows that an
envariant transformation $\widehat{u}_\mathcal{S} \otimes
\widehat{I}_\mathcal{E}$ acting on $\ket{\psi_\mathcal{SE}}$ leaves
any measurable properties of $\mathcal{S}$ unchanged, in particular
the probabilities associated with outcomes of measurements performed
on $\mathcal{S}$.

Let us now consider two different envariant transformations: A \emph{phase
  transformation} of the form
\begin{equation} 
\widehat{u}_\mathcal{S}(\xi_1,\xi_2) = e^{i\xi_1} \ketbra{s_1}{s_1} + e^{i\xi_2} \ketbra{s_2}{s_2}
\end{equation}
that changes the phases associated with the Schmidt product states
$\ket{s_k}\ket{e_k}$ in Eq.~\eqref{eq:schmidt}, and a \emph{swap
  transformation}
\begin{equation} \label{eq:swap}
\widehat{u}_\mathcal{S}(1 \leftrightarrow 2) = e^{i\xi_{12}} \ketbra{s_1}{s_2} +
e^{i\xi_{21}} \ketbra{s_2}{s_1} 
\end{equation}
that exchanges the pairing of the $\ket{s_k}$ with the $\ket{e_l}$.
Based on the assumptions (A1)--(A3) mentioned above, envariance of
$\ket{\psi_\mathcal{SE}}$ under these transformations means that
measurable properties of $\mathcal{S}$ cannot depend on the phases
$\varphi_k$ in the Schmidt expansion of $\ket{\psi_\mathcal{SE}}$,
Eq.~\eqref{eq:schmidt}. Similarily, it follows that a swap
$\widehat{u}_\mathcal{S}(1 \leftrightarrow 2)$ leaves the state of
$\mathcal{S}$ unchanged, and that the consequences of the swap cannot
be detected by any measurement that pertains to $\mathcal{S}$ alone.

\subsubsection{Deducing the Born rule}

Together with an additional assumption, this result can then be used
to show that the probabilities of the ``outcomes'' $\ket{s_k}$
appearing in the Schmidt decomposition of $\ket{\psi_\mathcal{SE}}$
must be equal, thus arriving at Born's rule for the special case of a
state-vector expansion with coefficients of equal magnitude.
\citet{Zurek:2004:yb} offers three possibilities for such an
assumption. Here we shall limit our discussion to one of these
possible assumptions \citep[see also the comments in][]{Schlosshauer:2003:ms}:

\begin{enumerate} 

\item[(A4)] The Schmidt product states $\ket{s_k}\ket{e_k}$ appearing
  in the state-vector expansion of $\ket{\psi_\mathcal{SE}}$ imply a
  direct and perfect correlation of the measurement outcomes
  associated with the $\ket{s_k}$ and $\ket{e_k}$.  That is, if an
  observable $\widehat{O}_\mathcal{S} = \sum s_{kl} \ketbra{s_k}{s_l}$
  is measured on $\mathcal{S}$ and $\ket{s_k}$ is obtained, a
  subsequent measurement of $\widehat{O}_\mathcal{E} = \sum e_{kl}
  \ketbra{e_k}{e_l}$ on $\mathcal{E}$ will yield $\ket{e_k}$ with
  certainty (i.e., with probability equal to one).

\end{enumerate}

This assumption explicitly introduces a probability concept into the
derivation. (Similarly, the two other possible assumptions suggested
by Zurek establish a connection between the state of $\mathcal{S}$ and
probabilities of outcomes of measurements on $\mathcal{S}$.)

Then, denoting the probability for the outcome $\ket{s_k}$ by
$p(\ket{s_k}, \ket{\psi_\mathcal{SE}})$ when the composite system
$\mathcal{SE}$ is described by the state vector
$\ket{\psi_\mathcal{SE}}$, this assumption implies that
\begin{equation} \label{eq:1} 
p(\ket{s_k}; \ket{\psi_\mathcal{SE}}) = p(\ket{e_k};
\ket{\psi_\mathcal{SE}}).
\end{equation}
After acting on $\ket{\psi_\mathcal{SE}}$ with the envariant swap
transformation $\widehat{U}_\mathcal{S} = \widehat{u}_\mathcal{S}(1
\leftrightarrow 2) \otimes \widehat{I}_\mathcal{E}$ [see
Eq.~\eqref{eq:swap}] and using assumption (A4) again, we get
\begin{equation} \label{eq:2} 
\begin{split}
p(\ket{s_1}; \widehat{U}_\mathcal{S}\ket{\psi_\mathcal{SE}}) &= p(\ket{e_2};
\widehat{U}_\mathcal{S}\ket{\psi_\mathcal{SE}}), \\ p(\ket{s_2};
\widehat{U}_\mathcal{S}\ket{\psi_\mathcal{SE}}) &= p(\ket{e_1};
\widehat{U}_\mathcal{S}\ket{\psi_\mathcal{SE}}). 
\end{split}
\end{equation}
Now, when a ``counterswap'' $\widehat{U}_\mathcal{E} =
\widehat{I}_\mathcal{S} \otimes u_\mathcal{E}(1 \leftrightarrow 2)$ is
applied to $\ket{\psi_\mathcal{SE}}$, the original state vector
$\ket{\psi_\mathcal{SE}}$ is restored, i.e., $\widehat{U}_\mathcal{E}
(\widehat{U}_\mathcal{S} \ket{\psi_\mathcal{SE}}) =
\ket{\psi_\mathcal{SE}}$. It then follows from assumptions (A2) and
(A3) listed above that
\begin{equation} \label{eq:3} 
  p(\ket{s_k}; \widehat{U}_\mathcal{E} \widehat{U}_\mathcal{S}
  \ket{\psi_\mathcal{SE}}) = p(\ket{s_k}; \ket{\psi_\mathcal{SE}}).
\end{equation}
Furthermore, assumptions (A1) and (A2) imply that the first and second
swap cannot have affected the measurable properties of $\mathcal{E}$
and $\mathcal{S}$, respectively, particularly not the probabilities
for outcomes of measurements on $\mathcal{E}$ ($\mathcal{S}$),
\begin{equation} \label{eq:4}
\begin{split}
  p(\ket{s_k}; \widehat{U}_\mathcal{E} \widehat{U}_\mathcal{S}
  \ket{\psi_\mathcal{SE}}) &= p(\ket{s_k}; \widehat{U}_\mathcal{S}
  \ket{\psi_\mathcal{SE}}),\\
  p(\ket{e_k}; \widehat{U}_\mathcal{S} \ket{\psi_\mathcal{SE}}) &=
  p(\ket{e_k}; \ket{\psi_\mathcal{SE}}).
\end{split}
\end{equation}
Combining Eqs.~\eqref{eq:1}--\eqref{eq:4} yields
\begin{eqnarray}
  p(\ket{s_1}; \ket{\psi_\mathcal{SE}}) &\stackrel{\eqref{eq:3}}{=}& p(\ket{s_1};
  \widehat{U}_\mathcal{E} \widehat{U}_\mathcal{S}
  \ket{\psi_\mathcal{SE}}) \nonumber \\ &\stackrel{\eqref{eq:4}}{=}&
  p(\ket{s_1}; \widehat{U}_\mathcal{S} 
  \ket{\psi_\mathcal{SE}}) \nonumber \\ &\stackrel{\eqref{eq:2}}{=}&
  p(\ket{e_2}; \widehat{U}_\mathcal{S} 
  \ket{\psi_\mathcal{SE}})  \nonumber \\ &\stackrel{\eqref{eq:4}}{=}& p(\ket{e_2};
  \ket{\psi_\mathcal{SE}})  \nonumber \\ &\stackrel{\eqref{eq:1}}{=} &
  p(\ket{s_2}; \ket{\psi_\mathcal{SE}}),
\end{eqnarray}
which establishes the desired result $p(\ket{s_1};
\ket{\psi_\mathcal{SE}})=p(\ket{s_2}; \ket{\psi_\mathcal{SE}})$.  The
general case of unequal coefficients in the Schmidt decomposition of
$\ket{\psi_\mathcal{SE}}$ can then be treated by means of a simple
counting method \citep{Zurek:2003:rv,Zurek:2004:yb}, leading to Born's
rule for probabilities that are rational numbers. Using a continuity
argument, this result can be further generalized to include
probabilities that cannot be expressed as rational numbers
\citep{Zurek:2004:yb}.

\subsubsection{Summary and outlook}

If one grants the stated assumptions, Zurek's development of the
theory of envariance offers a novel and promising way of deducing
Born's rule in a noncircular manner.  Compared to the relatively
well-studied field of decoherence, envariance and its consequences
have only begun to be explored. In this review, we have focused on
envariance in the context of a derivation of the Born rule, but other
far-reaching implications of envariance have recently been suggested
by \citet{Zurek:2004:yb}.  For example, envariance could also account
for the emergence of an environment-selected preferred basis (that is,
for environment-induced superselection) without an appeal to the trace
operation or to reduced density matrices. This could open up the
possibility of a redevelopment of the decoherence program based on
fundamental quantum-mechanical principles that do not require one to
presuppose the Born rule; this also might shed new light, for example,
on the interpretation of reduced density matrices that has led to much
controversy in discussions of decoherence (see Sec.~\ref{sec:redmat}).
As of now, the development of such ideas is at a very early stage, but
we can expect further interesting results derived from envariance in
the near future.

\section{\label{sec:interpret}The role of decoherence in
  interpretations of quantum mechanics}

It was not until the early 1970s that the importance of the
interaction of physical systems with their environments for a
realistic quantum-mechanical description of these systems was realized
and a proper viewpoint on such interactions was established
\citep{Zeh:1970:yt,Zeh:1973:wq}. It took another decade for the first
concise formulation of the theory of decoherence
\citep{Zurek:1981:dd,Zurek:1982:tv} to be worked out and for numerical
studies to be made that showed the ubiquity and effectiveness of
decoherence effects \citep{Joos:1985:iu}. Of course, by that time,
several interpretive approaches to quantum mechanics had already been
established, for example, Everett-style relative-state interpretations
\citep{Everett:1957:rw}, the concept of modal interpretations
introduced by \citet{Fraassen:1973:yb,Fraassen:1991:ys}, and the
pilot-wave theory of de~Broglie and Bohm \citep{Bohm:1952:rc}.

When the relevance of decoherence effects was recognized by (parts of)
the scientific community, decoherence provided a motivation for a
fresh look at the existing interpretations and for the introduction of
changes and extensions to these interpretations, as well as for new
interpretations.  Some of the central questions in this context were,
and still are, the following:
\begin{enumerate} 
  
\item Can decoherence by itself solve certain foundational issues at
  least for all practical purposes, such as to make certain
  interpretive additives superfluous? What, then, are the crucial
  remaining foundational problems?
  
\item Can decoherence protect an interpretation from empirical
  disproof?
  
\item Conversely, can decoherence provide a mechanism to
  exclude an interpretive strategy as incompatible with
  quantum mechanics and/or as empirically inadequate?
  
\item Can decoherence physically motivate some of the assumptions on
  which an interpretation is based and give them a more precise
  meaning?
  
\item Can decoherence serve as an amalgam that would unify and
  simplify a spectrum of different interpretations?

\end{enumerate}
These and other questions have been widely discussed, both in the
context of particular interpretations and with respect to the general
implications of decoherence for any interpretation of quantum
mechanics.  In particular, interpretations that uphold the universal
validity of the unitary Schr\"odinger time evolution, most notably
relative-state and modal interpretations, have frequently incorporated
environment-induced superselection of a preferred basis and
decoherence into their framework. It is the purpose of this section to
critically investigate the implications of decoherence for the
existing interpretations of quantum mechanics, with particular
attention to the questions outlined above.

\subsection{General implications of decoherence for interpretations}

When measurements are understood as ubiquitous interactions that lead
to the formation of quantum correlations, the selection of a preferred
basis becomes in most cases a fundamental requirement.  This also
corresponds, in general, to the question of what properties are being
ascribed to systems (or worlds, minds, etc.).  Thus the
preferred-basis problem is at the heart of any interpretation of
quantum mechanics. Some of the difficulties that must be faced in
solving the preferred-basis problem are 

\begin{enumerate}

\item[(i)] to decide whether the
selection of any preferred basis (or quantity or property) is
justified at all or only an artefact of our subjective experience;

\item[(ii)] if we decide on (i) in the positive, to select those determinate
quantity or quantities (what appears determinate to us does not
need to be appear determinate to other kinds of observers, nor does it
need to be the ``true'' determinate property); 

\item[(iii)] to avoid any \emph{ad hoc} character of the choice and
  any possible empirical inadequacy or inconsistency with the
  confirmed predictions of quantum mechanics;
  
\item[(iv)] if a multitude of quantities is selected that apply
  differently among different systems, to be able to formulate
  specific rules that specify the determinate quantity or quantities
  under every circumstance;
  
\item[(v)] to ensure that the basis is chosen such that if the system
  is embedded in a larger (composite) system, the principle of
  property composition holds, i.e., the property selected by the basis
  of the original system should also persist when the system is
  considered as part of a larger composite system.\footnote{This is a
    problem encountered in some modal interpretations
    \citep[see][]{Clifton:1996:op}.}

\end{enumerate}
  
The hope is then that environment-induced superselection of a
preferred basis can provide a universal mechanism that fulfills the
above criteria and solves the preferred-basis problem on strictly
physical grounds.

A popular reading of the decoherence program typically goes as
follows. First, the interaction of the system with the environment
selects a preferred basis, i.e., a particular set of quasiclassical
robust states that commute, at least approximately, with the
Hamiltonian governing the system--environment interaction.  Since the
form of the interaction Hamiltonians usually depends on familiar
``classical'' quantities, the preferred states will typically also
correspond to the small set of ``classical'' properties.  Decoherence
then quickly damps superpositions between the localized preferred
states when only the system is considered. This is taken as an
explanation of the appearance to a local observer of a ``classical''
world of determinate, ``objective'' (in the sense of being robust)
properties.  The tempting interpretation of these achievements is then
to conclude that this accounts for the observation of unique (via
environment-induced superselection) and definite (via decoherence)
pointer states at the end of the measurement, and the measurement
problem appears to be solved, at least for all practical purposes.

However, the crucial difficulty in the above reasoning is justifying
the second step: How is one to interpret the local suppression of
interference in spite of the fact that full coherence is retained in
the total state that describes the system-environment combination?
While the local destruction of interference allows one to infer the
emergence of an (improper) ensemble of individually localized
components of the wave function, one still needs to impose an
interpretive framework that explains why only one of the localized
states is realized and/or perceived. This has been done in various
interpretations of quantum mechanics, typically on the basis of the
decohered reduced density matrix to ensure consistency with the
predictions of the Schr\"odinger dynamics and thus empirical adequacy.

In this context, one might raise the question whether retention of
full coherence in the composite state of the system-environment
combination could ever lead to empirical conflicts with the ascription
of definite values to (mesoscopic and macroscopic) systems in some
decoherence-based interpretive approach.  After all, one could think
of enlarging the system so as to include the environment in such a way
that measurements could now actually reveal the persisting quantum
coherence even on a macroscopic level.  However, \citet{Zurek:1982:tv}
asserted that such measurements would be impossible to carry out in
practice, a statement that was supported by a simple model calculation
by \citet[p.~356]{Omnes:1992:gy} for a body with a macrosopic number
($10^{24}$) of degrees of freedom.

\subsection{The standard and the Copenhagen interpretation}

As is well known, the standard interpretation (``orthodox'' quantum
mechanics) postulates that every measurement induces a discontinuous
break in the unitary time evolution of the state through the collapse
of the total wave function onto one of its terms in the state-vector
expansion (uniquely determined by the eigenbasis of the measured
observable), which selects a single term in the superposition as
representing the outcome. The nature of the collapse is not at all
explained, and thus the definition of measurement remains unclear.
Macroscopic superpositions are not \emph{a priori} forbidden, but are
never observed since any observation would entail a measurementlike
interaction. In the following, we shall also consider a ``Copenhagen''
variant of the standard interpretation, which adds an additional key
element, postulating the necessity of classical concepts in order to
describe quantum phenomena, including measurements.

\subsubsection{The problem of definite outcomes}

The interpretive rule of orthodox quantum mechanics that tells us when
we can speak of outcomes is given by the {e-e} link.\footnote{It is
  not particularly relevant for the subsequent discussion whether the
  {e-e} link is assumed in its ``exact'' form, i.e., requiring the
  exact eigenstates of an observable, or a ``fuzzy'' form that allows
  the ascription of definiteness based on only approximate eigenstates
  or on wave functions with (tiny) ``tails.''}  This is an
``objective'' criterion since it allows us to infer the existence of a
definite state of the system to which a value of a physical quantity
can be ascribed. Within this interpretive framework (and without
presuming the collapse postulate) decoherence cannot solve the problem
of outcomes: Phase coherence between macroscopically different pointer
states is preserved in the state that includes the environment, and we
can always enlarge the system so as to include (at least parts of) the
environment. In other words, the superposition of different pointer
positions still exists, coherence is only ``\emph{delocalized} into
the larger system'' \citep[p.~5]{Kiefer:1998:rz}, that is, into the
environment---or, as \citet[p.~224]{Joos:1985:iu} put it, ``the
interference terms still exist, but they are not \emph{there}''---and
the process of decoherence could in principle always be reversed.
Therefore, if we assume the orthodox {e-e} link to establish the
existence of determinate values of physical quantities, decoherence
cannot ensure that the measuring device actually ever is in a definite
pointer state (unless, of course, the system is initially in an
eigenstate of the observable), or that measurements have outcomes at
all. Much of the general criticism directed against decoherence with
respect to its ability to solve the measurement problem (at least in
the context of the standard interpretation) has been centered on this
argument.

Note that, with respect to the global postmeasurement state vector,
given by the final step in Eq.~\eqref{eq:measurement2}, the
interaction with the environment has only led to additional
entanglement. it has not transformed the state vector in any way,
since the rapidly increasing orthogonality of the states of the
environment associated with the different pointer positions has not
influenced the state description at all. In brief, the entanglement
brought about by interaction with the environment could even be
considered as making the measurement problem worse.
\citet[Sec.~3.2]{Bacciagaluppi:2003:yz} puts it like this:
\begin{quote} {\small 
    Intuitively, if the environment is carrying out, without our
    intervention, lots of approximate position measurements, then the
    measurement problem ought to apply more widely, also to these
    spontaneously occurring measurements. (\dots) The state of the
    object and the environment could be a superposition of zillions of
    very well localised terms, each with slightly different positions,
    and which are collectively spread over a macroscopic distance,
    even in the case of everyday objects.  (\dots) If everything is in
    interaction with everything else, everything is entangled with
    everything else, and that is a worse problem than the entanglement
    of measuring apparatuses with the measured probes.
}\end{quote}
Only once we have formed the reduced pure-state density matrix
$\widehat{\rho}_\mathcal{SA}$, Eq.~\eqref{eq:rho-S+A}, can the
orthogonality of the environmental states have an effect; then,
$\widehat{\rho}_\mathcal{SA}$ dynamically evolves into the improper
ensemble $\widehat{\rho}_\mathcal{SA}^{\,\,d}$
[Eq.~\eqref{eq:rho-S+A-diagonal}]. However, as pointed out in our
general discussion of reduced density matrices in
Sec.~\ref{sec:redmat}, the orthodox rule of interpreting
superpositions prohibits regarding the components in the sum of
Eq.~\eqref{eq:rho-S+A-diagonal} as corresponding to individual
well-defined quantum states.

Rather than considering the postdecoherence state of the system (or,
more precisely, of the system-apparatus combination $\mathcal{SA}$),
we can instead analyze the influence of decoherence on the
\emph{expectation values} of observables pertaining to $\mathcal{SA}$;
after all, such expectation values are what local observers would
measure in order to arrive at conclusions about $\mathcal{SA}$.  The
diagonalized reduced density matrix, Eq.~\eqref{eq:rho-S+A-diagonal},
together with the trace relation, Eq.~\eqref{eq:global-to-local},
implies that, for all practical purposes, the statistics of the system
$\mathcal{SA}$ will be indistinguishable from that of a proper mixture
(ensemble) by any local observation on $\mathcal{SA}$. That is, given
(i) the trace rule $\langle \widehat{O} \rangle =
\text{Tr}(\widehat{\rho}\widehat{O})$ and (ii) the interpretation of
$\langle \widehat{O} \rangle$ as the expectation value of an
observable $\widehat{O}$, the expectation value of any observable
$\widehat{O}_\mathcal{SA}$ restricted to the local system
$\mathcal{SA}$ will be for all practical purposes identical to the
expectation value of this observable if $\mathcal{SA}$ had been in one
of the states $\ket{s_n} \ket{a_n}$ (as if $\mathcal{SA}$ were
described by an ensemble of states).  In other words, decoherence has
effectively removed any interference terms (such as $\ket{s_m}
\ket{a_m} \bra{a_n} \bra{s_n}$ where $m \not= n$) from the calculation
of the trace
$\text{Tr}(\widehat{\rho}_\mathcal{SA}\widehat{O}_\mathcal{SA})$ and
thereby from the calculation of the expectation value $\langle
\widehat{O}_\mathcal{SA} \rangle$.  It has therefore been claimed that
\emph{formal equivalence}---i.e., the fact that decoherence transforms
the reduced density matrix into a form identical to that of a density
matrix representing an ensemble of pure states---yields
\emph{observational equivalence} in the sense above, namely, the
(local) indistiguishability of the expectation values derived from
these two types of density matrices via the trace rule.

But we must be careful in interpreting the correspondence between the
mathematical formalism (such as the trace rule) and the common terms
employed in describing ``the world.'' In quantum mechanics, the
identification of the expression ``$\text{Tr}(\rho A)$'' as the
expectation value of a quantity relies on the mathematical fact that,
when writing out this trace, it is found to be equal to a sum over the
possible outcomes of the measurement, weighted by the Born
probabilities for the system to be ``thrown'' into a particular state
corresponding to each of these outcomes in the course of a
measurement. This certainly represents our common-sense intuition
about the meaning of expectation values as the sum over possible
values that can appear in a given measurement, multiplied by the
relative frequency of actual occurrence of these values in a series of
such measurements. This interpretation, however, presumes (i) that
measurements have outcomes, (ii) that measurements lead to definite
``values,'' (iii) that measurable physical quantities are identified as
operators (observables) in a Hilbert space, and (iv) that the modulus
square of the expansion coefficients of the state in terms of the
eigenbasis of the observable can be interpreted as representing
probabilities of actual measurement outcomes (Born rule).

Thus decoherence brings about an apparent (and approximate) mixture of
states that seem, based on the models studied, to correspond well to
those states that we perceive as determinate.  Moreover, our
observation tells us that this apparent mixture indeed looks like a
proper ensemble in a measurement situation, as we observe that
measurements lead to the ``realization'' of precisely one state in the
``ensemble.''  But within the framework of the orthodox
interpretation, decoherence cannot explain this crucial step from an
apparent mixture to the existence and/or perception of single
outcomes.

\subsubsection{Observables, measurements, and environment-induced superselection}

In the standard and Copenhagen interpretationS, property ascription is
determined by an observable that represents the measurement of a
physical quantity and that in turn defines the preferred basis.
However, any Hermitian operator can play the role of an observable,
and thus any given state has the potential for an infinite number of
different properties whose attribution is usually mutually exclusive
unless the corresponding observables commute (in which case they share
a common eigenbasis which preserves the uniqueness of the preferred
basis). What then determines the observable that is being measured? As
our discussion in Sec.~\ref{sec:pbprob} has demonstrated, the
derivation of the measured observable from the particular form of a
given state-vector expansion can lead to paradoxial results since this
expansion is in general nonunique, so the observable must be chosen by
other means. In the standard and Copenhagen interpretations, it is
essentially the ``user'' who simply ``chooses'' the particular
observable to be measured and thus determines which properties the
system possesses.

This positivist point of view has, of course, led to a lot of
controversy, since it runs counter to the attempt to establish an
observer-independent reality that has been the central pursuit of
natural science since its beginning.  Moreover, in practice, one
certainly does not have the freedom to choose any arbitrary observable
and measure it; instead, one has ``instruments'' (including one's
senses) that are designed to measure a particular observable. For most
(and maybe all) practical purposes, this will ultimately boil down to
a single relevant observable, namely, position. But what, then, makes
the instruments designed for such a particular observable?

Answering this crucial question essentially means abandoning the
orthodox view of treating measurements as a ``black box'' process that
has little, if any, relation to the workings of actual physical
measurements (where measurements can here be understood in the
broadest sense of a ``monitoring'' of the state of the system).  The
first key point, the formalization of measurements as a formation of
quantum correlations between system and apparatus, goes back to the
early years of quantum mechanics and is reflected in the measurement
scheme of von \citet{Neumann:1932:gq}, but it does not resolve the
issue of how the choice of observables is made. The second key point,
the explicit inclusion of the environment in a description of the
measurement process, was brought into quantum theory by the studies of
decoherence. Zurek's~\citeyearpar{Zurek:1981:dd} stability criterion
discussed in Sec.~\ref{sec:einsel} has shown that measurements must be
of such a nature as to establish stable records, where stability is to
be understood as preserving the system-apparatus correlations in spite
of the inevitable interaction with the surrounding environment.  The
``user'' cannot choose the observables arbitrarily, but must design a
measuring device whose interaction with the environment is such as to
ensure stable records in the sense above (which, in turn, defines a
measuring device for this observable). In the reading of orthodox
quantum mechanics, this can be interpreted as the environment
determining the properties of the system.

In this sense, the decoherence program has embedded the rather formal
concept of measurement as proposed by the standard and Copenhagen
interpretations---with its vague notion of observables that are
seemingly freely chosen by the observer---in a more realistic and
physical framework. This is accomplished via the specification of
observer-free criteria for the selection of the measured observable
through the physical structure of the measuring device and its
interaction with the environment, which is, in most cases, needed to
amplify the measurement record and thereby to make it accessible to
the external observer.

\subsubsection{The concept of classicality in the Copenhagen interpretation}

The Copenhagen interpretation additionally postulates that
classicality is not to be derived from quantum mechanics, for example,
as the macroscopic limit of an underlying quantum structure (as is in
some sense assumed, but not explicitely derived, in the standard
interpretation), but instead that it be viewed as an indispensable and
irreducible element of a complete quantum theory---and, in fact, be
considered as a concept prior to quantum theory.  In particular, the
Copenhagen interpretation assumes the existence of macroscopic
measurement apparatuses that obey classical physics and that are not
supposed to be described in quantum mechanical terms (in sharp
contrast to the von Neumann measurement scheme, which rather belongs
to the standard interpretation); such a classical apparatus is
considered necessary in order to make quantum-mechanical phenomena
accessible to us in terms of the ``classical'' world of our
experience. This strict dualism between the system $\mathcal{S}$, to
be described by quantum mechanics, and the apparatus $\mathcal{A}$,
obeying classical physics, also entails the existence of an
essentially fixed boundary between $\mathcal{S}$ and $\mathcal{A}$,
which separates the microworld from the macroworld (the ``Heisenberg
cut''). This boundary cannot be moved significantly without destroying
the observed phenomenon (i.e., the full interacting compound
$\mathcal{SA}$).

Especially in the light of the insights gained from decoherence it
seems impossible to uphold the notion of a fixed quantum--classical
boundary on a fundamental level of the theory.  Environment-induced
superselection and suppression of interference have demonstrated how
quasiclassical robust states can emerge, or remain absent, using the
quantum formalism alone and over a broad range of microscopic to
macroscopic scales, and have established the notion that the boundary
between $\mathcal{S}$ and $\mathcal{A}$ is to a large extent movable
towards $\mathcal{A}$. Similar results have been obtained from the
general study of quantum nondemolition measurements \citep[see, for
example, Chap.~19 of][]{Auletta:2000:rv} which include the monitoring
of a system by its environment. Also note that since the apparatus is
described in classical terms, it is macroscopic by definition; but not
every apparatus must be macrosopic: the actual ``instrument'' could
well be microscopic. Only the ``amplifier'' must be macrosopic. As an
example, consider Zurek's \citeyearpar{Zurek:1981:dd} toy model of
decoherence, outlined in Sec.~\ref{sec:zurekmodel}, in which the
instrument can be represented by a bistable atom while the environment
plays the role of the amplifier; a more realistic example is a
macrosopic detector of gravitational waves that is treated as a
quantum-mechanical harmonic oscillator.

Based on the progress already achieved by the decoherence program, it
is reasonable to anticipate that decoherence embedded in some
additional interpretive structure could lead to a complete and
consistent derivation of the classical world from quantum mechanical
principles. This would make the assumption of an intrinsically
classical apparatus (which has to be treated outside of the realm of
quantum mechanics) appear as neither a necessary nor a viable
postulate. \citet[p.~22]{Bacciagaluppi:2003:az} refers to this
strategy as ``having Bohr's cake and eating it'': acknowledging the
correctness of Bohr's notion of the necessity of a classical world
(``having Bohr's cake''), but being able to view the classical world
as part of and as emerging from a purely quantum-mechanical world.

\subsection{Relative-state interpretations \label{sec:everett}}

Everett's original~\citeyearpar{Everett:1957:rw} proposal of a
relative-state interpretation of quantum mechanics has motivated
several strands of interpretation, presumably owing to the fact that
Everett himself never clearly spelled out how his theory was supposed
to work. The system-observer duality of orthodox quantum mechanics
introduces into the theory external ``observers'' who are not
described by the deterministic laws of quantum systems but instead
follow a stochastic indeterminism. This approach obviously runs into
problems when the universe as a whole is considered: by definition,
there cannot be any external observers. The central idea of Everett's
proposal is then to abandon duality and instead (i) to assume the
existence of a total state $\ket{\Psi}$ representing the state of the
entire universe and (ii) to uphold the universal validity of the
Schr\"odinger evolution, while (iii) postulating that all terms in the
superposition of the total state at the completion of the measurement
actually correspond to physical states. Each such physical state can
be understood as relative (a) to the state of the other part in the
composite system \citep[as in Everett's original proposal; also
see][]{Rovelli:1996:rq,Mermin:1998:ii}, (b) to a particular ``branch''
of a constantly ``splitting'' universe \citetext{the {\em many-worlds
    interpretations}, popularized by \citealp{DeWitt:1970:pl} and
  \citealp{Deutsch:1985:rx}}, or (c) to a particular ``mind'' in the
set of minds of the conscious observer \citep[the {\em many-minds
  interpretation}; see, for example,][]{Lockwood:1996:pu}. In other
words, every term in the final-state superposition can be viewed as
representing an equally ``real'' physical state of affairs that is
realized in a different ``branch of reality.''

Decoherence adherents have typically been inclined towards
relative-state interpretations \citep[for
instance][]{Zeh:1970:yt,Zeh:1973:wq,Zeh:1993:lt,Zurek:1998:re},
presumably because the Everett approach takes unitary quantum
mechanics essentially ``as is,'' with a minimum of added interpretive
elements. This matches well the spirit of the decoherence program,
which attempts to explain the emergence of classicality purely from
the formalism of basic quantum mechanics. It may also seem natural to
identify the decohering components of the wave function with different
Everett branches. Conversely, proponents of relative-state
interpretations have frequently employed the mechanism of decoherence
in solving the difficulties associated with this class of
interpretations \citep[see, for
example,][]{Saunders:1995:zx,Saunders:1997:za,Saunders:1998:rc,%
Deutsch:1985:rx,Deutsch:1996:fz,Deutsch:2001:aq,Vaidmain:1998:zp,%
Wallace:2003:iq,Wallace:2003:iz}.

There are many different readings and versions of relative-state
interpretations, especially with respect to what defines the
``branches,'' ``worlds,'' and ``minds''; whether we deal with one, a
multitude, or an infinity of such worlds and minds; and whether there
is an actual (physical) or only perspectival splitting of the worlds
and minds into different branches corresponding to the terms in the
superposition. Does the world or mind split into two separate copies
(thus somehow doubling all the matter contained in the orginal
system), or is there just a ``reassignment'' of states to a multitude
of worlds or minds of constant (typically infinite) number, or is
there only one physically existing world or mind in which each branch
corresponds to different ``aspects'' (whatever they are). Regardless,
in the following discussion of the key implications of decoherence,
the precise details and differences of these various strands of
interpretation will, for the most part, be largely irrelevant.

Relative-state interpretations face two core difficulties. First, the
preferred-basis problem: If states are only relative, the question
arises, relative to what? What determines the particular basis terms
that are used to define the branches, which in turn define the worlds
or minds in the next instant of time? When precisely does the
``splitting'' occur?  Which properties are made determinate in each
branch, and how are they connected to the determinate properties of
our experience? Second, what is the meaning of probabilities, since
\emph{every} outcome actually occurs in some world or mind, and how
can Born's rule be motivated in such an interpretive framework?

\subsubsection{Everett branches and the preferred-basis problem}

\citet[p.~1043]{Stapp:2002:pc} stated the requirement that ``a
many-worlds interpretation of quantum theory exists only to the extent
that the associated basis problem is solved.'' In the context of
relative-state interpretations, the preferred-basis problem is not
only much more severe than in the orthodox interpretation, but also
more fundamental for several reasons: (i) The branching occurs
continuously and essentially everywhere; in the general sense of
measurements understood as the formation of quantum correlations,
every newly formed correlation, whether it pertains to microscopic or
macroscopic systems, corresponds to a branching. (ii) The ontological
implications are much more drastic, at least in those relative-state
interpretations, which assume an actual ``splitting'' of worlds or
minds, since the choice of the basis determines the resulting
``world'' or ``mind'' as a whole.

The environment-based basis superselection criteria of the decoherence
program have frequently been employed to solve the preferred-basis
problem of relative-state interpretations \citep[see, for
example,][]{Zurek:1998:re,Butterfield:2001:ua,Wallace:2003:iq,Wallace:2003:iz}.
There are several advantages in a decoherence-related approach to
selecting the preferred Everett bases: First, no \emph{a priori}
existence of a preferred basis needs to be postulated, but instead the
preferred basis arises naturally from the physical criterion of
robustness.  Second, the selection will be likely to yield empirical
adequacy, since the decoherence program is derived solely from the
well-confirmed Schr\"odinger dynamics (modulo the possibility that
robustness may not be the universally valid criterion).  Lastly, the
decohered components of the wave function evolve in such a way that
they can be reidentified over time (forming ``trajectories'' in the
preferred state spaces) and thus can be used to define stable,
temporally extended Everett branches. Similarly, such trajectories can
be used to ensure robust observer record states and/or environmental
states that make information about the state of the system of interest
widely accessible to observers (see, for example, Zurek's
``existential interpretation,'' outlined in
Sec.~\ref{sec:exist-interpret} below).

While the idea of directly associating the environment-selected basis
states with Everett worlds seems natural and straightforward, it has
also been subject to criticism. \citet{Stapp:2002:pc} has argued that
an Everett-type interpretation must aim at determining a denumerable
set of distinct branches that correspond to the apparently discrete
events of our experience. Among these branches one must be able to
assign determinate values and finite probabilities according to the
usual rules and therefore one would need to be able to specify a
denumerable set of mutually orthogonal projection operators.  It is
well known, however \citep{Zurek:1998:re}, that the preferred states
chosen through the interaction with the environment via the stability
criterion frequently form an overcomplete set of states---often a
continuum of narrow Gaussian-type wave packets, for example, the
coherent states of harmonic-oscillator models that are not necessarily
orthogonal \citetext{i.e., the Gaussians may overlap; see
  \citealp{Kubler:1973:ux,Zurek:1993:qq}}.  Stapp therefore considers
this approach to the preferred-basis problem in relative-state
interpretations to be unsatisfactory. \citet{Zurek:2003:uu} has
rebutted this criticism by pointing out that a collection of harmonic
oscillators that would lead to such overcomplete sets of Gaussians
cannot serve as an adequate model of the human brain, and it is
ultimately only in the brain where the perception of denumerability
and mutual exclusiveness of events must be accounted for (see
Sec.~\ref{sec:objsubj}); when neurons are more appropriately modeled
as two-state systems, the issue raised by Stapp disappears (for a
discussion of decoherence in a simple two-state model, see
Sec.~\ref{sec:zurekmodel}).\footnote{For interesting quantitative
  results on the role of decoherence in neuronal processes, see
  \citet{Tegmark:2000:wz}.}

The approach of using environment-induced superselection and
decoherence to define the Everett branches has also been critized on
grounds of being ``conceptually approximate,'' since the stability
criterion generally leads only to an approximate specification of a
preferred basis and therefore cannot give an ``exact'' definition of
the Everett branches \citetext{see, for example, the comments of
  \citealp{Zeh:1973:wq,Kent:1990:nm}, and also the well-known
  ``anti-FAPP'' position of \citealp{Bell:1982:ag}}.
\citet[pp.~90--91]{Wallace:2003:iz} has argued against such an
objection as
\begin{quote} {\small
    (\dots) arising from a view implicit in much discussion of
    Everett-style interpretations: that certain concepts and objects
    in quantum mechanics must either enter the theory formally in its
    axiomatic structure, or be regarded as illusion. (\dots) [Instead]
    the emergence of a classical world from quantum mechanics is to be
    understood in terms of the emergence from the theory of certain
    sorts of structures and patterns, and \dots this means that we have
    no need (as well as no hope!) of the precision which Kent [in his
    \citeyearpar{Kent:1990:nm} critique] and others (\dots) demand.}
\end{quote}
Accordingly, in view of our argument in Sec.~\ref{sec:objsubj} for
considering subjective solutions to the measurement problem as
sufficient, there is no {\em a priori} reason to doubt that an
``approximate'' criterion for the selection of the preferred basis can
give a meaningful definition of the Everett branches---one that is
empirically adequate and that accounts for our experiences. The
environment-superselected basis emerges naturally from the physically
very reasonable criterion of robustness together with the purely
quantum mechanical effect of decoherence. It would be rather difficult
to imagine rgar ab axiomatically introduced ``exact'' rule could be
able to select preferred bases in a manner that is similarly
physically motivated and capable of ensuring empirical adequacy.

Besides using the environment-superselected pointer states to describe
the Everett branches, various authors have directly used the
instantaneous Schmidt decomposition of the composite state (or,
equivalently, the set of orthogonal eigenstates of the reduced density
matrix) to define the preferred basis (see also
Sec.~\ref{sec:schmidt}). This approach is easier to implement than the
explicit search for dynamically stable pointer states since the
preferred basis follows directly from a simple mathematical
diagonalization procedure at each instant of time. Furthermore, it has
been favored by some \citep[e.g.,][]{Zeh:1973:wq} since it gives an
``exact'' rule for basis selection in relative-state interpretations;
the consistently quantum origin of the Schmidt decomposition, which
matches well the ``pure quantum-mechanics'' spirit of Everett's
proposal (where the formalism of quantum mechanics supplies its own
interpretation), has also been counted as an advantage
\citep{Barvinsky:1995:pa}. In an earlier work, \citet{Deutsch:1985:rx}
attributed a fundamental role to the Schmidt decomposition in
relative-state interpretations as defining an ``interpretation basis''
that imposes the precise structure that is needed to give meaning to
Everett's basic concept.

However, as pointed out in Sec.~\ref{sec:schmidt}, emerging basis
states based on the instantaneous Schmidt states will frequently have
properties that are very different from those selected by the
stability criterion and that are undesirably nonclassical. For
example, they may lack the spatial localization of the
robustness-selected Gaussians \citep{Stapp:2002:pc}. The question to
what extent the Schmidt basis states correspond to classical
properties in Everett-style interpretations was investigated in detail
by \citet{Barvinsky:1995:pa}. The authors study the similarity of the
states selected by the Schmidt decomposition to coherent states (i.e.,
minimum-uncertainty Gaussians) that are chosen as the ``yardstick
states'' representing classicality \citep[see also][]{Eisert:2003:ib}.
For the investigated toy models it is found that only subsets of the
Everett worlds corresponding to the Schmidt decomposition exhibit
classicality in this sense; furthermore, the degree of robustness of
classicality in these branches is very sensitive to the choice of the
initial state and the interaction Hamiltonian, such that classicality
emerges typically only temporarily, and the Schmidt basis generally
lacks robustness under time evolution.  Similar difficulties with the
Schmidt basis approach have been described by \citet{Kent:1997:oz}.

These findings indicate that a selection criterion based on robustness
provides a much more meaningful, physically transparent, and general
rule for the selection of quasiclassical branches in relative-state
interpretations, especially with respect to its ability to lead to
wave-function components representing quasiclassical properties that
can be reidentified over time (which a simple diagonalization of the
reduced density matrix at each instant of time does not, in general,
allow for).

\subsubsection{Probabilities in Everett interpretations}

Various attempts unrelated to decoherence have been made to find a
consistent derivation of the Born probabilities \citep[for
instance,][]{Everett:1957:rw,Hartle:1968:gg,DeWitt:1971:pz,%
  Graham:1973:ww,Geroch:1984:yt,Deutsch:1999:tz} in the explicit or
implicit context of a relative-state interpretation, but several
arguments have been presented that show that these approaches
fail.\footnote{See, for example, the critiques of
  \citet{Stein:1984:uu,Kent:1990:nm,Squires:1990:lz,Barnum:2000:oz};
  however, also note the arguments of \citet{Wallace:2003:zr} and
  \citet{Gill:2003:tz}, defending the approach of
  \citet{Deutsch:1999:tz}; see also \citet{Saunders:2002:tz}.}  When
the effects of decoherence and environment-induced superselection are
included, it seems natural to identify the diagonal elements of the
decohered reduced density matrix (in the environment-superselected
basis) with the set of possible elementary events and to interpret the
corresponding coefficients as relative frequencies of worlds (or
minds, etc.) in the Everett theory, assuming a typically infinite
multitude of worlds, minds, etc.  Since decoherence enables one to
reidentify the individual localized components of the wave function
over time (describing, for example, observers and their measurement
outcomes attached to individual well-defined ``worlds''), this leads
to a natural interpretation of the Born probabilities as empirical
frequencies.

However, decoherence cannot yield an actual derivation of the Born
rule \citep[for attempts in this direction,
see][]{Deutsch:1999:tz,Zurek:1998:re}.  As mentioned before, this is
so because the key elements of the decoherence program, the formalism
and the interpretation of reduced density matrices and the trace rule,
\emph{presume} the Born rule.  Attempts to consistently derive
probabilities from reduced density matrices and from the trace rule
are therefore subject to the charge of circularity
\citep{Zeh:1996:gy,Zurek:2002:ii}. In Sec.~\ref{sec:envar}, we
outlined a recent proposal by \citet{Zurek:2003:rv} that evades this
circularity and deduces the Born rule from envariance, a symmetry
property of entangled systems, and from certain assumptions about the
connection between the state of the system $\mathcal{S}$ of interest,
the state vector of the composite system $\mathcal{SE}$ that includes
an environment $\mathcal{E}$ entangled with $\mathcal{S}$, and
probabilities of outcomes of measurements performed on $\mathcal{S}$.
Decoherence combined with this approach provides a framework in which
quantum probabilities and the Born rule can be given a rather natural
motivation, definition, and interpretation in the context of
relative-state interpretations.

\subsubsection{\label{sec:exist-interpret}The ``existential
  interpretation''}

A well-known Everett-type interpretation that relies heavily on
decoherence has been proposed by Zurek
\citetext{\citeyear{Zurek:1993:pu}, \citeyear{Zurek:1998:re}; see also
  the recent reevaluation in \citealp{Zurek:2004:yb}}.  This approach,
termed the ``existential interpretation,'' defines the reality, or
objective existence, of a state as the possibility of finding out what
the state is and simultaneously leaving it unperturbed, similar to a
classical state.\footnote{This intrinsically requires the notion of
  open systems, since in isolated systems, the observer would need to
  know in advance what observables commute with the state of the
  system, in order to perform a nondemolition measurement that avoids
  repreparing the state of the system.} Zurek assigns a ``relative
objective existence'' to the robust states (identified with elementary
``events'') selected by the environmental stability criterion. By
measuring properties of the system-environment interaction
Hamiltonian and employing the robustness criterion, the observer can,
at least in principle, determine the set of observables that can be
measured on the system without perturbing it and thus find out its
``objective'' state.  Alternatively, the observer can take advantage
of the redundant records of the state of the system as monitored by
the environment. By intercepting parts of this environment, for
example, a fraction of the surrounding photons, he can determine the
state of the system essentially without perturbing it \citep[\cf also
the related recent ideas of ``quantum Darwinism'' and the role of the
environment as a ``witness,''
see][]{Zurek:2000:tr,Zurek:2002:ii,Zurek:2003:pl,Ollivier:2003:za}.\footnote{The
  partial ignorance is necessary to avoid  redefinition of the
  state of the system.}

Zurek emphasizes the importance of stable records for observers, i.e.,
robust correlations between the environment-selected states and the
memory states of the observer. Information must be represented
physically, and thus the ``objective'' state of the observer who has
detected one of the potential outcomes of a measurement must be
physically distinct and objectively different from the state of an
observer who has recorded an alternative outcome (since the record
states can be determined from the outside without perturbing
them---see the previous paragraph). The different objective states
of the observer are, via quantum correlations, attached to different
branches defined by the environment-selected robust states; they thus
ultimately label the different branches of the universal state
vector. This is claimed to lead to the perception of classicality; the
impossibility of perceiving arbitrary superpositions is explained via
the quick suppression of interference between different memory states
induced by decoherence, where each (physically distinct) memory state
represents an individual observer identity.

A similar argument has been given by \citet{Zeh:1993:lt} who
employs decoherence together with an (implicit) branching process to
explain the perception of definite outcomes: 
\begin{quote} {\small
    [A]fter an observation one need not necessarily conclude that only
    one component now \emph{exists} but only that only one component
    \emph{is observed}.  (\dots) Superposed world components
    describing the registration of different macroscopic properties by
    the ``same'' observer are dynamically entirely independent of one
    another: they describe different observers. (\dots) He who
    considers this conclusion of an indeterminism or splitting of the
    observer's identity, derived from the Schr\"odinger equation in
    the form of dynamically decoupling (``branching'') wave packets on a
    fundamental global configuration space, as unacceptable or
    ``extravagant'' may instead dynamically formalize the superfluous
    hypothesis of a disappearance of the ``other'' components by
    whatever method he prefers, but he should be aware that he may
    thereby also create his own problems: Any deviation from the
    global Schr\"odinger equation must in principle lead to observable
    effects, and it should be recalled that none have ever been
    discovered.
}\end{quote}
The existential interpretation has recently been connected to the
theory of envariance \citep[see][and
Sec.~\ref{sec:envar}]{Zurek:2004:yb}. In particular, the derivation of
Born's rule based on envariance as outlined in Sec.~\ref{sec:envar}
can be recast in the framework of the existential interpretation such
that probabilities refer explicitly to the future record state of an
observer. Such a concept of probability  bears similarities with
classical probability theory \citep[for more details on these ideas,
see][]{Zurek:2004:yb}.

The existential interpretation continues Everett's goal of
interpreting quantum mechanics using the quantum-mechanical formalism
itself. Zurek takes the standard no-collapse quantum theory ``as is''
and explores to what extent the incorporation of environment-induced
superselection and decoherence (and recently also envariance) could
form a viable interpretation that would, with a minimal additional
interpretive framework, be capable of accounting for the perception of
definite outcomes and of explaining the origin and nature of
probabilities.

\subsection{Modal interpretations}

The first type of modal interpretation was suggested by
\citet{Fraassen:1973:yb,Fraassen:1991:ys}, based on his program of
``constructive empiricism,'' which proposes to take only empirical
adequacy, but not necessarily ``truth,'' as the goal of science. Since
then, a large number of interpretations of quantum mechanics have been
suggested that can be considered as modal \citep[for a review and
discussion of some of the basic properties and problems of such
interpretations, see][]{Clifton:1996:op}.

In general, the approach of modal interpretations consists in
weakening the orthodox {e-e} link by allowing for the assignment of
definite measurement outcomes even if the system is not in an
eigenstate of the observable representing the measurement.  In this
way, one can preserve a purely unitary time evolution without the need
for an additional collapse postulate to account for definite
measurement results. Of course, this immediately raises the question
of how physical properties that are perceived through measurements and
measurement results are connected to the state, since the
bidirectional link is broken between the eigenstate of the observable
(which corresponds to the physical property) and the eigenvalue (which
represents the manifestation of the value of this physical property in
a measurement). The general goal of modal interpretations is then to
specify rules that determine a catalog of possible properties of a
system described by the density matrix $\rho$ at time $t$. Two
different views are typically distinguished: a semantic approach that
only changes the way of talking about the connection between
properties and state; and a realistic view that provides a different
specification of what the possible properties of a system really are,
given the state vector (or the density matrix).

Such an attribution of possible properties must fulfill certain
requirements. For instance, probabilities for outcomes of measurements
should be consistent with the usual Born probabilities of standard
quantum mechanics; it should be possible to recover our experience of
classicality in the perception of macroscopic objects; and an explicit
time evolution of properties and their probabilities should be
definable that is consistent with the results of the Schr\"odinger
equation. As we shall see in the following, decoherence has frequently
been employed in modal interpretations to motivate and define rules
for property ascription. \citet{Dieks:1994:oz,Dieks:1994:rc} has
argued that one of the central goals of modal approaches is to provide
an interpretation for decoherence.

\subsubsection{Property assignment based on environment-induced superselection}

The intrinsic difficulty of modal interpretations is to avoid any
\emph{ad hoc} character of the property assignment, yet to find
generally applicable rules that lead to a selection of possible
properties that include the determinate properties of our experience.
To solve this problem, various modal interpretations have embraced the
results of the decoherence program. A natural approach would be to
employ the environment-induced superselection of a preferred
basis---since it is based on an entirely physical and very general
criterion (namely, the stability requirement) and has, for the cases
studied, been shown to give results that agree well with our
experience, thus matching van Fraassen's goal of empirical
adequacy---to yield sets of possible quasiclassical properties
associated with the correct probabilities.

Furthermore, since the decoherence program is based solely on
Schr\"odinger dynamics, the task of defining a time evolution of the
``property states'' and their associated probabilities that is in
agreement with the results of unitary quantum mechanics would
presumably be easier than in a model of property assignment in which
the set of possibilities does not arise dynamically via the
Schr\"odinger equation alone \citep[for a detailed proposal for modal
dynamics of the latter type, see][]{Bacciagaluppi:1999:iz}. The need
for explicit dynamics of property states in modal interpretations is
controversial.  One can argue that it suffices to show that at each
instant of time, the set of possibly possessed properties that can be
ascribed to the system is empirically adequate, in the sense of
containing the properties of our experience, especially with respect
to the properties of macroscopic objects \citep[this is essentially
the view of, for example,][]{Fraassen:1973:yb,Fraassen:1991:ys}.  On
the other hand, this cannot ensure that these properties behave over
time in agreement with our experience (for instance, that macroscopic
objects that are left undisturbed do not change their position in
space spontaneously in an observable manner). In other words, the
emergence of classicality is to be tied not only to determinate
properties at each instant of time, but also to the existence of
quasiclassical ``trajectories'' in property space.  Since decoherence
allows one to reidentify components of the decohered density matrix
over time, this could be used to derive property states with
continuous, quasiclassical trajectorylike time evolution based on
Schr\"odinger dynamics alone. For some discussions of this approach,
see \citet{Hemmo:1996:fz} and \citet{Bacciagaluppi:1999:iz}.

The fact that the states emerging from decoherence and the stability
criterion are sometimes nonorthogonal or form a continuum will
presumably be of even less relevance in modal interpretations than in
Everett-style interpretations (see Sec.~\ref{sec:everett}) since the
goal here is solely to specify sets of possible properties, of which
only one set actually gets assigned to the system. Hence an
``overlap'' of the sets is not necessarily a problem (modulo the
potential difficulty of a straightforward assignment of probabilities
in such a situation).

\subsubsection{Property assignment based on instantaneous Schmidt decompositions}

Since it is usually rather difficult to determine explicitly the
robust ``pointer states'' through the stability (or a similar)
criterion, it would not be easy to specify a general rule for property
assignment based on environment-induced superselection.  To simplify
this situation, several modal interpretations have restricted
themselves to the orthogonal decomposition of the density matrix to
define the set of properties that can be assigned \citep[see, for
instance,][]{Kochen:1985:po,Healey:1989:cd,Dieks:1989:rm,Vermaas:1995:gd,Bub:1997:iq}.
For example, the approach of \citet{Dieks:1989:rm} recognizes, by
referring to the decoherence program, the relevance of the environment
by considering a composite system-environment state vector and its
diagonal Schmidt decomposition, $\ket{\psi} = \sum_k \sqrt{p_k} \,\,
\ket{\phi^\mathcal{S}_k} \ket{\phi^\mathcal{E}_k}$, which always
exists. Possible properties that can be assigned to the system are
then represented by the Schmidt projectors $\widehat{P}_k = \lambda_k
\ketbra{\phi^\mathcal{S}_k}{\phi^\mathcal{S}_k}$. Although all terms
are present in the Schmidt expansion (that Dieks calls the
``mathematical state''), the ``physical state'' is postulated to be
given by only one of the terms, with probability $p_k$. A
generalization of this approach to a decomposition into any number of
subsystems has been described by \citet{Vermaas:1995:gd}. In this
sense, the Schmidt decomposition itself is taken to define an
interpretation of quantum mechanics. \citet{Dieks:1995:aa} suggested a
physical motivation for the Schmidt decomposition in modal
interpretations based on the assumed requirement of a one-to-one
correspondence between the properties of the system and its
environment. For a comment on the violation of the property
composition principle in such interpretations, see the analysis of
\citet{Clifton:1996:op}.

A central problem associated with the approach of orthogonal
decomposition is that it is not at all clear that the properties
determined by the Schmidt diagonalization represent the determinate
properties of our experience. As outlined in Sec.~\ref{sec:schmidt},
the states selected by the (instantaneous) orthogonal decomposition of
the reduced density matrix will in general differ from the robust
``pointer states'' chosen by the stability criterion of the
decoherence program and may have distinctly nonclassical properties.
That this will be the case especially when the states selected by the
orthogonal decomposition are close to degeneracy has already been
indicated in Sec.~\ref{sec:schmidt}. It has also been explored in more
detail in the context of modal interpretations by
\citet{Bacciagaluppi:1995:zx} and \citet{Donald:1998:xz}, who showed
that in the case of near degeneracy (as it typically occurs for
macroscopic systems with many degrees of freedom), the resulting
projectors will be extremely sensitive to the precise form of the
state \citep{Bacciagaluppi:1995:zx}. Clearly such sensitivity is 
undesired since the projectors, and thus the properties of the system,
will not be well behaved under the inevitable approximations employed
in physics \citep{Donald:1998:xz}.

\subsubsection{Property assignment based on decompositions of the decohered density matrix}

Other authors therefore have appealed to the orthogonal decomposition
of the decohered reduced density matrix (instead of the decomposition
of the instantaneous density matrix) which has led to noteworthy
results.  When the system is represented by only a finite-dimensional
Hilbert space, a discrete model of decoherence, the resulting states
were indeed found to be typically close to the robust states selected
by the stability criterion (for macroscopic systems, this typically
meant localization in position space), unless again the final
composite state was very nearly degenerate
\citetext{\citealp{Bacciagaluppi:1996:po,Bene:2001:po}; see also
  Sec.~\ref{sec:schmidt}}. Thus, in sufficiently nondegenerate cases,
decoherence can ensure that the definite properties selected by modal
interpretations of the Dieks type will be appropriately close to the
properties corresponding to the ideal pointer states if the modal
properties are based on the orthogonal decomposition of the reduced
decohered density matrix.

On the other hand, \citet{Bacciagaluppi:2000:yz} showed that in the
more general and realistic case of an infinite-dimensional state space
of the system, when one employs a continuous model of decoherence
\citep[namely, that of][]{Joos:1985:iu}, the predictions of the modal
approach \citep{Dieks:1989:rm,Vermaas:1995:gd} and those of
decoherence can differ significantly. It was demonstrated that the
definite properties obtained from the orthogonal decomposition of the
decohered density matrix were highly delocalized (that is, smeared out
over the entire spread of the state), although the coherence length of
the density matrix itself was shown to be very small, so that
decoherence indicated very localized properties.  Thus, based on these
results \citep[and similar ones of][]{Donald:1998:xz}, decoherence can
be used to argue for the physical inadequacy of the rule for the
assignment of definite properties as proposed by \citet{Dieks:1989:rm}
and \citet{Vermaas:1995:gd}.

More generally, if the definite properties selected by the modal
interpretation fail to mesh with the results of decoherence (in
particular, when they also lack the desired classicality and
correspondence to the determinate properties of our experience), we
are given reason to doubt whether the proposed rules for property
assignment have sufficient physical motivation, legitimacy, or
generality.

\subsubsection{Concluding remarks}

There are many different proposals that can be grouped under the
heading of modal interpretations. They all share the problem of
motivating and verifying a consistent system of property assignment.
Using the robust pointer states selected by interaction with the
environment and by the stability criterion is a step in the right
direction, but the difficulty remains to derive a general rule for
property assignment from this method that would yield explicitly the
sets of possibilities in every situation.  In certain cases, for
example, close to degeneracy and in Hilbert-state spaces of infinite
dimension, the simpler approach of deriving the possible properties
from the orthogonal decomposition of the decohered reduced density
matrix fails to yield the sharply localized, quasiclassical pointer
states as selected by environmental robustness criteria.  These are
the cases in which decoherence can play a vital role in helping to
identify inadequate rules for property assignment in modal
interpretations.

\subsection{Physical collapse theories}

The basic idea of physical collapse theories is to introduce an
explicit modification of the Schr\"odinger time evolution to achieve a
physical mechanism for state-vector reduction \citep[for an extensive
recent review, see][]{Bassi:2003:yb}. This is in general motivated by
a ``realist'' interpretation of the state vector, that is, the state
vector is directly identified with a physical state, which then
requires reduction to one of the terms in the superposition to
establish equivalence to the observed determinate properties of
physical states, at least as far as the macroscopic realm is
concerned.

The first proposals for theories of this type were made by
\citet{Pearle:1976:on,Pearle:1979:rq,Pearle:1982:rv} and
\citet{Gisin:1984:qs}, who developed dynamical reduction models that
modify unitary dynamics such that a superposition of quantum states
evolves continuously into one of its terms \citep[see also the review
by][]{Pearle:1999:cr}. Typically, terms representing external white
noise are added to the Schr\"odinger equation, causing the squared
amplitudes $|c_n(t)|^2$ in the state-vector expansion
$\ket{\Psi(t)}=\sum_n c_n(t) \ket{\psi_n}$ to fluctuate randomly in
time, while maintaining the normalization condition $\sum_n
|c_n(t)|^2=1$ for all $t$. This process is known as {\em stochastic
  dynamical reduction}.  Eventually one amplitude $|c_n(t)|^2
\rightarrow 1$, while all other squared coefficients $\rightarrow 0$
(the ``gambler's ruin game'' mechanism), where $|c_n(t)|^2 \rightarrow
1$ with probability $|c_n(t=0)|^2$ (the squared coefficients in the
initial precollapse state-vector expansion) in agreement with the Born
probability interpretation of the expansion coefficients.
   
These early models exhibit two main difficulties.  First, the
preferred-basis problem: What determines the terms in the state-vector
expansion into which the state vector gets reduced? Why does reduction
lead to precisely the distinct macroscopic states of our experience
and not superpositions thereof? Second, how can one account for the
fact that the effectiveness of collapsing superpositions increases
when going from microscopic to macroscopic scales?

These problems motivated {\em spontaneous localization} models,
initially proposed by Ghirardi, Rimini, and Weber
\citep[GRW;][]{Ghirardi:1986:ud}. Here state-vector reduction is not
implemented as a dynamical process (i.e., as a continuous evolution
over time), but instead occurs instantaneously and spontaneously,
leading to a spatial localization of the wave function. To be precise,
the $N$-particle wave function $\psi(\mathbf{x}_1, \hdots,
\mathbf{x}_N)$ is at random intervals multiplied by a Gaussian of the
form $\exp \bigl[ -(\mathbf{X}-\mathbf{x}_k)^2 / 2\Delta^2 \bigr]$
(this process is often called a ``hit'' or a ``jump''), and the
resulting product is subsequently normalized.  The occurrence of these
hits is not explained, but rather postulated as a new fundamental
physical mechanism. Both the coordinate $\mathbf{x}_k$ and the
``center of the hit'' $\mathbf{X}$ are chosen at random, but the
probability for a specific $\mathbf{X}$ is postulated to be given by
the squared inner product of $\psi(\mathbf{x}_1, \hdots,
\mathbf{x}_N)$ with the Gaussian (and therefore hits are more likely
to occur where $|\psi|^2$, viewed as a function of $\mathbf{x}_k$
only, is large).  The mean frequency $\nu$ of hits for a single
microscopic particle is chosen so as to effectively preserve unitary
time evolution for microscopic systems, while ensuring that for
macroscopic objects containing a very large number $N$ of particles
the localization occurs rapidly (on the order of $N\nu$), in such a
way as to preclude the persistence of spatially separated macroscopic
superpositons (such as the pointer's being in a superpositon of ``up''
and ``down'') on time scales shorter than realistic observations could
resolve. \citet{Ghirardi:1986:ud} chose $\nu \approx
10^{-16}~\text{s}^{-1}$, so a macrosopic system with $N \approx
10^{23}$ particles undergoes localization on average every
$10^{-7}$~s.  Inevitable coupling to the environment can in
general be expected to lead to a further drastic increase of $N$ and
therefore to an even higher localization rate. Note, however, that the
localization process itself is independent of any interaction
with environment, in sharp contrast to the decoherence approach.

Subsequently, the ideas of the stochastic dynamical reduction and GRW
theory were combined into {\em continuous spontaneous localization}
models \cite{Pearle:1989:cs,Ghirardi:1990:lm} in which localization of
the GRW type can be shown to emerge from a nonunitary, nonlinear It\^o
stochastic differential equation, namely, the Schr\"odinger equation
augmented by spatially correlated Brownian motion terms \citep[see
also][]{Diosi:1988:wx,Diosi:1989:yb}. The particular choice of 
stochastic term determines the preferred basis. Frequently, the
stochastic term has been based on the mass density which yields a
GRW-type spatial localization
\citep{Diosi:1989:yb,Pearle:1989:cs,Ghirardi:1990:lm}, but stochastic
terms driven by the Hamiltonian, leading to a reduction on an energy
basis, have also been studied
\citep{Bedford:1975:un,Bedford:1977:un,Milburn:1991:tv,Percival:1995:om,%
Percival:1998:om,Hughston:1996:in,Fivel:1997:ia,%
Adler:2002:rc,Adler:2001:tr,Adler:2000:gv}.  If we focus on the first
type of term, the Ghirardi-Rimini-Weber theory and continuous
spontaneous localization become phenomenologically similar, and we
shall refer to them jointly as ``spontaneous localization'' models in
the following discussion whenever it is unnecessary to distinguish
them explicitly.

\subsubsection{The preferred-basis problem}
  
Physical reduction theories typically remove wave-function collapse
from the restrictive context of the orthodox interpretation (where the
external observer arbitrarily selects the measured observable and thus
determines the preferred basis), and rather understand reduction as a
universal mechanism that acts constantly on every state vector
regardless of an explicit measurement situation. In view of this it is
particularly important to provide a definition for the states into
which the wave function collapses.

As mentioned before, the original stochastic dynamical reduction
models suffer from this preferred-basis problem. Taking into account
environment-induced superselection of a preferred basis could help
resolve this issue.  Decoherence has been shown to occur, especially
for mesoscopic and macroscopic objects, on extremely short time
scales, and thus would presumably be able to bring about basis
selection much faster than the time required for dynamical
fluctuations to establish a ``winning'' expansion coefficient.

In contrast, the GRW theory solves the preferred-basis problem by
postulating a mechanism that leads to reduction to a particular state
vector in an expansion on a position basis, i.e., position is assumed
to be the universal preferred basis. State-vector reduction then
amounts to simply modifying the functional shape of the projection of
the state vector $\ket{\psi}$ onto the position basis
$\bra{\mathbf{x}_1, \hdots, \mathbf{x}_N}$.  This choice can be
motivated by the insight that essentially all (human) observations
must be grounded in a position measurement.\footnote{This measurement
  may ultimately occur only in the brain of the observer; see the
  objection to the GRW model by \citet{Albert:1989:ps}. With respect
  to the general preference for position as the basis of measurements,
  see also the comment by \citet{Bell:1982:ag}.}

On the one hand, the selection of position as the preferred basis is
supported by the decoherence program, since physical interactions
frequently are governed by distance-dependent laws. Given the
stability criterion or a similar requirement, this leads to position
as the preferred observable. In this sense, decoherence provides a
physical motivation for the assumption of the GRW model.  On the other
hand, it makes this assumption appear as too restrictive as it cannot
account for cases in which position is not the preferred basis---for
instance, in microscopic systems where typically energy is the robust
observable, or in the superposition of (macroscopic) currents in
SQUIDs. The GRW model simply excludes such cases by choosing the
parameters of the spontaneous localization process such that
microscopic systems remain generally unaffected by any state vector
reduction.  The basis selection approach proposed by the decoherence
program is therefore much more general and also avoids the \emph{ad
  hoc} character of the GRW theory by allowing for a range of
preferred observables and motivating their choice on physical grounds.

A similar argument can be made with respect to the continuous
spontaneous localization approach. Here, one essentially preselects a
preferred basis through the particular choice of the stochastic terms
added to the Schr\"odinger equation.  This allows for a greater range
of possible preferred bases, for instance by combining terms driven by
the Hamiltonian and by the mass density, leading to a competition
between localization in energy and position space (corresponding to
the two most frequently observed eigenstates).  Nonetheless, any
particular choice of terms will again be subject to the charge of
possessing an \emph{ad hoc} flavor, in contrast to the physical
definition of the preferred basis derived from the structure of the
unmodified Hamiltonian as suggested by environment-induced selection.

\subsubsection{Simultaneous presence of decoherence and spontaneous
  localization}  

Since decoherence can be considered as an omnipresent phenomenon that
has been extensively verified both theoretically and experimentally,
the assumption that a physical collapse theory holds means that the
evolution of a system must be guided by both decoherence effects
\emph{and} the reduction mechanism.

Let us first consider the situation in which decoherence and the
localization mechanism act constructively in the same direction, i.e.,
towards a common preferred basis. This raises the question in which
order these two effects influence the evolution of the system
\citep{Bacciagaluppi:2003:yz}. If localization occurs on a shorter
time scale than environment-induced superselection of a preferred
basis and suppression of local interference, decoherence will in most
cases have very little influence on the evolution of the system, since
typically the system will already have evolved into a reduced state.
Conversely, if decoherence effects act more quickly on the system than
the localization mechanism, the interaction with the environment will
presumably lead to the preparation of quasiclassical robust states
that are subsequently chosen by the localization mechanism. As pointed
out in Sec.~\ref{sec:interf}, decoherence usually occurs on extremely
short time scales, which can be shown to be significantly smaller than
the action of the spontaneous localization process for most cases
\citetext{for studies related to the GRW model, see
  \citealp{Tegmark:1993:uz} and \citealp{Benatti:1995:re}}. This
indicates that decoherence will typically play an important role even
in the presence of physical wave-function reduction.

The second case occurs when decoherence leads to the selection of a
different preferred basis than the reduction basis specified by the
localization mechanism. As remarked by
\citet{Bacciagaluppi:2003:yz,Bacciagaluppi:2003:az} in the context of
the GRW theory, one might then imagine the collapse either to occur
only at the level of the environment (which would then serve as an
amplifying and recording device with different localization properties
than the system under study), or to lead to an explicit competition
between decoherence and localization effects.

\subsubsection{The tails problem}

The clear advantage of physical collapse models over the consideration
of decoherence-induced effects alone for a solution to the measurement
problem lies in the fact that an actual state reduction is achieved
such that one may be tempted to conclude that at the conclusion of the
reduction process the system actually is in a determinate state.
However, all collapse models achieve only an approximate (``for all
practical purposes'') reduction of the wave function.  In the case of
dynamical reduction models, the state will always retain small
interference terms for finite times.  Similarly, in the GRW theory the
width $\Delta$ of the multiplying Gaussian cannot be made arbitrarily
small, and therefore the reduced wave packet cannot be made infinitely
sharply localized in position space, since this would entail an
infinitely large energy gain by the system via the time-energy
uncertainty relation, which would certainly show up experimentally
\citetext{\citealp{Ghirardi:1986:ud}, chose $\Delta \approx
  10^{-5}~\text{cm}$}. This need for only an approximate reduction
leads to wave function ``tails'' \citep{Albert:1996:po}, that is, in
any region in space and at any time $t>0$, the wave function will
remain nonzero if it has been nonzero at $t=0$ (before the collapse),
and thus there will be always a part of the system that is not
``here.''

Physical collapse models that achieve reduction only ``for all
practical purposes'' require a modification, namely, a weakening, of
the orthodox {e-e} link to allow one to speak of the system's actually
being in a definite state, and thereby to ensure the objective
attribution of determinate properties to the system.\footnote{It
  should be noted, however, that such ``fuzzy'' {e-e} links may in
  turn lead to difficulties, as the discussion of Lewis's ``counting
  anomaly'' has shown \citep{Lewis:1997:ta}.} In this sense, collapse
models are as much ``fine for all practical purposes'' \citep[to
paraphrase][]{Bell:1990:po} as decoherence is, where perfect
orthogonality of the environment states is only attained as $t
\rightarrow \infty$.  The severity of the consequences, however, is
not equivalent for the two strategies. Since collapse models directly
change the state vector, a single outcome is at least approximately
selected, and it only requires a ``slight'' weakening of the {e-e}
link to make this state of affairs correspond to the (objective)
existence of a determinate physical property. In the case of
decoherence, the lack of a precise destruction of interference terms
is not the main problem; even if exact orthogonality of the
environment states were ensured at all times, the resulting reduced
density matrix would represent an improper mixture, with no outcome
having been singled out according to the {e-e} link. This would be the
case regardless of whether the {e-e} link is expressed in the strong
or weakened form, and we would still have to supply some additional
interpretative framework to explain our perception of outcomes
\citep[see also the comment by][]{Ghirardi:1987:po}.

\subsubsection{Connecting decoherence and collapse models} 

It was realized early that there exists a striking formal similarity
of the equations that govern the time evolution of density matrices in
the GRW approach and in models of decoherence. For example, the GRW
equation for a single free mass point reads
[\citealp{Ghirardi:1986:ud}, Eq.~(3.5)]
\begin{equation} 
i \frac{\partial
  \rho(x,x',t)}{\partial t} = \frac{1}{2m} \bigg[
\frac{\partial^2}{\partial x^2} - \frac{\partial^2}{\partial x'^2}
\bigg] \rho - i \Lambda (x-x')^2 \rho, 
\end{equation}
where the second term on the right-hand side accounts for the
destruction of spatially separated interference terms. A simple model
for environment-induced decoherence yields a very similar equation
[\citealp{Joos:1985:iu}, Eq.~(3.75); see also the comment by
\citealp{Joos:1987:yu}]. Thus the physical justification for an
\emph{ad hoc} postulate of an explicit reduction-inducing mechanism
could be questioned (of course modulo the important interpretive
difference between the approximately proper ensembles arising from
collapse models and the improper ensembles resulting from decoherence;
see also \citealp{Ghirardi:1987:po}). More constructively, the
similarity of the governing equations might enable one to choose the
free parameters in collapse models on physical grounds rather than on
the basis of empirical adequacy.  Conversely, this similiarity can
also be viewed as leading to a ``protection'' of physical collapse
theories from empirical disproof.  This is so because the inevitable
and ubiquitous interaction with the environment will always, for all
practical purposes of observation (that is, of statistical
prediction), result in (local) density matrices that are formally very
similar to those of collapse models. What is measured is not the state
vector itself, but the probability distribution of outcomes, i.e.,
values of a physical quantity and their frequency, and this
information is equivalently contained in the state vector and the
density matrix.  Measurements with their intrinsically local character
will presumably be unable to distinguish between the probability
distribution given by the decohered reduced density matrix and the
probability distribution defined by an (approximately) proper mixture
obtained from a physical collapse. In other words, as long as the free
parameters of collapse theories are chosen in agreement with those
determined from decoherence, models for state-vector reduction can be
expected to be empirically adequate since decoherence is an effect
that will be present with near certainty in every realistic
(especially macroscopic) physical system.

One might of course speculate that the simultaneous presence of both
decoherence and reduction effects might actually allow for an
experimental disproof of collapse theories by preparing states that
differ in an observable manner from the predictions of the reduction
models.\footnote{For proposed experiments to detect the GRW collapse,
  see for example \citet{Squires:1991:az} and \citet{Rae:1990:wa}. For
  experiments that could potentially demonstrate deviations from the
  predictions of quantum theory when dynamical state-vector reduction
  is present, see \citet{Pearle:1984:qz,Pearle:1986:po}.} If we
acknowledge the existence of interpretations of quantum mechanics that
employ only decoherence-induced suppression of interference to explain
the perception of apparent collapses \citetext{as is, for example,
  claimed by the ``existential interpretation'' of
  \citealp{Zurek:1993:pu,Zurek:1998:re}; see
  Sec.~\ref{sec:exist-interpret}}, we will not be able to distinguish
experimentally between a ``true'' collapse and a mere suppression of
interference as explained by decoherence.  Instead, an experimental
situation is required in which the collapse model predicts a collapse,
but in which no suppression of interference through decoherence
arises. Again, the problem in the realization of such an experiment is
that it is very difficult to shield a system from decoherence effects,
especially since we will typically require a mesoscopic or macroscopic
system in which the reduction is efficient enough to be observed.  For
example, based on explicit numerical estimates,
\citet{Tegmark:1993:uz} has shown that decoherence due to scattering
of environmental particles such as air molecules or photons will have
a much stronger influence than the proposed GRW effect of spontaneous
localization \citetext{see also
  \citealp{Bassi:2003:yb,Benatti:1995:re}; for different results for
  energy-driven reduction models, \cf \citealp{Adler:2002:rc}}.

\subsubsection{Summary and outlook} 

Decoherence has the distinct advantage of being derived directly from
the laws of standard quantum mechanics, whereas current collapse
models are required to postulate their reduction mechanism as a new
fundamental law of nature. On the other hand, collapse models yield,
at least for all practical purposes, proper mixtures, so they are
capable of providing an ``objective'' solution to the measurement
problem. The formal similarity between the time evolution equations of
the collapse and decoherence models nourishes hopes that the
postulated reduction mechanisms of collapse models could possibly be
derived from the ubiquituous and inevitable interaction of every
physical system with its environment and the resulting decoherence
effects. We may therefore regard collapse models and decoherence not
as mutually exclusive alternatives for a solution to the measurement
problem, but rather as potential candidates for a fruitful
unification. For a vague proposal along these lines, see
\citet{Pessoa:1998:yl}; \cf also \citet{Diosi:1989:yb} and
\citet{Pearle:1999:cr} for speculations that quantum gravity might act
as a collapse-inducing universal ``environment.''

\subsection{Bohmian mechanics}

Bohm's approach \citep{Bohm:1952:rc,Bohm:1966:ps,Bohm:1993:ll} is a
modification of de~Broglie's~\citeyearpar{DeBroglie:1930:yt} original
``pilot-wave'' proposal. In Bohmian mechanics, a system containing $N$
(nonrelativistic) particles is described by a wave function $\psi(t)$
and the configuration $\mathcal{Q}(t)=\bigl(\mathbf{q}_1(t), \hdots,
\mathbf{q}_N(t)\bigr) \in \mathbb{R}^{3N}$ of particle positions
$\mathbf{q}_i(t)$, i.e., the state of the system is represented by
$(\psi, \mathcal{Q})$ for each instant $t$. The evolution of the
system is guided by two equations. The wave function $\psi(t)$
is transformed as usual via the standard Schr\"odinger equation, $i
\hbar (\partial / \partial t) \psi = \widehat{H}\psi$, while the
particle positions $\mathbf{q}_i(t)$ of the configuration
$\mathcal{Q}(t)$ evolve according to the ``guiding equation''
\begin{equation} \frac{d\mathbf{q}_i}{dt}  =
\mathbf{v}_i^\psi (\mathbf{q}_1, \hdots, \mathbf{q}_N) \equiv
\frac{\hbar}{m_i} \, \mathrm{Im} \frac{\psi^* \nabla_{\mathbf{q}_i}
  \psi}{\psi^*\psi} (\mathbf{q}_1, \hdots, \mathbf{q}_N), 
\end{equation}
where $m_i$ is the mass of the $i$th particle. Thus the particles
follow determinate trajectories described by $\mathcal{Q}(t)$, with
the distribution of $\mathcal{Q}(t)$ being given by the quantum
equilibrium distribution $\rho = |\psi|^2$.

\subsubsection{Particles as fundamental entities} 

Bohm's theory has been critized for ascribing fundamental ontological
status to particles. General arguments against particles on a
fundamental level of any relativistic quantum theory have been
frequently given \citetext{see, for instance,
  \citealp{Malament:1996:gt}, and
  \citealp{Halvorson:2002:xz}}.\footnote{On the other hand, there are
  proposals for a ``Bohmian mechanics of quantum fields,'' i.e., a
  theory that embeds quantum field theory into a Bohmian-style
  framework \citep{Durr:2003:gu,Durr:2002:gs}.}  Moreover, and this is
the point we would like to discuss in this section, it has been argued
that the appearance of particles (``discontinuities in space'') could
be derived from the continuous process of decoherence, leading to
claims that no fundamental role need be attributed to particles
\citep{Zeh:1993:lt,Zeh:1999:rr,Zeh:2003:pp}.  Based on decohered
density matrices of mesoscopic and macroscopic systems that
essentially always represent quasi-ensembles of narrow wave packets in
position space, \citet[p.~190]{Zeh:1993:lt} holds that such wave
packets can be viewed as representing individual ``particle''
positions:\footnote{\citet{Schrodinger:1926:pz} had made an attempt
  into a similar direction but had failed since the Schr\"odinger
  equation tends to continuously spread out any localized wavepacket
  when it is considered as describing an isolated system. The
  inclusion of an interacting environment and thus decoherence
  counteracts the spread and opens up the possibility of maintaining
  narrow wave packets over time \citep{Joos:1985:iu}.}
\begin{quote} {\small 
    All particle aspects observed in measurements of quantum fields
    (like spots on a plate, tracks in a bubble chamber, or clicks of a
    counter) can be understood by taking into account this decoherence
    of the relevant local (\emph{i.e.}, subsystem) density matrix.}
\end{quote}
The first question is then whether a narrow wave packet in position
space can be identified with the subjective experience of a
``particle.'' The answer appears to be yes: our notion of
``particles'' hinges on the property of localizability, i.e., the
definition of a region of space $\mathbf{\Omega} \in \mathbb{R}^3$ in
which the system (that is, the support of the wave function) is
entirely contained. Although the nature of the Schr\"odinger dynamics
implies that any wave function will have nonvanishing support
(``tails'') outside of any finite spatial region $\mathbf{\Omega}$ and
therefore exact localizatibility will never be achieved, we only need
to demand approximate localizability to account for our experience of
particle aspects.

However, to interpret the ensembles of narrow wave packets resulting
from decoherence as leading to the perception of individual particles,
we must embed standard quantum mechanics (with decoherence) into an
additional interpretive framework that explains why only one of the
wavepackets is perceived;\footnote{Zeh himself, like
  \citet{Zurek:1998:re}, adheres to an Everett-style branching to
  which distinct observers are attached \citep{Zeh:1993:lt}; see also
  the quote in Sec.~\ref{sec:everett}.}  that is, we do need to add
some interpretive rule to get from the improper ensemble emerging from
decoherence to the perception of individual terms, so decoherence
alone does not necessarily make Bohm's particle concept superfluous.
But it suggests that the postulate of particles as fundamental
entities could be unnecessary, and taken together with the
difficulties in reconciling such a particle theory with a relativistic
quantum field theory, Bohm's \emph{a priori} assumption of particles
at a fundamental level of the theory appears seriously challenged.

\subsubsection{Bohmian trajectories and decoherence}
 
A well-known property of Bohmian mechanics is the fact that its
trajectories are often highly nonclassical \citep[see, for
example,][]{Bohm:1993:ll,Holland:1993:fc,Appleby:1999:uy}.  This poses
the serious problem of how Bohm's theory can explain the existence of
quasiclassical trajectories on a macroscopic level.

\citet{Bohm:1993:ll} considered the scattering of a beam of
environmental particles on a macroscopic system, a process that is
known to give rise to decoherence \citep{Joos:1985:iu,Joos:2003:jh}.
The authors demonstrate that this scattering yields quasiclassical
trajectories for the system. It has further been shown that for
isolated systems, the Bohm theory will typically not give the correct
classical limit \citep{Appleby:1999:uy}.  It was thus suggested that
the inclusion of the environment and of the resulting decoherence
effects might be helpful in recovering quasiclassical trajectories in
Bohmian mechanics
\citep{Appleby:1999:zs,Zeh:1999:rr,Allori:2001:po,Allori:2001:tc,Allori:2001:vl,Sanz:2003:za}.

We mentioned before that the interaction between a macroscopic system
and its environment will typically lead to a rapid approximate
diagonalization of the reduced density matrix in position space, and
thus to spatially localized wave packets that follow (approximately)
Hamiltonian trajectories. [This observation also provides a physical
motivation for the choice of position as the fundamental preferred
basis in Bohm's theory, in agreement with Bell's
\citeyearpar{Bell:1982:ag} well-known comment that ``in physics the
only observations we must consider are position observations, if only
the positions of instrument pointers.''] The intuitive step is then to
associate these trajectories with the particle trajectories
$\mathcal{Q}(t)$ of the Bohm theory. As pointed out by
\citet{Bacciagaluppi:2003:az}, a great advantage of this strategy lies
in the fact that the same approach would allow for a recovery of both
quantum and classical phenomena.

However, a careful analysis by \citet{Appleby:1999:zs} showed that
this decoherence-induced diagonalization in the position basis alone
will in general not suffice to yield quasiclassical trajectories in
Bohm's theory; only under certain additional assumptions will
processes that lead to decoherence also give correct quasiclassical
Bohmian trajectories for macroscopic systems (Appleby described the
example of the long-time limit of a system that has initially been
prepared in an energy eigenstate).  Interesting results were also
reported by Allori and co-workers
\citep{Allori:2001:po,Allori:2001:tc,Allori:2001:vl}. They
demonstrated that decoherence effects can play the role of
preserving classical properties of Bohmian trajectories. Furthermore,
they showed that while in standard quantum mechanics it is important
to maintain narrow wave packets to account for the emergence of
classicality, the Bohmian description of a system by both its wave
function and its configuration allows for the derivation of
quasiclassical behavior from highly delocalized wave functions.
\citet{Sanz:2003:za} studied the double-slit experiment in the
framework of Bohmian mechanics and in the presence of decoherence and
showed that even when coherence is fully lost, and thus interference is
absent, nonlocal quantum correlations remain that influence the
dynamics of the particles in the Bohm theory, demonstrating that in
this example decoherence does not suffice to achieve the classical
limit in Bohmian mechanics.

In conclusion, while the basic idea of employing decoherence-related
processes to yield the correct classical limit of Bohmian trajectories
seems reasonable, many details of this approach still need to be
worked out.

\subsection{Consistent histories interpretations}

The consistent- (or decoherent-) histories approach was introduced by
\citet{Griffiths:1984:tr,Griffiths:1993:ll,Griffiths:1996:re} and
further developed by
\citet{Omnes:1988:lz,Omnes:1988:lq,Omnes:1988:zg,Omnes:1990:ww,%
  Omnes:1992:gy,Omnes:1994:pz,Omnes:2003:tt},
\citet{GellMann:1990:uz,GellMann:1991:wk,GellMann:1991:pp,GellMann:1993:oi},
\citet{Dowker:1992:vz}, and others.  Reviews of the program can be
found in the papers by \citet{Omnes:1992:gy} and
\citet{Halliwell:1993:ds,Halliwell:1995:gh}, as well as in the recent
book by \citet{Griffiths:2002:tr}. Thoughtful critiques investigating
key features and assumptions of the approach have been given, for
example, by \citet{Espagnat:1989:fl},
\citet{Dowker:1995:pa,Dowker:1996:ch}, \citet{Kent:1998:th}, and
\citet{Bassi:1999:zp}. The basic idea of the consistent-histories
approach is to eliminate the fundamental role of measurements in
quantum mechanics, and instead study quantum histories, defined as
sequences of events represented by sets of time-ordered projection
operators, and to assign probabilities to such histories. The approach
was originally motivated by quantum cosmology, i.e., the study of the
evolution of the entire universe, which, by definition, represents a
closed system. Therefore no external observer (which is, for example,
an indispensable element of the Copenhagen interpretation) can be
invoked.

\subsubsection{Definition of histories} 

We assume that a physical system $\mathcal{S}$ is described by a
density matrix $\rho_0$ at some initial time $t_0$ and define a sequence
of arbitrary times $t_1 < t_2 < \cdots < t_n$ with $t_1 > t_0$. For
each time point $t_i$ in this sequence, we consider an exhaustive set
$\mathcal{P}^{(i)} = \{ \widehat{P}^{(i)}_{\alpha_i}(t_i) \, | \,
\alpha_i = 1 \cdots m_i \}$, $1 \le i \le n$, of mutually orthogonal
Hermitian projection operators $\widehat{P}^{(i)}_{\alpha_i}(t_i)$,
obeying
\begin{equation} 
\sum_{\alpha_i} \widehat{P}^{(i)}_{\alpha_i}(t_i) = 1, \quad
\widehat{P}^{(i)}_{\alpha_i}(t_i) \widehat{P}^{(i)}_{\beta_i}(t_i) =
\delta_{\alpha_i,\beta_i} \widehat{P}^{(i)}_{\alpha_i}(t_i), 
\end{equation}
and evolving, using the Heisenberg picture, according to
\begin{equation}
\widehat{P}^{(i)}_{\alpha_i}(t) = U^\dagger(t_0,t)
\widehat{P}^{(i)}_{\alpha_i}(t_0) U(t_0,t),
\end{equation}
where $U(t_0,t)$ is the operator that dynamically propagates the state
vector from $t_0$ to $t$.

A possible, ``maximally fine-grained'' history is defined by the
sequence of times $t_1 < t_2 < \cdots < t_n$ and by the choice of one
projection operator in the set $\mathcal{P}^{(i)}$ for each time point
$t_i$ in the sequence, i.e., by the set
\begin{equation} \label{eq:hist-fine}
\mathcal{H}_{\{\alpha\}} = \{
\widehat{P}^{(1)}_{\alpha_1}(t_1), \widehat{P}^{(2)}_{\alpha_2}(t_2),
\dots, \widehat{P}^{(n)}_{\alpha_n}(t_n) \}.  
\end{equation}
We also define the set $\mathfrak{H}= \{\mathcal{H}_{\{\alpha\}} \}$
of all possible histories for a given time sequence $t_1 < t_2 <
\cdots < t_n$. The natural interpretation of a history
$\mathcal{H}_{\{\alpha\}}$ is then to take it as a series of
propositions of the form ``the system $\mathcal{S}$ was, at time
$t_i$, in a state of the subspace spanned by
$\widehat{P}^{(i)}_{\alpha_i}(t_i)$.''

Maximally fine-grained histories can be combined to form
``coarse-grained'' sets which assign to each time point $t_i$ a linear
combination
\begin{equation} 
\widehat{Q}^{(i)}_{\beta_i}(t_i) = \sum_{\alpha_i} \pi^{(i)}_{\alpha_i}
\widehat{P}^{(i)}_{\alpha_i}(t_i), \quad \pi^{(i)}_{\alpha_i} \in \{0,1\} 
\end{equation}
of the original projection operators $\widehat{P}^{(i)}_{\alpha_i}(t_i)$.

So far, the projection operators $\widehat{P}^{(i)}_{\alpha_i}(t_i)$
chosen at a certain instant $t_i$ in time in order to form a history
$\mathcal{H}_{\{\alpha\}}$ were independent of the choice of the
projection operators at earlier times $t_0 < t < t_i$ in
$\mathcal{H}_{\{\alpha\}}$. This situation was generalized by
\citet{Omnes:1988:lz,Omnes:1988:lq,Omnes:1988:zg,Omnes:1990:ww,Omnes:1992:gy}
to include ``branch-dependent'' histories of the form \citep[see
also][]{GellMann:1993:oi}
\begin{equation} \label{eq:hist-fine-branch}
\mathcal{H}_{\{\alpha\}} = \{
\widehat{P}^{(1)}_{\alpha_1}(t_1), \widehat{P}^{(2,\alpha_1)}_{\alpha_2}(t_2),
\dots, \widehat{P}^{(n,\alpha_1,\hdots,\alpha_{n-1})}_{\alpha_n}(t_n) \}.  
\end{equation}

\subsubsection{Probabilities and consistency} 

In standard quantum mechanics, we can always assign probabilities to
single events, represented by the eigenstates of some projection
operator $\widehat{P}^{(i)}(t)$, via the rule
\begin{equation}
p(i,t)=\text{Tr} \bigl[ \widehat{P}^{(i)\dagger}(t) \rho(t_0)
\widehat{P}^{(i)}(t) \bigr].
\end{equation}
The natural extension of this formula to the calculation of the
probability $p(\mathcal{H}_{\{\alpha\}})$ of a history
$\mathcal{H}_{\{\alpha\}}$ is given by
\begin{equation}  \label{eq:cprob}
p(\mathcal{H}_{\{\alpha\}}) = \mathcal{D}(\alpha,\alpha), 
\end{equation}
where the so-called {\em decoherence functional}
$\mathcal{D}(\alpha,\beta)$ is defined by \citep{GellMann:1990:uz} 
\begin{equation} \label{eq:df}
\mathcal{D}(\alpha,\beta) = \text{Tr} \big[ \widehat{P}^{(n)}_{\alpha_n}(t_n) 
 \cdots \widehat{P}^{(1)}_{\alpha_{1}}(t_{1}) \rho_0 
\widehat{P}^{(1)}_{\beta_{1}}(t_{1}) \cdots
\widehat{P}^{(n)}_{\beta_{n}}(t_n) \big].
\end{equation}
If we instead work in the Schr\"odinger picture, the decoherence
functional is

\begin{widetext}

\begin{equation} \label{eq:dfs}
\mathcal{D}(\alpha,\beta) = \text{Tr} \big[
\widehat{P}^{(n)}_{\alpha_n} U(t_{n-1},t_n) 
 \cdots \widehat{P}^{(1)}_{\alpha_{1}}  \rho(t_1) 
\widehat{P}^{(1)}_{\beta_{1}} \cdots
U^\dagger(t_{n-1},t_n) \widehat{P}^{(n)}_{\beta_{n}}(t_n) \big].
\end{equation}
Consider now the coarse-grained history that arises from a combination
of the two maximally fine-grained histories $\mathcal{H}_{\{\alpha\}}$
and $\mathcal{H}_{\{\beta\}}$,
\begin{equation} \label{eq:hist-combined}
\mathcal{H}_{\{\alpha \lor \beta\}} = \{
\widehat{P}^{(1)}_{\alpha_1}(t_1)+ \widehat{P}^{(1)}_{\beta_1}(t_1),\,
\widehat{P}^{(2)}_{\alpha_2}(t_2) + \widehat{P}^{(2)}_{\beta_2}(t_2),\,
\dots,  \widehat{P}^{(n)}_{\alpha_n}(t_n) +
\widehat{P}^{(n)}_{\beta_n}(t_n) \}.   
\end{equation}
We interpret each combined projection operator
$\widehat{P}^{(i)}_{\alpha_i}(t_i)+ \widehat{P}^{(i)}_{\beta_i}(t_i)$
as stating that, at time $t_i$, the system was in the range described
by the union of $\widehat{P}^{(i)}_{\alpha_i}(t_i)$ and
$\widehat{P}^{(i)}_{\beta_i}(t_i)$. Accordingly, we would like to
require that the probability for a history containing such a combined
projection operator be equivalently calculable from the sum of the
probabilities of the two histories containing the individual
projectors $\widehat{P}^{(i)}_{\alpha_i}(t_i)$ and
$\widehat{P}^{(i)}_{\beta_i}(t_i)$, that is,
\begin{equation}
\begin{split}
\text{Tr} \big[ &
\widehat{P}^{(n)}_{\alpha_n}(t_n) 
\cdots \bigl(\widehat{P}^{(i)}_{\alpha_i}(t_i)+
\widehat{P}^{(i)}_{\beta_i}(t_i)\bigr) \cdots 
\widehat{P}^{(1)}_{\alpha_1}(t_1) \rho_0 \nonumber 
\widehat{P}^{(1)}_{\alpha_1}(t_1) \cdots \bigl(\widehat{P}^{(i)}_{\alpha_i}(t_i)+
\widehat{P}^{(i)}_{\beta_i}(t_i)\bigr) \cdots
\widehat{P}^{(n)}_{\alpha_n}(t_n) \big] \nonumber \\ &\stackrel{!}{=}
\text{Tr} \big[
\widehat{P}^{(n)}_{\alpha_n}(t_n) 
\cdots \widehat{P}^{(i)}_{\alpha_i}(t_i) \cdots 
\widehat{P}^{(1)}_{\alpha_1}(t_1) \rho_0\nonumber 
\widehat{P}^{(1)}_{\alpha_1}(t_1) \cdots \widehat{P}^{(i)}_{\alpha_i}(t_i) \cdots
\widehat{P}^{(n)}_{\alpha_n}(t_n)  \big] \nonumber \\ &\quad +
\text{Tr} \big[
\widehat{P}^{(n)}_{\alpha_n}(t_n) 
\cdots \widehat{P}^{(i)}_{\beta_i}(t_i) \cdots 
\widehat{P}^{(1)}_{\alpha_1}(t_1) \rho_0 \nonumber 
\widehat{P}^{(1)}_{\alpha_1}(t_1) \cdots \widehat{P}^{(i)}_{\beta_i}(t_i) \cdots
\widehat{P}^{(n)}_{\alpha_n}(t_n) \big].
\end{split}
\end{equation}
It can be easily shown that this relation holds if and only if
\begin{equation}
\text{Re} \big\{ \text{Tr} \big[
\widehat{P}^{(n)}_{\alpha_n}(t_n) 
\cdots \widehat{P}^{(i)}_{\alpha_i}(t_i) \cdots 
\widehat{P}^{(1)}_{\alpha_1}(t_1) \rho_0 
\widehat{P}^{(1)}_{\alpha_1}(t_1) \cdots \widehat{P}^{(i)}_{\beta_i}(t_i) \cdots
\widehat{P}^{(n)}_{\alpha_n}(t_n)  \big] \big\} = 0 \quad \text{if $\alpha_i \not= \beta_i$}.
\end{equation}
\end{widetext}
\noindent Generalizing this two-projector case to the coarse-grained history
$\mathcal{H}_{\{\alpha \lor \beta\}}$ of Eq.~\eqref{eq:hist-combined},
we arrive at the (sufficient and necessary) {\em consistency condition}
for two histories $\mathcal{H}_{\{\alpha\}}$ and
$\mathcal{H}_{\{\beta\}}$
\citep{Griffiths:1984:tr,Omnes:1990:ww,Omnes:1992:gy},
\begin{equation} \label{eq:cons-weak}
\text{Re}[\mathcal{D}(\alpha,\beta)] = \delta_{\alpha,\beta}
\mathcal{D}(\alpha,\alpha).
\end{equation}
If this relation is violated, the usual sum rule for calculating
probabilities does not apply. This situation arises when quantum
interference between the two combined histories
$\mathcal{H}_{\{\alpha\}}$ and $\mathcal{H}_{\{\beta\}}$ is present.
Therefore, to ensure that the standard laws of probability theory also
hold for coarse-grained histories, the set $\mathfrak{H}$ of possible
histories must be consistent in the above sense.

However, \citet{GellMann:1990:uz} have pointed out that when
decoherence effects are included to model the emergence of
classicality, it is more natural to require
\begin{equation} \label{eq:cons-med}
\mathcal{D}(\alpha,\beta) = \delta_{\alpha,\beta}
\mathcal{D}(\alpha,\alpha).
\end{equation}
Condition \eqref{eq:cons-weak} has often been referred to as {\em weak
  decoherence}, and Eq.~\eqref{eq:cons-med} as {\em medium
  decoherence} \citep[for a proposal of a criterion for {\em strong
  decoherence}, see][]{GellMann:1998:xy}. The set $\mathfrak{H}$ of
histories is called consistent (or decoherent) when all its members
$\mathcal{H}_{\{\alpha\}}$ fulfill the consistency condition,
Eqs.~\eqref{eq:cons-weak} or \eqref{eq:cons-med}, i.e., when they can
be regarded as independent (noninterfering).

\subsubsection{Selection of histories and classicality}

Even when the stronger consistency criterion \eqref{eq:cons-med} is
imposed on the set $\mathfrak{H}$ of possible histories, the number of
mutually incompatible consistent histories remains relatively large
\citep{Espagnat:1989:fl,Dowker:1996:ch}. It is not at all clear
\emph{a priori} that at least some of these histories should represent
any meaningful set of propositions about the world of our observation.
Even if a collection of such ``meaningful'' histories is found, it
leaves open the question how to select such histories and which
additional criteria would need to be invoked.

Griffith's \citeyearpar{Griffiths:1984:tr} original aim in formulating
the consistency criterion was only to allow for a consistent
description of sequences of events in closed quantum systems without
running into logical contradictions.\footnote{However,
  \citet{Goldstein:1998:yu} used a simple example to argue that the
  consistent-histories approach can lead to contradictions with
  respect to a combinination of joint probabilities, even if the
  consistency criterion is imposed; see also the subsequent exchange
  of letters in the February 1999 issue of {\em Physics Today}.}
Commonly, however, consistency has also been tied to the emergence of
classicality. For example, the consistency criterion corresponds to
the demand for the absence of quantum interference---a property of
classicality---between two combined histories.  It has become clear
that most consistent histories are in fact flagrantly nonclassical
\citep{GellMann:1990:uz,GellMann:1991:pp,Zurek:1993:pu,Paz:1993:ww,%
  Albrecht:1993:pq,Dowker:1995:pa,Dowker:1996:ch}.  For instance, when
the projection operators $\widehat{P}^{(i)}_{\alpha_i}(t_i)$ are
chosen to be the time-evolved eigenstates of the initial density
matrix $\rho(t_0)$, the consistency condition will automatically be
fulfilled, yet the histories composed of these projection operators
have been shown to result in highly nonclassical macroscopic
superpositions when applied to standard examples such as quantum
measurement or Schr\"odinger's cat. This demonstrates that the
consistency condition cannot serve as a sufficient criterion for
classicality.

\subsubsection{Consistent histories of open systems}

Various authors have appealed to  interaction with the environment
and the resulting decoherence effects in defining additional criteria
that would select quasiclassical histories and would also lead to a
physical motivation for the consistency criterion \citep[see, for
example,][]{GellMann:1990:uz,Dowker:1992:vz,Albrecht:1992:rz,%
Albrecht:1993:pq,Zurek:1993:pu,Paz:1993:ww,Twamley:1993:bz,%
Finkelstein:1993:gc,Anastopoulos:1996:kl,GellMann:1998:xy,Halliwell:2001:qp}.
This approach intrinsically requires the notion of local, open systems
and the split of the universe into subsystems, in contrast to the
original aim of the consistent-histories approach to describe the
evolution of a single closed, undivided system (typically the entire
universe). The decoherence-based studies then assume the usual
decomposition of the total Hilbert space $\mathcal{H}$ into a space
$\mathcal{H}_\mathcal{S}$, corresponding to the system $\mathcal{S}$,
and $\mathcal{H}_\mathcal{E}$ of an environment $\mathcal{E}$. One
then describes the histories of the system $\mathcal{S}$ by employing
projection operators that act only on the system, i.e., that are of
the form $\widehat{P}^{(i)}_{\alpha_i}(t_i) \otimes
\widehat{I}_\mathcal{E}$, where $\widehat{P}^{(i)}_{\alpha_i}(t_i)$
acts only on $\mathcal{H}_\mathcal{S}$ and $\widehat{I}_\mathcal{E}$
is the identity operator in $\mathcal{H}_\mathcal{E}$.

This raises the question of when, i.e., under which circumstances, the
reduced density matrix $\rho_\mathcal{S} = \text{Tr}_\mathcal{E}
\rho_\mathcal{SE}$ of the system $\mathcal{S}$ suffices to calculate
the decoherence functional. The reduced density matrix arises from a
nonunitary trace over $\mathcal{E}$ at every time point $t_i$, whereas
the decoherence functional of Eq.~\eqref{eq:dfs} employs the full,
unitarily evolving density matrix $\rho_\mathcal{SE}$ for all times
$t_i < t_f$ and only applies a nonunitary trace operation (over both
$\mathcal{S}$ and $\mathcal{E}$) at the final time $t_f$.
\citet{Paz:1993:ww} have answered this (rather technical) question by
showing that the decoherence functional can be expressed entirely in
terms of the reduced density matrix if the time evolution of the
reduced density matrix is independent of the correlations dynamically
formed between the system and the environment. A necessary (but not
always sufficient) condition for this requirement to be satisfied is
given by demanding that the reduced dynamics be governed by a master
equation that is local in time.

When a ``reduced'' decoherence functional exists, at least to a good
approximation, i.e., when the reduced dynamics are sufficiently
insensitive to the formation of system-environment correlations, the
consistency of whole-universe histories, described by a unitarily
evolving density matrix $\rho_\mathcal{SE}$ and sequences of
projection operators of the form $\widehat{P}^{(i)}_{\alpha_i}(t_i)
\otimes \widehat{I}_\mathcal{E}$, will be directly related to that of
open-system histories, represented by a nonunitarily evolving reduced
density matrix $\rho_\mathcal{S}(t_i)$ and ``reduced'' projection
operators $\widehat{P}^{(i)}_{\alpha_i}(t_i)$ \citep{Zurek:1993:pu}.

\subsubsection{Schmidt states vs pointer basis as projectors}

The ability of the instantaneous eigenstates of the reduced density
matrix (Schmidt states; see also Sec.~\ref{sec:schmidt}) to serve as
projectors for consistent histories and possibly to lead to the
emergence of quasiclassical histories has been studied in much detail
\citep{Albrecht:1992:rz,Albrecht:1993:pq,Zurek:1993:pu,Paz:1993:ww,Kent:1997:oz}.
\citet{Paz:1993:ww} have shown that Schmidt projectors
$\widehat{P}^{(i)}_{\alpha_i}$, defined by their commutativity with
the evolved, path-projected reduced density matrix, 
\begin{multline} 
\big[\widehat{P}^{(i)}_{\alpha_i}, U(t_{i-1},t_i) \{ \cdots
U(t_1,t_2) \widehat{P}^{(1)}_{\alpha_1} \rho_\mathcal{S}(t_1)
\\ \times \, \widehat{P}^{(1)}_{\alpha_1} U^\dagger(t_1,t_2) \cdots \}
U^\dagger(t_{i-1},t_i)\big] = 0, 
\end{multline}
will always give rise to an infinite number of sets of consistent
histories (``Schmidt histories''). However, these histories are
branchdependent [see Eq.~\eqref{eq:hist-fine-branch}] and usually
extremely unstable, since small modifications of the time sequence
used for the projections (for instance by deleting a time point) will
typically lead to drastic changes in the resulting history, indicating
that Schmidt histories are usually very nonclassical
\citep{Zurek:1993:pu,Paz:1993:ww}.

This situation is changed when the time points $t_i$ are chosen such
that the intervals $(t_{i+1}-t_i)$ are larger than the typical
decoherence time $\tau_D$ of the system over which the reduced density
matrix becomes approximately diagonal in the preferred pointer basis
chosen through environment-induced superselection (see also the
discussion in Sec.~\ref{sec:schmidt}). When the resulting pointer
states, rather than the instantaneous Schmidt states, are used to
define the projection operators, stable quasiclassical histories will
typically emerge \citep{Zurek:1993:pu,Paz:1993:ww}.  In this sense, it
has been suggested that interaction with the environment can
provide the missing selection criterion that ensures the
quasiclassicality of histories, i.e., their stability (predictability),
and the correspondence of the projection operators (the pointer basis)
to the preferred determinate quantities of our experience.

The approximate noninterference, and thus consistency, of histories
based on local density operators (energy, number, momentum, charge
etc.) as quasiclassical projectors \citep[the so-called {\em
  hydrodynamic observables},
see][]{GellMann:1991:pp,Dowker:1992:vz,Halliwell:1998:fg} has been
attributed to the dynamical stability exhibited by the eigenstates of
the local density operators. This stability leads to decoherence in
the corresponding basis \citep{Halliwell:1998:fg,Halliwell:1999:ii}.
It has been argued by \citet{Zurek:2002:ii} that this behavior and
thus the special quasiclassical properties of hydrodynamic observables
can be explained by the fact that these observables obey the
commutativity criterion, Eq.~\eqref{eq:commut}, of the
environment-induced superselection approach.

\subsubsection{Exact vs approximate consistency}

In the idealized case where the pointer states lead to an exact
diagonalization of the reduced density matrix, histories composed of
the corresponding {\em pointer projectors} will automatically be
consistent. However, under realistic circumstances decoherence will
typically lead only to approximate diagonality in the pointer basis.
This implies that the consistency criterion will not be fulfilled
exactly and that hence the probability sum rules will only hold
approximately---although usually, due to the efficiency of
decoherence, to a very good approximation
\citep{Griffiths:1984:tr,GellMann:1991:pp,Omnes:1992:gy,%
Omnes:1994:pz,Albrecht:1992:rz,Albrecht:1993:pq,Zurek:1993:pu,%
Paz:1993:ww,Twamley:1993:bz}.  Hence, the consistency
criterion has been viewed both as overly restrictive, since the
quasiclassical pointer projectors rarely obey the consistency
equations exactly, and as insufficient, because it does not give rise
to constraints that would single out quasiclassical histories.

A relaxation of the consistency criterion has therefore been
suggested, leading to ``approximately consistent histories'' whose
decoherence functional would be allowed to contain nonvanishing
off-diagonal terms (corresponding to a violation of the probability
sum rules) as long as the net effect of all the off-diagonal terms was
``small'' in the sense of remaining below the experimentally
detectable level \citep[see, for
example,][]{GellMann:1991:pp,Dowker:1992:vz}.
\citet{GellMann:1991:pp} have even ascribed a fundamental role to such
approximately consistent histories, a move that has sparked much
controversy and has been considered as unnecessary and irrelevant by
some \citep{Dowker:1995:pa,Dowker:1996:ch}. Indeed, if only
approximate consistency is demanded, it is difficult to regard this
condition as a fundamental concept of a physical theory, and the
question of how much consistency is required will inevitably arise.

\subsubsection{Consistency and environment-induced superselection}

The relationship between consistency and environment-induced
superselection, and therefore the connection between the decoherence
functional and the diagonalization of the reduced density matrix
through environmental decoherence, has been investigated by various
authors. The basic idea, promoted, for example, by
\citet{Zurek:1993:pu} and \citet{Paz:1993:ww}, is to suggest
that if the interaction with the environment leads to rapid
superselection of a preferred basis, which approximately diagonalizes
the local density matrix, coarse-grained histories defined in this
basis will automatically be (approximately) consistent.

This approach has been explored by \citet{Twamley:1993:bz}, who
carried out detailed calculations in the context of a quantum optical
model of phase-space decoherence and compared the results with
two-time projected phase-space histories of the same model system. It
was found that when the parameters of the interacting environment were
changed such that the degree of diagonality of the reduced density
matrix in the emerging preferred pointer basis was increased,
histories in that basis also became more consistent. For a similar
model, however, \citet{Twamley:1993:cg} also showed that consistency
and diagonality in a pointer basis as possible criteria for the
emergence of quasiclassicality may exhibit a very different dependence
on the initial conditions.

Extensive studies on the connection between Schmidt states, pointer
states and consistent quasiclassical histories have also been made by
\citet{Albrecht:1992:rz,Albrecht:1993:pq}, based on analytical
calculations and numerical simulations of toy models for decoherence,
including detailed numerical results on the violation of the sum rule
for histories composed of different (Schmidt and pointer) projector
bases. It was demonstrated that the presence of stable
system-environment correlations (``records''), as demanded by the
criterion for the selection of the pointer basis, was of great
importance in making certain histories consistent. The relevance of
``records'' for the consistent-histories approach in ensuring the
``permanence of the past'' has also been emphasized by other authors,
for example, by \citet{Paz:1993:ww} and
\citet{Zurek:1993:pu,Zurek:2002:ii}, and in the ``strong decoherence''
criterion by \citet{GellMann:1998:xy}.  The redundancy with which
information about the system is recorded in the environment and can
thus be found out by different observers without measurably disturbing
the system itself has been suggested to allow for the notion of
``objectively existing histories,'' based on environment-selected
projectors that represent sequences of ``objectively existing''
quasiclassical events
\citep{Zurek:1993:pu,Paz:1993:ww,Zurek:2002:ii,Zurek:2003:pl}.

In general, damping of quantum coherence caused by decoherence will
necessarily lead to a loss of quantum interference between individual
histories \citep[but not vice versa---see the discussion
by][]{Twamley:1993:bz}, since the final trace operation over the
environment in the decoherence functional will make the off-diagonal
elements very small due to the decoherence-induced approximate mutual
orthogonality of the environmental states.
\citet{Finkelstein:1993:gc} has used this observation to propose a new
decoherence condition that coincides with the original definition,
Eqs.~\eqref{eq:df} and \eqref{eq:dfs}, except for restricting the
trace to $\mathcal{E}$, rather than tracing over both $\mathcal{S}$
and $\mathcal{E}$. It was shown that this condition not only implies
the consistency condition of Eq.~\eqref{eq:cons-med}, but also
characterizes those histories that decohere due to interaction with
the environment and that lead to the formation of ``records'' of the
state of the system in the environment.

\subsubsection{Summary and discussion}

The core difficulty associated with the consistent-histories approach
has been to explain the emergence of the classical world of our
experience from the underlying quantum nature. Initially, it was hoped
that classicality could be derived from the consistency criterion
alone.  Soon, however, the status and the role of this criterion in
the formalism and its proper interpretation became rather unclear and
diffuse, since exact consistency was shown to provide neither a
necessary nor a sufficient criterion for the selection of
quasiclassical histories.

The inclusion of decoherence effects into the consistent histories
approach, leading to the emergence of stable quasiclassical pointer
states, has been found to yield a highly efficient mechanism and a
sensitive criterion for singling out quasiclassical observables that
simultaneously fulfill the consistency criterion to a very good
approximation due to the suppression of quantum coherence in the state
of the system. The central question is then: What is the meaning and
the remaining role of an explicit consistency criterion in the light
of such ``natural'' mechanisms for the decoherence of histories? Can
one dispose of this criterion as a key element of the fundamental
theory by noting that for all ``realistic'' histories consistency will
be likely to arise naturally from environment-induced decoherence
alone?

The answer to this question may actually depend on the viewpoint one
takes with respect to the aim of the consistent-histories approach. As
we have noted before, the original goal was simply to provide a
formalism in which one could, in a measurement-free context, assign
probabilities to certain sequences of quantum events without logical
inconsistencies. The more recent and rather opposite aim would be to
provide a formalism that selects only a very small subset of
``meaningful'' quasiclassical histories, all of which are consistent
with our world of experience, and whose projectors can be directly
interpreted as objective physical events.

The consideration of decoherence effects that can give rise to
effective superselection of possible quasiclassical (and consistent)
histories certainly falls into the latter category. It is interesting
to note that this approach has also led to a departure from the
original ``closed systems only'' view to the study of local open
quantum systems and thus to the decomposition of the total Hilbert
space into subsystems, within the consistent-histories formalism.
Besides the fact that decoherence intrinsically requires the openness
of systems, this move might also reflect the insight that the notion
of classicality itself can be viewed as only arising from a conceptual
division of the universe into parts (see the discussion in
Sec.~\ref{sec:division}).

Therefore environment-induced decoherence and superselection have
played a remarkable role in consistent-histories interpretations: a
practical one by suggesting a physical selection mechanism for
quasiclassical histories; and a conceptual one by contributing to a
shift in our view of originally rather fundamental concepts, such as
consistency, and of the aims of the consistent-histories approach,
like the focus on description of closed systems.

\section{Concluding remarks}

We have presented an extensive discussion of the role of decoherence
in the foundations of quantum mechanics, with a particular focus on
the implications of decoherence for the measurement problem in the
context of various interpretations of quantum mechanics.

A key achievement of the decoherence program is the recognition that
openness in quantum systems is important for their realistic
description. The well-known phenomenon of quantum entanglement had
already, early in the history of quantum mechanics, demonstrated that
correlations between systems can lead to ``paradoxical'' properties of
the composite system that cannot be composed from the properties of
the individual systems.  Nonetheless, the viewpoint of classical
physics that the idealization of isolated systems is necessary to
arrive at an ``exact description'' of physical systems has influenced
quantum theory for a long time.  It is the great merit of the
decoherence program to have emphasized the ubiquity and essential
inescapability of system-environment correlations and to have
established the important role of such correlations as factors in the
emergence of ``classicality'' from quantum systems. Decoherence also
provides a realistic physical modeling and a generalization of the
quantum measurement process, thus enhancing the ``black-box'' view of
measurements in the standard (``orthodox'') interpretation and
challenging the postulate of fundamentally classical measuring devices
in the Copenhagen interpretation.

With respect to the preferred-basis problem of quantum measurement,
decoherence provides a very promising definition of preferred pointer
states via a physically meaningful requirement, namely, the robustness
criterion, and it describes methods for selecting operationally such
states, for example, via the commutativity criterion or by extremizing
an appropriate measure such as purity or von Neumann entropy.  In
particular, the fact that macroscopic systems virtually always
decohere into position eigenstates gives a physical explanation for
why position is the ubiquitous determinate property of the world of
our experience.

We have argued that, within the standard interpretation of quantum
mechanics, decoherence cannot solve the problem of definite outcomes
in quantum measurement: We are still left with a multitude of (albeit
individually well-localized quasiclassical) components of the wave
function, and we need to supplement or otherwise to interpret this
situation in order to explain why and how single outcomes are
perceived. Accordingly, we have discussed how environment-induced
superselection of quasiclassical pointer states together with the
local suppression of interference terms can be put to great use in
physically motivating, or potentially disproving, rules and
assumptions of alternative interpretive approaches that change (or
altogether abandon) the strict orthodox eigenvalue-eigenstate link
and/or modify the unitary dynamics to account for the perception of
definite outcomes and classicality in general. For example, to name
just a few applications, decoherence can provide a universal criterion
for the selection of the branches in relative-state interpretations
and a physical argument for the noninterference between these branches
from the point of view of an observer; in modal interpretations, it
can be used to specify empirically adequate sets of properties that
can be ascribed to systems; in collapse models, the free parameters
(and possibly even the nature of the reduction mechanism itself) might
be derivable from environmental interactions; decoherence can also
assist in the selection of quasiclassical particle trajectories in
Bohmian mechanics, and it can serve as an efficient mechanism for
singling out quasiclassical histories in the consistent-histories
approach. Moreover, it has become clear that decoherence can ensure
the empirical adequacy and thus empirical equivalence of different
interpretive approaches, which has led some to the claim that the
choice, for example, between the orthodox and the Everett
interpretation becomes ``purely a matter of taste, roughly equivalent
to whether one believes mathematical language or human language to be
more fundamental'' \cite[p.~855]{Tegmark:1998:qq}.

It is fair to say that the decoherence program sheds new light on many
foundational aspects of quantum mechanics.  It paves a physics-based
path towards motivating solutions to the measurement problem; it
imposes constraints on the strands of interpretations that seek such a
solution and thus makes them also more and more similar to each other.
Decoherence remains an ongoing field of intense research, in both the
theoretical and experimental domain, and we can expect further
implications for the foundations of quantum mechanics from such
studies in the near future.

\begin{acknowledgments}

The author would like to thank A.~Fine for many valuable discussions
and comments throughout the process of writing this article.  He
gratefully acknowledges thoughtful and extensive feedback on this
manuscript from S.~L.~Adler, V.~Chaloupka, H.-D.~Zeh, and W.~H.~Zurek.

\end{acknowledgments}

\bibliographystyle{apsrmp}

\bibliography{references}

\end{document}